\documentclass[%
aip,
 amsmath,amssymb,
preprint,%
]{revtex4-1}
\usepackage{accents}
\usepackage{xcolor}
\usepackage{graphicx}% Include figure files
\usepackage{dcolumn}% Align table columns on decimal point
\usepackage{bm}% bold math
\usepackage{appendix}
\usepackage{color}
\usepackage{float}
\usepackage{subfig}
 \usepackage{siunitx}

\usepackage[a4paper,top=3cm,bottom=2cm,left=3cm,right=3cm,marginparwidth=1.75cm]{geometry}

\usepackage{amsmath,amssymb,amsthm}
\usepackage{graphicx}
\usepackage[colorlinks=true, allcolors=blue]{hyperref}

\begin{document}

%\preprint{APS/123-QED}

   \title{Ion kinetic  effects and instabilities in the plasma flow in the magnetic mirror}

\author{M. Jimenez}%
 \email{marilynj8j8@hotmail.com }
\affiliation{%
 University of Saskatchewan, Saskatchewan, Saskatoon SK S7N 5E2, Canada 
}%

\author{A. I. Smolyakov}%
 %\email{andrei.smolyakov@usask.ca}
\affiliation{%
University of Saskatchewan, Saskatchewan, Saskatoon SK S7N 5E2, Canada 
}%

\author{O. Chapurin}%
 %\email{ }
\affiliation{%
 University of Saskatchewan, Saskatchewan, Saskatoon SK S7N 5E2, Canada 
}%

\author{P. Yushmanov}%
\affiliation{%
 TAE Technologies, 19631 Pauling, Foothill Ranch, CA, 92610, United States 
}%

\begin{abstract}
Kinetic effects in plasma flow due to a finite ion temperature and ion reflections  in a converging-diverging magnetic nozzle are investigated with collisionless quasineutral hybrid simulations with kinetic ions and isothermal Boltzmann electrons. It is shown that in the cold ions limit the velocity profile of the particles agrees well with the analytical theory predicting the formation of the global accelerating potential due to the magnetic mirror with the maximum of the magnetic field and resulting in the transonic ion velocity profile. The global transonic ion velocity profile is  also obtained for warm ions with isotropic and anisotropic distributions. Partial ion reflections are observed due to a combined effect of the magnetic mirror and time-dependent fluctuations of the potential as a result of the wave breaking and instabilities in the regions when the fluid  solutions become multi-valued. Despite partial reflections, the flow of the passing ions still follows the global accelerating profile defined by the magnetic field profile. In simulations with  reflecting boundary condition imitating the plasma source and allowing the transitions between trapped and passing ions, the global nature of the transonic accelerating solution is revealed as a constrain on the plasma exhaust velocity that ultimately defines plasma density in the source region.

\end{abstract}

\maketitle

%\preprint{APS/123-QED}

%Use showkeys class option if keyword

% \title {Plasma flow and ion acceleration in the magnetic nozzle of the Variable Specific 
% Impulse Magnetoplasma Rocket (VASIMR) and NORMAN devices.}
\title {Kinetic effects in plasma flow in the magnetic nozzle} 

\maketitle

\section{Introduction}

Magnetic nozzle  in converging-diverging configuration is used in many applications, including electric propulsion \cite{diaz2000vasimr,eberson2012magnetic,MerinoPSST2020} and mirror  fusion devices \cite{onofri2017magnetohydrodynamic,BinderbauerAIP2016}.  In case of fully magnetized plasma,  magnetic field forms  the  magnetic nozzle that accelerates plasma  similar to Laval nozzle by the conversion of thermal  plasma energy into   the kinetic energy of the directed flow. This process is  used to create thrust in space propulsion applications. In open mirror fusion devices \cite{MirnovNF1972} the converging-diverging magnetic field is used to partially confine the plasma, while at the same time, the diverging (expander) region of the magnetic mirror is used as a divertor to spread the exhaust energy over the  larger area \cite{RyutovAIP2016}.
 
Physical processes involved in plasma acceleration and flows in magnetic nozzle  configurations have been discussed in different  contexts \cite{ArefievPoP2008,AhedoPSST2020,BoldyrevPNAS2020,WethertonPoP2021}, but the topic remains an active research area.  In this study,   we confirm with kinetic simulations our earlier analytical results \cite{smolyakov2021quasineutral,SaboPoP2022} that the accelerating
ion velocity profile obtained in fluid theory is indeed a global robust solution that persist in  presence of  kinetic effects. We also reveal a new mechanism for instabilities and fluctuations in the flow of plasma accelerated by the magnetic nozzle due to the kinetic effects of wave breaking and ion trapping.  
 
Fluid solutions for plasma flow in the magnetic nozzle \cite{smolyakov2021quasineutral}, demonstrate the necessity of a magnetic barrier with a maximum of the magnetic field to achieve acceleration and a smooth transition through ion sound velocity in the quasineutral plasma. It was also shown that the unique global solution is fully determined by the  regularization of the sonic singular point  where the plasma flow velocity becomes equal to the ion sound velocity, $V_\Vert=c_s$,  so that the smooth transition is possible without a shock. Such global accelerating solution is fully defined by the magnetic field profile. This behaviour is similar to the gas acceleration in Laval nozzle as well as the solar wind acceleration (Parker thermal wind solution) where the combined effect of  gravity force and radial expansion effectively create the converging-diverging  configuration   \cite{ParkerAJ1958,TajimaBook2002}.  

\begin{figure}[H]
    \centering
    \includegraphics[height=0.4\textwidth]{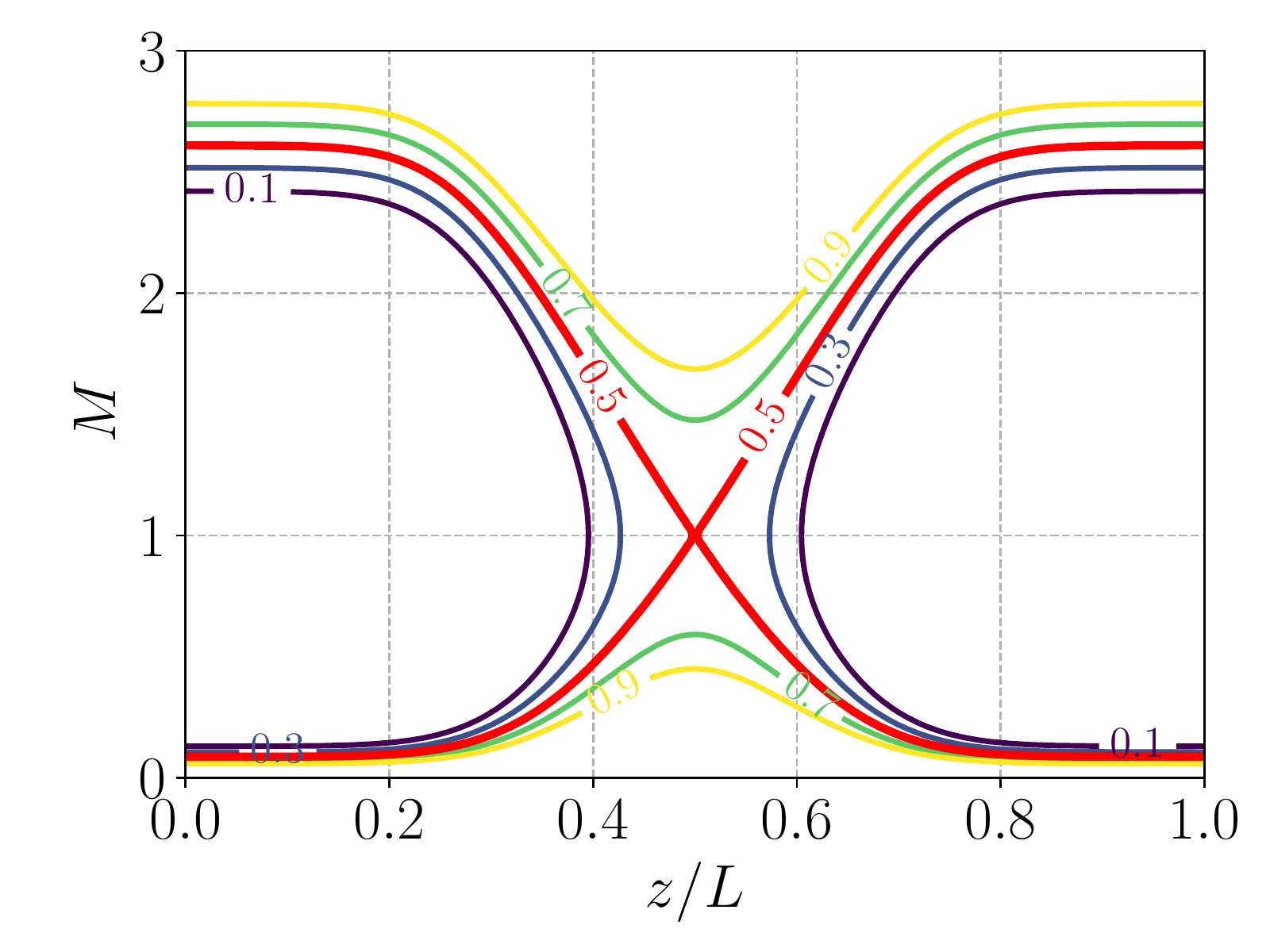}
    \caption{Fluid solutions diagram for cold ions from Eq.~(\ref{fluid_mach}) for the magnetic field in  Eq.~(\ref{mag_field}), $R=7$. The separatrices correspond to the accelerating and decelerating solutions with $M_a =  0.0869$ and $M_b =  2.6096$.}
    \label{fluid_solu}
\end{figure}

To formulate the goals of our study here, we briefly review the main results of the  fluid theory  of collisionless flow of fully magnetized plasma in the magnetic mirror, which is nearly isomorphic to the gas flow in the Laval nozzle.
Fluid equations for cold magnetized  ions and isothermal electrons can be integrated  \cite{ManheimerIEEE2001,FruchtmanPRL2006,smolyakov2021quasineutral} resulting in the equation  
 \begin{eqnarray}\label{fluid_mach}
\frac{M^2}{2} = \ln{\left(M\frac{ B_{max}}{B(z)}\right)}+C_0,
\end{eqnarray}
where $M = v_{iz}/c_s$ is the Mach number for the ion flow in the $z$-direction, $c_s^2=T_e/m_i$, $T_e$ is the electron temperature assumed uniform. 
The exact solutions of Eq.~(\ref{fluid_mach}) can be  written \cite{smolyakov2021quasineutral} in terms of the Lambert function \cite{DubinovJPP2005}.  General diagram of the solutions of this equation for various values of $C_0$ is shown in Fig.~\ref{fluid_solu}. For $C_0=1/2$ one obtains two global
(accelerating and decelerating) solutions corresponding to the separatrices in Fig.~\ref{fluid_solu}. 
 
\begin{figure}[H]
    \centering
    \includegraphics[width=0.49\linewidth]{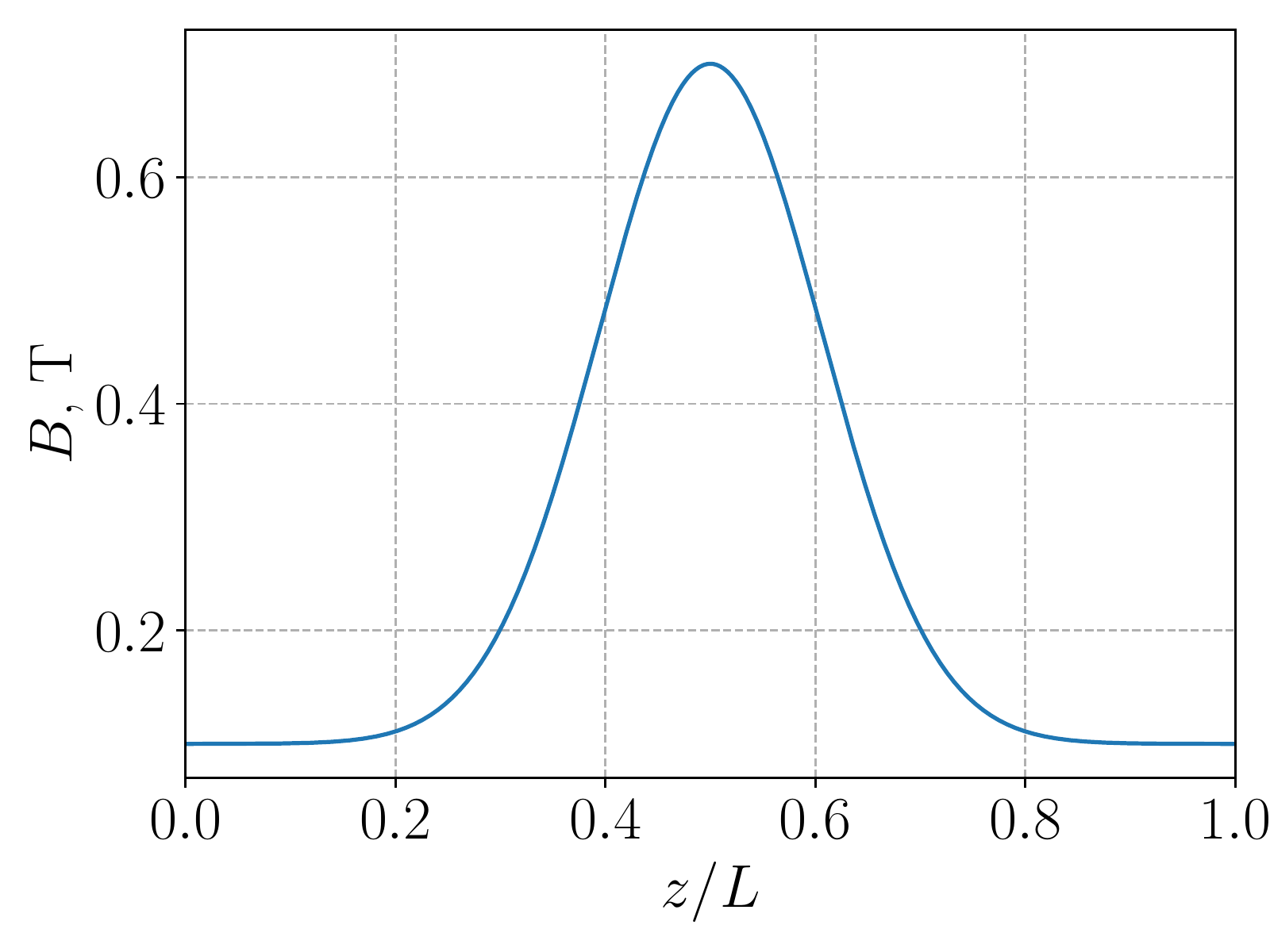}
    \caption{Converging-diverging magnetic field profile, with mirror ratio $R=7$, given by Eq.~(\ref{mag_field}).}
    \label{fig:profileB0}
\end{figure}

 As a model of the magnetic mirror, we have used here a symmetric magnetic field profile in the form:
\begin{equation}\label{mag_field}
    \mathrm{B(z)}= \mathrm{A} + \mathrm{B} \exp \frac{-(z-z_m)^2}{\delta^2 L^2}
\end{equation}
with the following parameters: $\mathrm{B}=\SI{0.6}{T}$, $\mathrm{A}=\SI{0.1}{T}$, $\delta=0.15$ $z_m=L/2$, where $L$ is the system length, $z=[0\ldots1]$, as  shown in  Fig.~\ref{fig:profileB0}. The magnetic mirror ratio for these parameters $R \equiv \mathrm{B}_{max}/\mathrm{B}(0) = 7$.

The global accelerating (decelerating) solutions are unique and have fixed values of the velocity at the boundaries;   $v/c_s=M_a$ for  $z=0$ and $v/c_s=M_b$ for  $z=L$.  
For a given profile of the magnetic field in Eq.~\ref{fluid_mach}, the values of plasma velocity for accelerating profile are $M_a =  0.0869$ and $M_b =  2.6096$. 
 For symmetric magnetic field profile, the decelerating and accelerating solutions are  symmetric. In what follows, the $z$ index for the ion velocity is omitted, and we will use $v_0$ to denote the initial ion velocity at the left boundary at  $z=0$.  The value of the ion velocity at the nozzle entrance fully defines the global solution across the nozzle region. The initial velocities above and below $v_0/c_s<M_a$, $v_0/c_s>M_b$ produce   respectively solutions that stay subsonic and supersonic  in the whole region. The formal solutions with values of  $v_0$ in the interval   $M_a<v_0/c_s<M_b$ are multi-valued and cannot exist within the fluid model. 

The solutions shown in Fig.~\ref{fluid_solu} are obtained assuming cold ions.  Furthermore, it was shown, also with the fluid model  \cite{SaboPoP2022}, that finite ion pressure  modifies the accelerating solutions and that ion pressure anisotropy is important, e.g.\ the perpendicular pressure enhances the acceleration.      One simple question addressed in our paper is what happens if the ions are injected with the initial velocity in the ``forbidden'' region of  $M_a<v_0/c_s<M_b$ on the cold ion solution diagram in Fig.~\ref{fluid_solu}. The goal of this paper is to investigate the plasma acceleration in the magnetic nozzle using the kinetic theory and taking into account  the finite (energy) thermal spread and  effects of the anisotropy  of ions injected into the nozzle at $z=0$. We compare the results of the cold and warm ion fluid theory  with the kinetic results that include effects of the ion reflections by the magnetic barrier and electric field. 

This paper is organized as follows. In Section II we describe general features of  particle motion in a converging-diverging magnetic field including the self-consistent electrostatic potential  and describe our quasineutral hybrid kinetic model.  Section III presents the main results of this work demonstrating the formation of global accelerating profile and excitation of instabilities due to the wave breaking  in different conditions and for different  distribution function of the injected ions. In Section IV, we discuss effects of the reflecting  wall imitating  particle mixing in the  real source and allowing transitions between trapped and passing particle. Summary and discussion are provided in Section V.  

\section{Basic properties of particle motion in the magnetic mirror and quasineutral hybrid model}\label{sec2}

In our model, ions are assumed magnetized
in the so called drift-kinetic limit when the time scale of the considered  processes are much slower than the ion  gyrofrequency,  $\omega \ll \omega_{ci}$ and spatial scales are  larger than the Larmor radius, $k_\perp \ll 1/\rho_i$. 
With these assumptions, we describe the  ion dynamics by the drift kinetic equation in the form   \cite{Sivukhin}
\begin{equation}\label{vlasov-ions}
\frac{\partial f}{\partial t} + \nabla \cdot \left( \frac{ d \mathbf{R}}{ dt } f\right)+ \frac{\partial}{\partial v_{\Vert}}\left( \frac{d v_{\Vert}}{dt} f \right)+       \frac{\partial}{\partial v_{\perp}^2}\left( \frac{d {v_{\perp}}^2}{dt} f \right)=0,
\end{equation}
where $f = f(\mathbf{R},v_\Vert,v_{\perp})$. The drift equations for particle motion and evolution of $v_{\Vert }$ and $v_{\bot }^{2}$ are
\begin{equation}
\frac{d\mathbf{R}}{dt}=v_{\Vert }\mathbf{b+}\frac{v_{\bot }^{2}/2+v_{\Vert
}^{2}}{\omega _{c}}\mathbf{b}\times \nabla \ln B  \label{dke},
\end{equation}%
\begin{equation}
\frac{dv_{\Vert }}{dt}=\frac{e}{m_i} E_\Vert+\frac{v_{\bot }^{2}}{2}\nabla \cdot \mathbf{b}=\frac{e}{m_i} E_\Vert-\frac{%
v_{\bot }^{2}}{2}\frac{\nabla _{\Vert }B}{B},
\end{equation}%
\begin{equation}
\frac{dv_{\bot }^{2}}{dt}=-v_{\bot }^{2}v_{\Vert }\nabla \cdot \mathbf{b}=v_{\bot }^{2}v_{\Vert }\frac{\nabla _{\Vert }B}{B},
\label{dvp}
\end{equation} where  $\mathbf{b}= \mathbf{B}/B$ is the unit vector along the magnetic field, and $\nabla_{\Vert}= \mathbf{b}\cdot \nabla $. Equations (\ref{dke}-\ref{dvp})  conserve  the phase space volume in the form
\begin{equation}
\nabla \cdot \frac{d\mathbf{R}}{dt}+\frac{\partial }{\partial v_{\Vert }}%
\frac{dv_{\Vert }}{dt}+\frac{\partial }{\partial v_{\bot }^{2}}\frac{%
dv_{\bot }^{2}}{dt}=0.
\end{equation}

 Near the axis of the magnetic mirror,  a slender flux tube can be considered in the paraxial  approximation so the problem becomes one-dimensional and  Eq.~(\ref{vlasov-ions}) takes the form
\begin{equation}\label{f-parax}
    \frac{\partial f}{\partial t}+ v_{z}\frac{\partial f}{\partial z}+ \frac{dv_{z}}{dt} \frac{\partial f}{\partial v_{z}}=0,
   \end{equation}
where $v_z$ is the particle velocity along the axis of the mirror (the z-direction). The equation of the ion motion  is 
\begin{equation}\label{mo_ions}
    \frac{dv_{z}}{dt}=\frac{e}{m_i}E_z-\frac{v_{\perp}^2(z)}{2 \mathrm{B}(z)}\frac{\partial \mathrm{B}}{\partial z},
\end{equation}
where $e$ is the absolute value of elementary charge, $m_i$ is the ion mass. Here $E_z = -\partial \phi / \partial z$ is the axial electric field,  and  the last term is the mirror force due to the inhomogeneous magnetic field, $\mathbf {B}=B_z(z) \mathbf{e_z}+B_r\mathbf{e_r}$, $B_r\ll B_z$. 
In the drift-kinetic approximation the ion magnetic moment is an adiabatic invariant: $\mu=m_iv_\perp^2/2B=\text{const}$. 

In absence of the electric field, the conservation of the magnetic moment and  energy  result in particles reflection by the magnetic mirror with the  well known condition for the loss cone in the $v_{\perp}-v_{z}$ space:
\begin{eqnarray}
    v_{z0}^2 > v_{\perp 0}^2 (R - 1),
\end{eqnarray}
where $R= B_m/B_0$, and $v_{\perp 0}$ and $v_{z0}$ are respectively the parallel and perpendicular velocities at the nozzle entrance.

In presence of the stationary electrostatic electric field the energy conservation for ions can be written in the form  
\begin{equation}
  \frac{m v_z^2}{2} = E_0 - (\mu B(z) + e\phi(z)) = E_0 - U_{Y}(z),
\end{equation}
where $E_0$ is the total energy, and $U_{Y}$  is the  effective potential energy, the so-called Yushmanov potential. 
The reflection occurs  when $v_z^2/2 = 0$, and the reflection point is found from $E_0 - U_{Y}(z^*) = 0$,
where $z^*$ is the location of the maximum $U_{Y}$ which depends on the magnetic moment value $\mu$. Then the condition for the  loss cone is modified:
\begin{eqnarray}\label{bound_cone}
    v_{z0}^2 > v_{\perp0}^2 ((B(z^*)/B_0)-1) + \frac{2e}{m} \phi(z^*),
\end{eqnarray}
For large values of the magnetic moment, the reflections  and the loss cone boundary are   dominated by the mirror force. For lower $\mu$, the electric field modifies the reflections.

The electric field has to be found self-consistently with the evolution of the ion and electron densities. We use a hybrid approach where the kinetic equation is solved for ions but electrons are fluid and assumed isothermal. The electron density  is given by Boltzmann relation that follows  from the inertialess  electron momentum balance equation  describing the equilibrium balance between the pressure gradient  and the electric field forces:
\begin{equation}\label{boltzmann}
    0= e n_e \frac{\partial \phi}{\partial z} - T_e\frac{\partial n_e}{\partial z}. 
\end{equation}

Equations (\ref{f-parax},\ref{mo_ions}, and \ref{boltzmann}) constitute our quasineutral hybrid kinetic model. Eq.~(\ref{f-parax}) is solved  via the particle-in-cell method by integrating along the characteristics according to Eq.~\ref{mo_ions}.  Particle density is obtained from the ion distribution function  $n = \int f dv_z 2 \pi v_{\perp} dv_{\perp}$, then the Eq.~(\ref{boltzmann}) is solved for the electric field, and all particles are advanced along the characteristics.
The effects of inhomogeneous magnetic field  are modeled with the cell area variation (magnetic flux conservation) \cite{ebersohn2017kinetic}.
 The ions  are injected from the left boundary with the specified distribution function thus imitating the particle source at $z=0$, Xenon mass is used as an example from plasma propulsion applications \cite{diaz2000vasimr}.   For all simulations the system length is $L$ = $\SI{3}{m}$, with a constant electron temperature $T_e=\SI{200}{eV}$. Since some particles are reflected back by the mirror and electric forces,  we consider two options. For absorbing boundary condition, we remove all particles returning to the boundary and continue  to inject new particles with a given distribution function and specified flux.  For reflecting boundary condition, all particles returning to the boundary are reflected back to the system. The reflections can be diffusive (re-injected particles velocities are randomly sampled), specular (mirror reflection), or their combination. Some other technical details of the simulations are given in Appendix A.

\begin{figure}[H]
    \centering
      \includegraphics[width=0.52\linewidth]{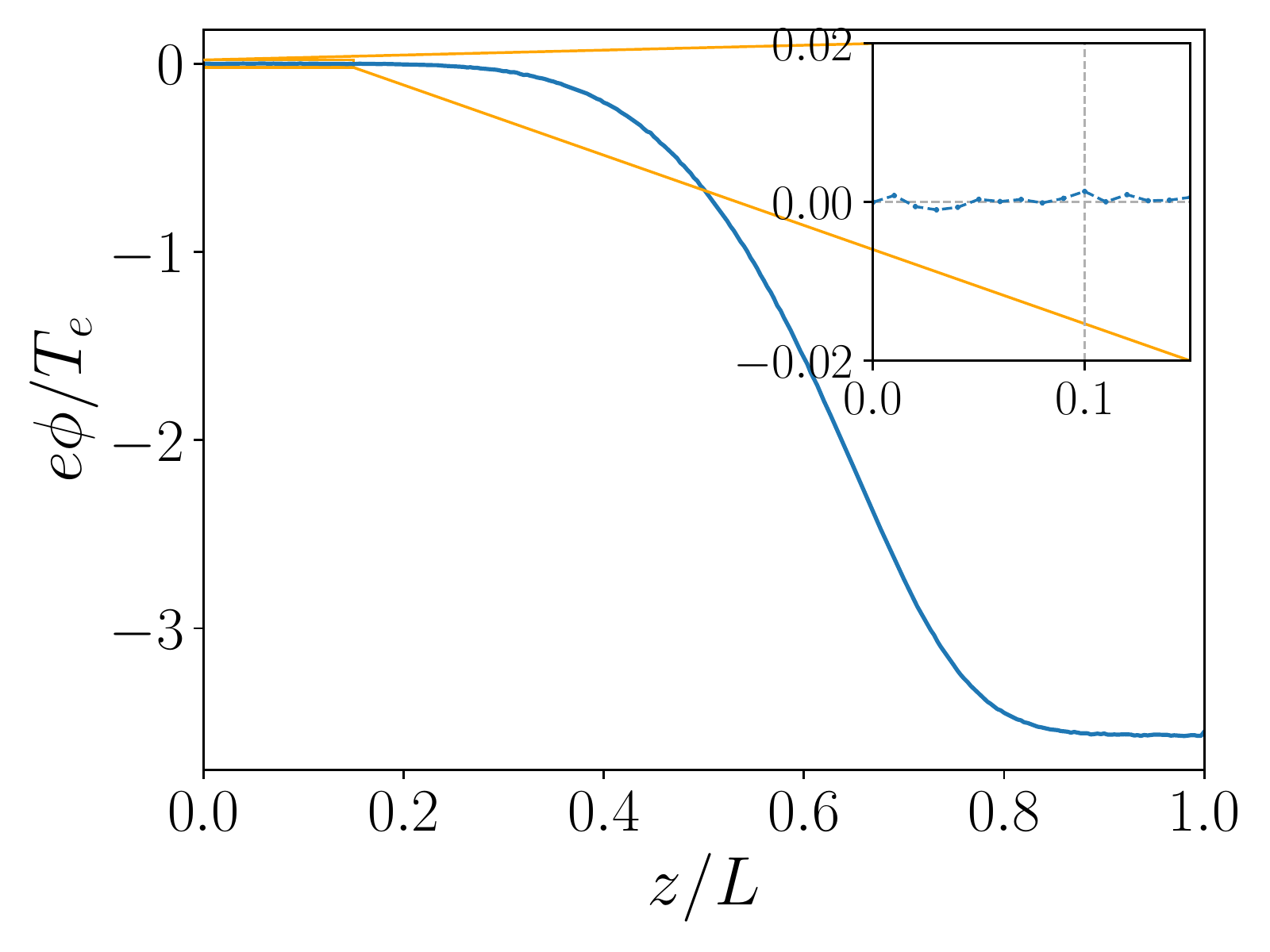}
    \caption{ Spatial profile of self-consistent electrostatic potential for the isotropic case with $T_i=\SI{50}{eV}$. The zoomed-in insert shows the region near $z=0$. } 
    \label{potCone}
   \end{figure}

\begin{figure}[H]
    \centering
    \subfloat[]{ 
      \includegraphics[width=0.46\linewidth]{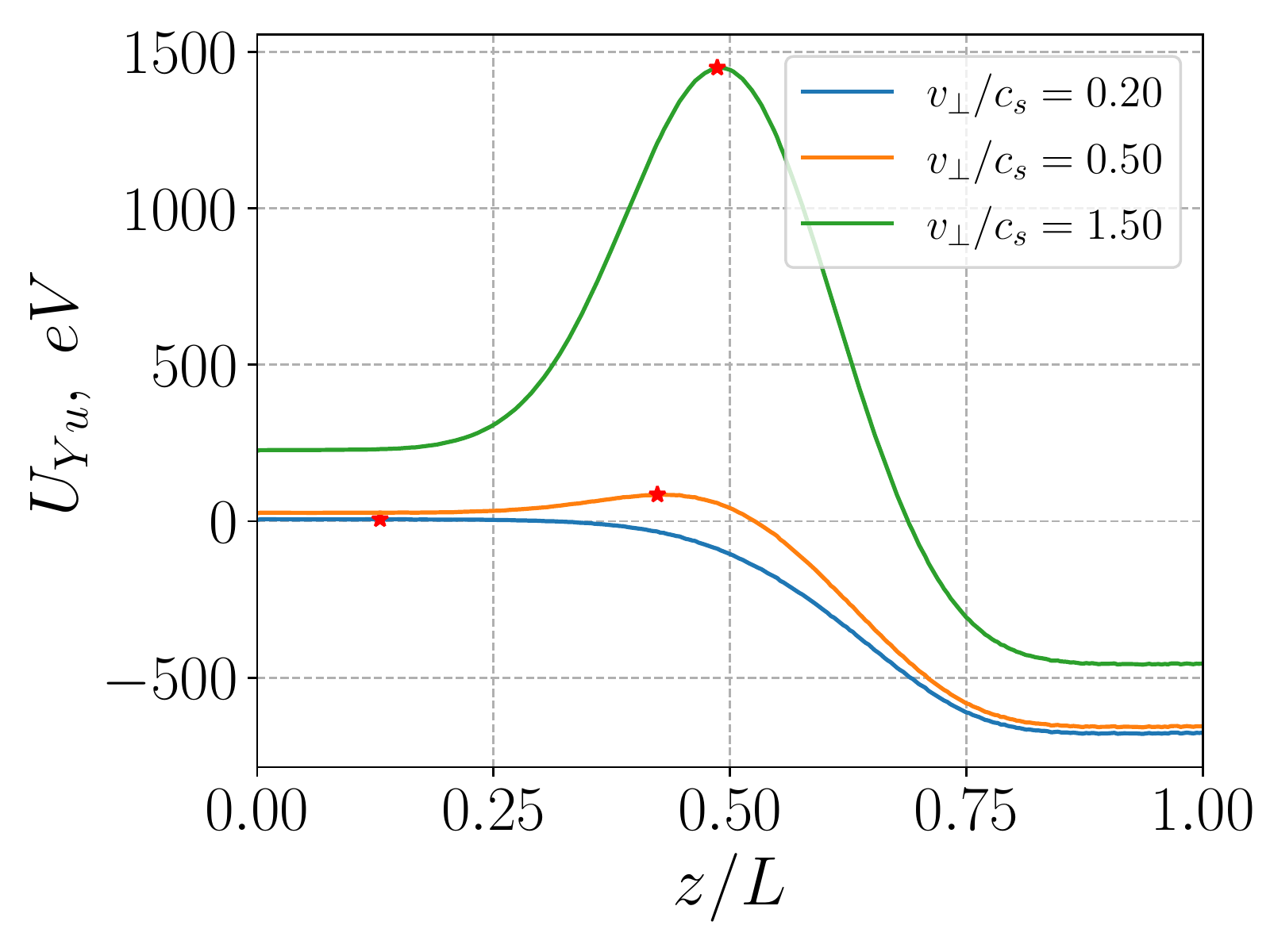}} 
    \subfloat[]{ \includegraphics[width=0.50\linewidth]{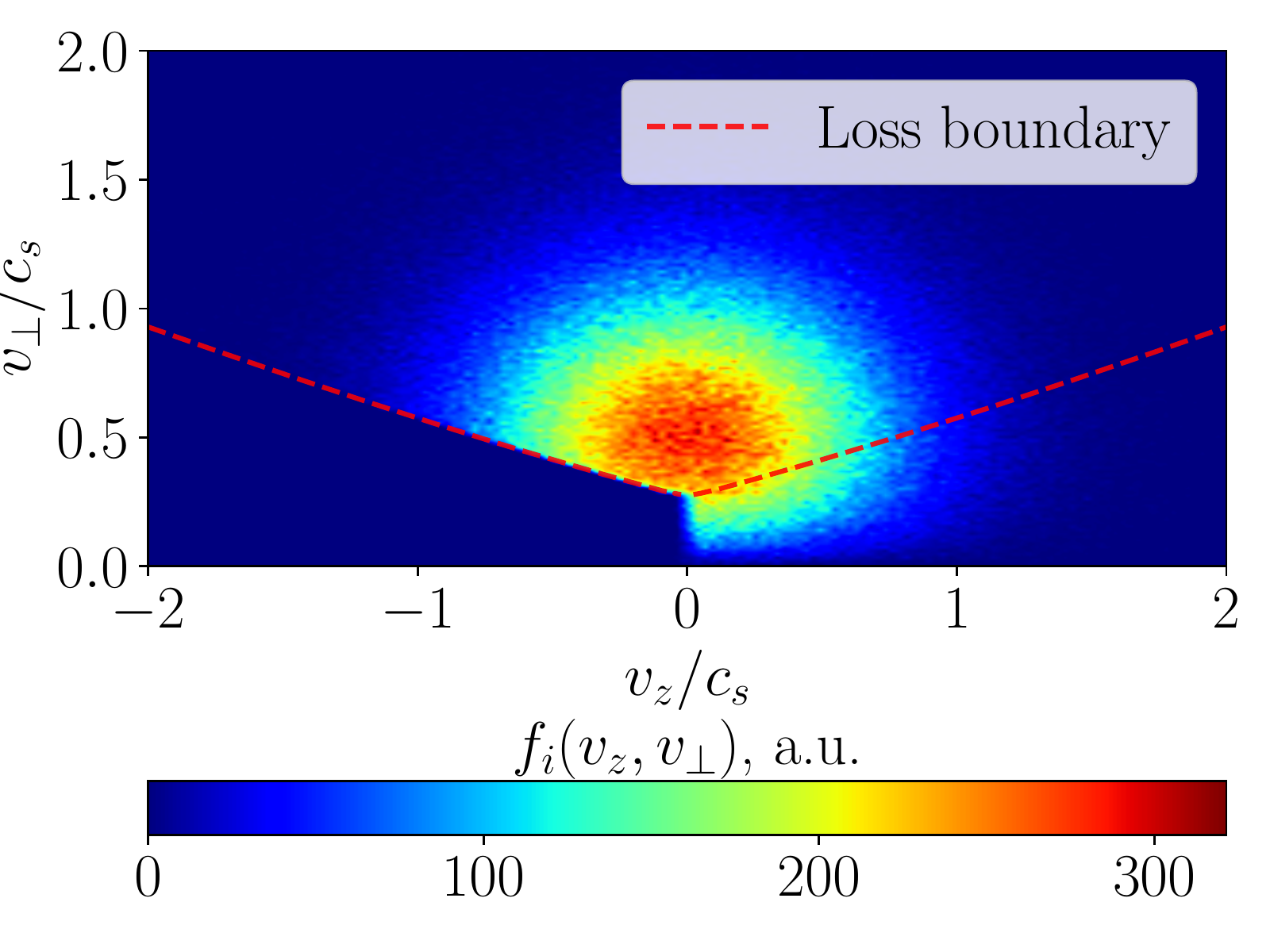}}
    \caption{(a) Effective Yushmanov potential for various values of the adiabatic magnetic moment $\mu$ and the potential shown in Fig.~\ref{potCone}, the maximum is shown by the red star. 
    %fluid theory\cite{smolyakov2021quasineutral};
    (b) Ion velocity distribution function in velocity space at $z=0$,  showing both injected ($v_z>0$) and reflected ($v_z<0$) particles. Loss boundary is evaluated from Eq.~(\ref{bound_cone}) for the electric field shown in Fig.~\ref{potCone}.}
    \label{loss-cone}
\end{figure}

  An example of a typical accelerating electrostatic potential is shown in  Fig.~\ref{potCone} for $T_e=\SI{200}{eV}$ and isotropic ions with $T_i=\SI{50}{eV}$. More details on the profile of the electric field obtained in simulations are given in Section III. It is assumed that $\phi (0) = {0}$.  
The insert in  Fig.~\ref{potCone} shows small fluctuations in the electrostatic potential near the injection (wall) point that occur due to the inherent particle-based noise. 
 \vspace{-5mm}
In general, monotonous accelerating potential makes loss cone wider. Examples of Yushmanov potential for various values of  $\mu$ are shown in Fig.~\ref{loss-cone}a for the electric field in Fig.~\ref{potCone}.
As an example, and for the purpose of the code verification, in Fig.~\ref{loss-cone}b we show the ion distribution function for the case of the injection with an isotropic temperature $T_i=\SI{50}{eV}$.
\\  %Fig.~\ref{vx_vp-ti200} shows the velocity distribution function at fixed location $z=0$ in the simulation with isotropic particle injection, $T_i = \SI{200}{eV}$, when the steady state is reached. It can be seen that the reflected particles ($v_z<0$) satisfy the loss condition given by Eq.~(\ref{bound_cone}).

Note that the condition (\ref{bound_cone}) for the loss cone boundary  is an implicit equation in $v_z-v_\perp$ space. The boundary becomes a straight line for large $\mu$ when the electric field can be neglected. For smaller values of $\mu$, it is deformed. For monotonically decreasing electric field potential, with an decrease of $\mu$, the maximum of the Yushmanov potential shifts to the origin, Fig.~\ref{loss-cone}a, and for some critical value the maximum disappears. Thus, the particles with $\mu$ below some critical value are not reflected as shown in   Fig.~\ref{loss-cone}b and this region of the phase space  is empty for $v_z<0$.

%\vspace{-2ex}
%The flux of the injected ions into the system corresponds to
%\begin{equation}
%    \Gamma _{0} = n V_0 = \SI{5.655e23}{m^{-2} s^{-1}}.
%    \label{g01}
%\end{equation}     

%\textbf{General parameters}
%\begin{table}[h]
%\centering % centering table
%\begin{tabular}{l| p{4cm} p{2cm} p{2cm}} % creating 10 columns
%\hline\hline % inserting double-line
%  & \multicolumn{3}{c}{General parameters} \\[0.5ex]

%Specie & mass flow, $\si{kg/s}$ & $n_0$ & $T_0,\si{eV}$ \\ [1ex]
%\hline % inserts single-line 
%\hline \\ [0.1ex]
% ion & $5\cdot 10^{-4}$ & source & 20-100-200\\[2ex]
% electron  & $5\cdot 10^{-4}$  & source  & 200\\[2ex]
% neutral & &       & \\[2ex]
%\hline % inserts single-line
%\end{tabular}
%\label{tab:PPer}
%\end{table}

%  \begin{figure}[H]
%    \centering
%    \includegraphics[width=0.45\linewidth]{img/cone_cold.png}
 %   \caption{ Loss cone, T for cold ions}
 %   \label{cone_cold}
%\end{figure}
 
% \begin{figure}[H]
%\centering
%\subfloat[]
%{\includegraphics[height=0.42\textwidth]{img/}}    
%\subfloat[]{\includegraphics[height=0.42
%\textwidth]{img/}}    
%\label{3}
%\caption{Cold ions, initialization}
%\end{figure}
 
%\textbf{Warm ions in loss cone}
%Imposing the condition $v_\perp < \frac{3 v_{th}}{\sqrt{(B_{max}/B_{min})-1)}} $\textcolor{red}{Where does this condition comes from?}

% \newpage

\section{Wave breaking,  instabilities, and ion reflections in the accelerated plasma flow}\label{sec3}

As it was noted, within the cold ions fluid theory \cite{smolyakov2021quasineutral} the magnetic field fully defines the velocity profile across the whole nozzle region. For the injection velocities $v_0/c_s> M_b$ and $v_0/c_s<M_a$ one obtains respectively the fully supersonic and subsonic solutions, which never cross the sonic point $v=c_s$. The unique accelerating (decelerating)  solution crossing the point $v=c_s$ is obtained for exact value $v_0/c_s=M_a$ ($v_0/c_s=M_b$). Solutions with $M_a<v_0/c_s<M_b$ are multi-valued and therefore do not exist in the fluid theory. In this section, we investigate injection with different initial velocities $v_0$ and different distribution and  show that, in general, the unique accelerating solution is rather robust.   Some injected particle are reflected, but the passing particles  get accelerated and their  net (fluid) velocity  follows the fluid solution even if the initial velocity at the injection point is not ``correct''. The adjustment occurs  at the expense of the particles reflected by temporal fluctuations of plasma potential as a result of wave breaking and subsequent turbulent pulsations of the ion-sound type.  This phenomena is  especially notable in the ``prohibited''  region $M_a<v_0/c_s<M_b$ where the reflections are most pronounced the time average potential is non monotonous with the region of the negative electric field, as in Fig. 5a.  In case of  finite ion temperature and anisotropy, the reflections also occur as a result of the magnetic mirror force  (for ions outside of the loss cone modified by the self-consistent potential). We note that turbulent ion-sound fluctuations can be easily excited by counter-streaming beams of passing and reflected ions. In general, the turbulent fluctuations remain present in majority of the cases we have considered.    In this section we provide more details of such processes for the injection with different distribution functions.

\subsection{ Mono-energetic injection with finite parallel  and zero perpendicular velocities: comparison with the cold ion fluid theory.}
\label{cold}
Here we use the injection of ions with the cold beam distribution and zero perpendicular velocity $v_\perp = 0$ exactly corresponding to the assumptions  of cold ion theory,
%\begin{equation}\label{f_cold}
%f=\frac{n_{0} }{ 2 \pi v_\perp}   \delta  \left({v_{z}-V_0}\right)
%\delta  \left( {v_{\perp}}\right),
%\end{equation}

\begin{equation}\label{f_cold}
f=\frac{n_{0} }{{\pi} v_\perp }\delta  \left({v_{z}-v_0}\right)
\delta  \left( {v_{\perp}}\right).
\end{equation}

The distribution function is normalized such that the density of the injected beam is $n_0$ and the particle flux $\Gamma_0$ in the injected beam is
\begin{equation}
\Gamma _{0}=n_{0}V_{0}= \int \limits_{-\infty} ^{+\infty} v_{z} \mathrm{d}v_z \int \limits_{0}^{+\infty} f 2 \pi v_{\perp} \mathrm{d}v{_\perp},
\end{equation}
where $V_0$ is the average flow velocity.
%\begin{equation}\label{f_cold}
%f=\frac{4 n_{0} }{\sqrt{\pi} V_{0}^{3}}  \delta  \left(({v_{z}-V_0)}\right)
%\delta  \left( {v_{\perp}}\right),
%\end{equation}

\begin{figure}[H]
    \centering
    \subfloat[]{\includegraphics[width=0.45\linewidth]{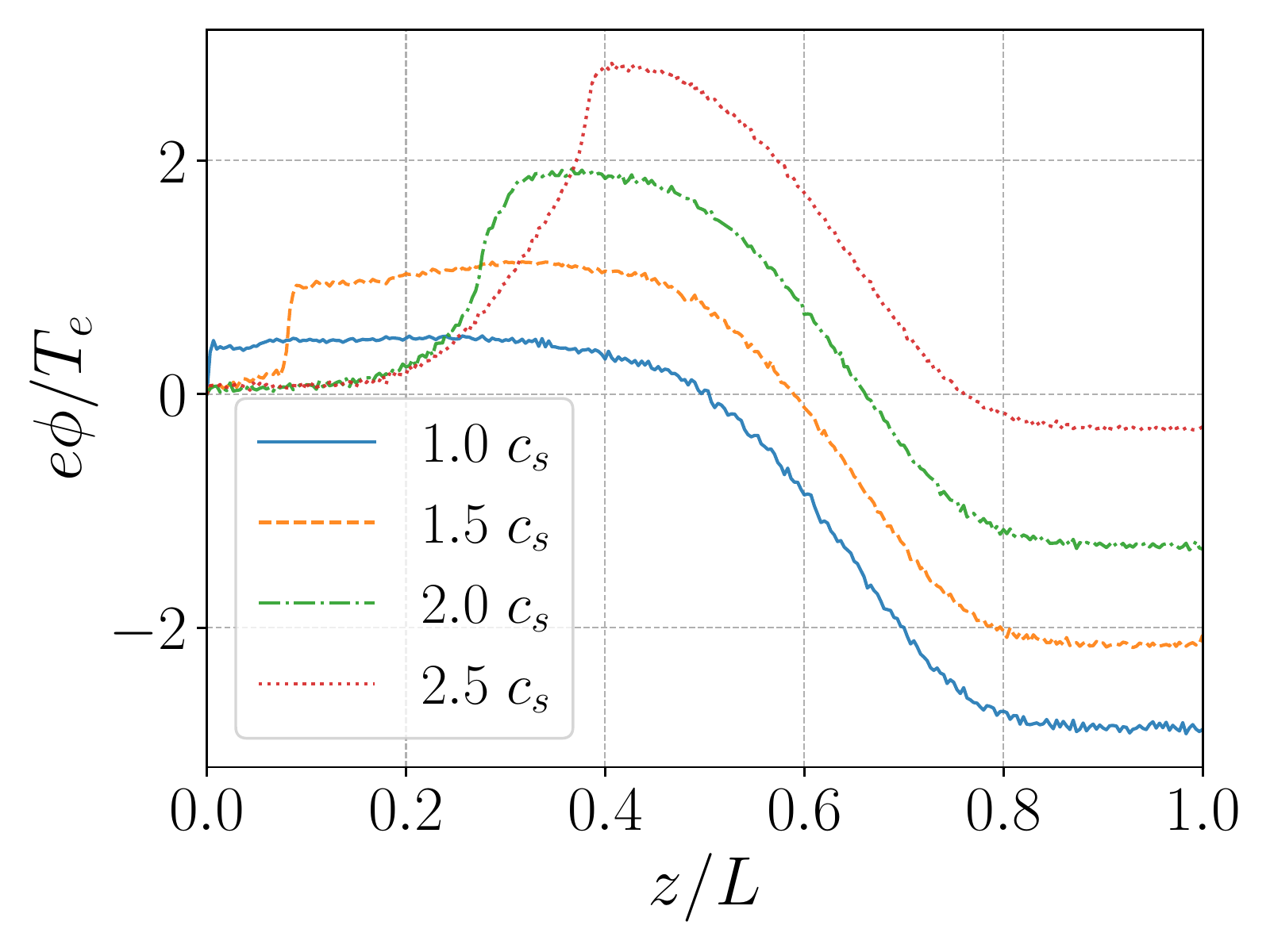}\label{cold1}}
    \subfloat[]{\includegraphics[width=0.45\linewidth]{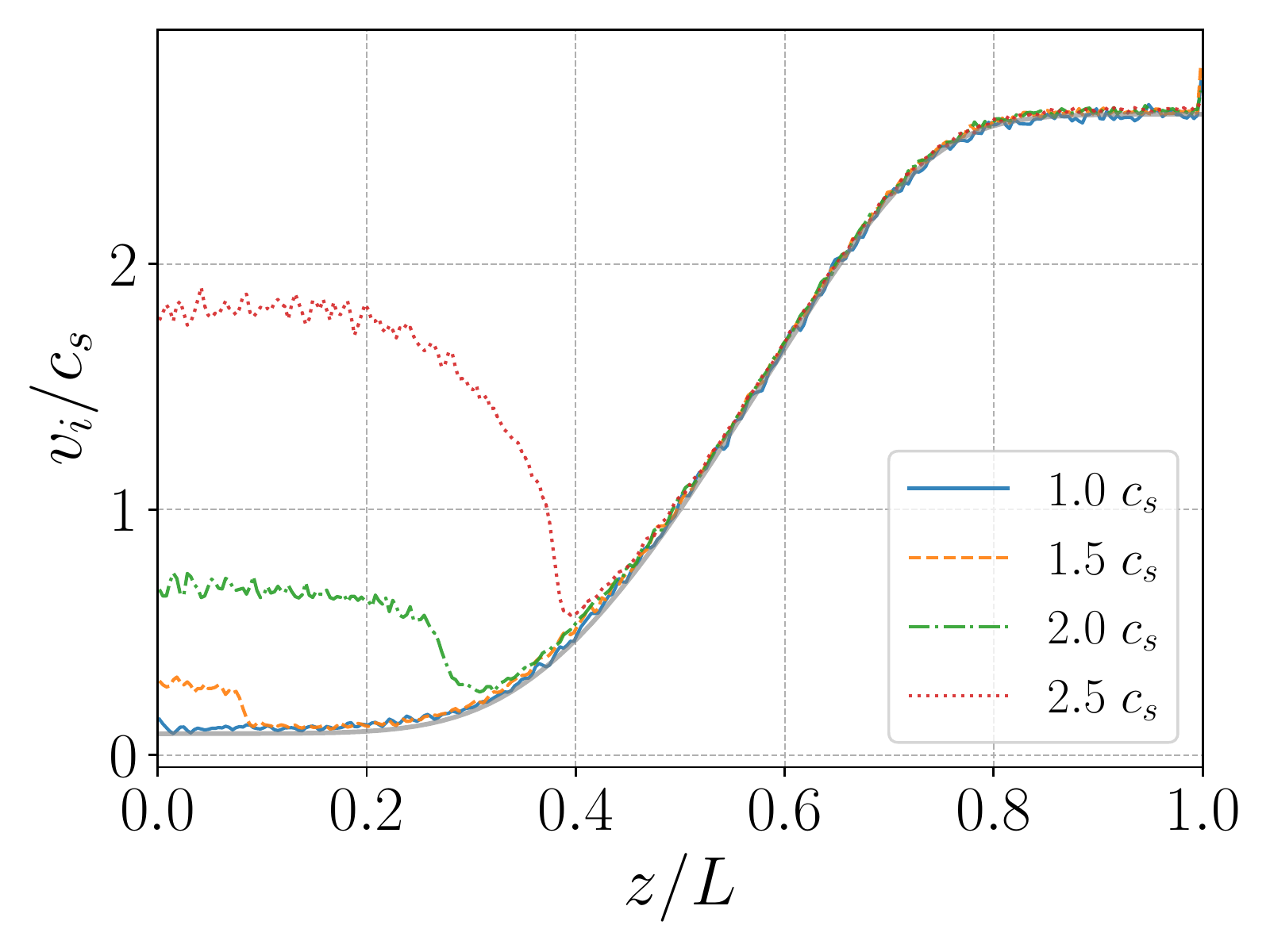}\label{cold2}} \\
    \subfloat[]{\includegraphics[width=0.45\linewidth]{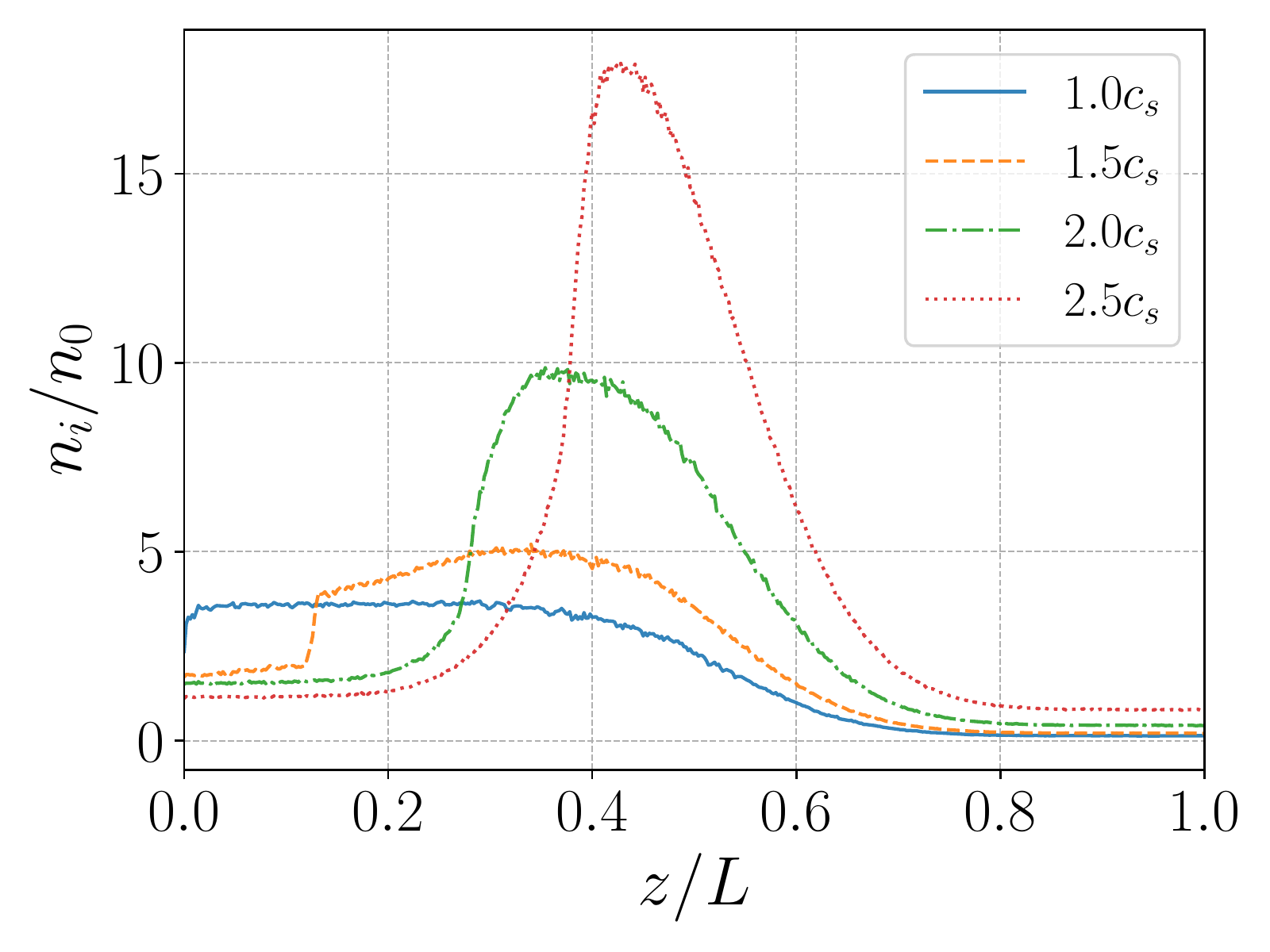}\label{cold3}}
   % \subfloat[]{\includegraphics[width=0.45\l%inewidth]{pic/}\label{cold4}}
   \caption{Spatial profiles of  electrostatic potential (a),  flow velocity (b), ion density (c), for injection velocity $M_a=0.0869 <  v_0/c_s<\mathrm{M}_b=2.6096$. Grey solid line in (b) shows the accelerating fluid solution.}
\end{figure}

\begin{figure}[H]
    \centering
    \subfloat[$v_0 = c_s$]{\includegraphics[width=0.45\linewidth]{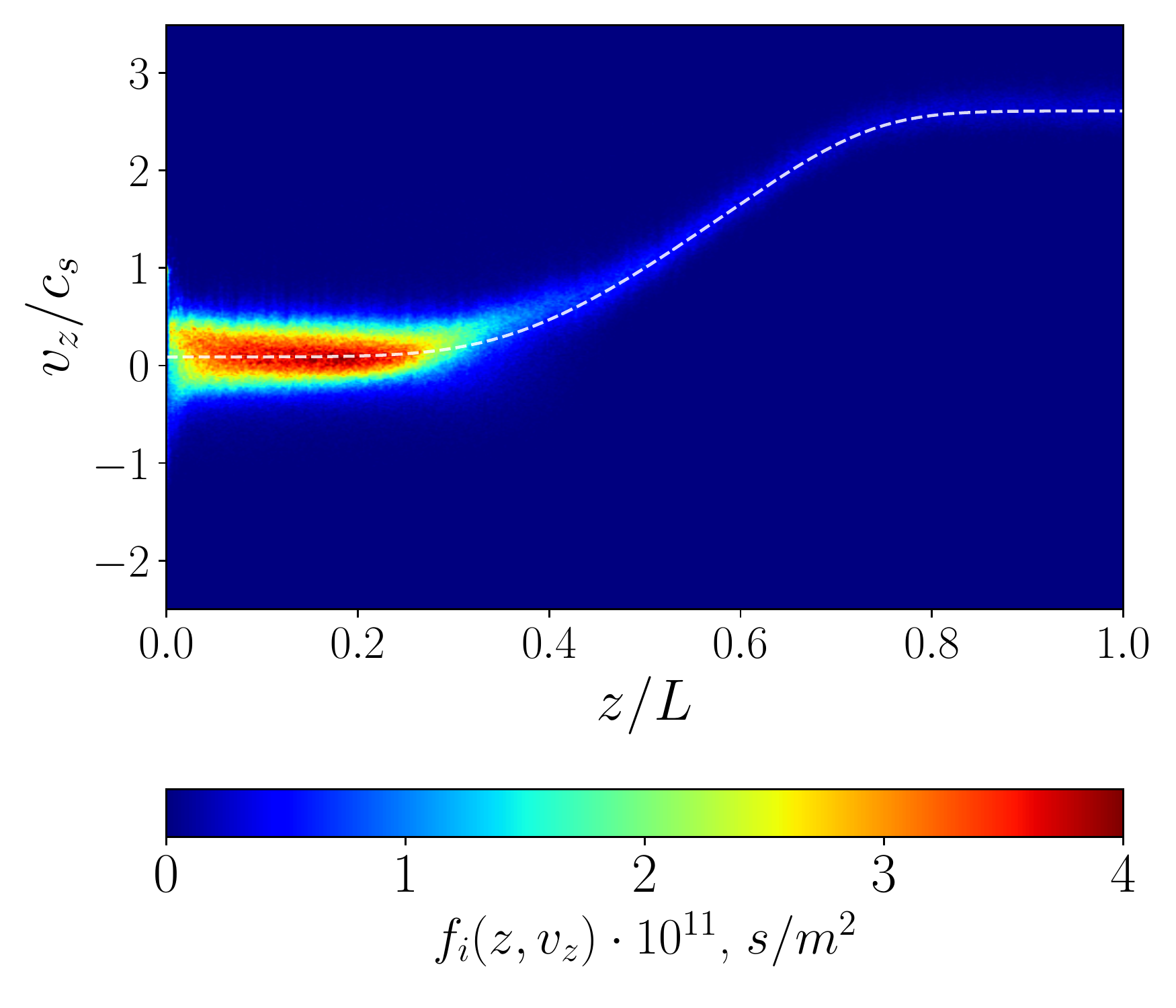}\label{ivdf_mono_cold1}}
    \subfloat[$v_0 = 1.5 c_s$]{\includegraphics[width=0.45\linewidth]{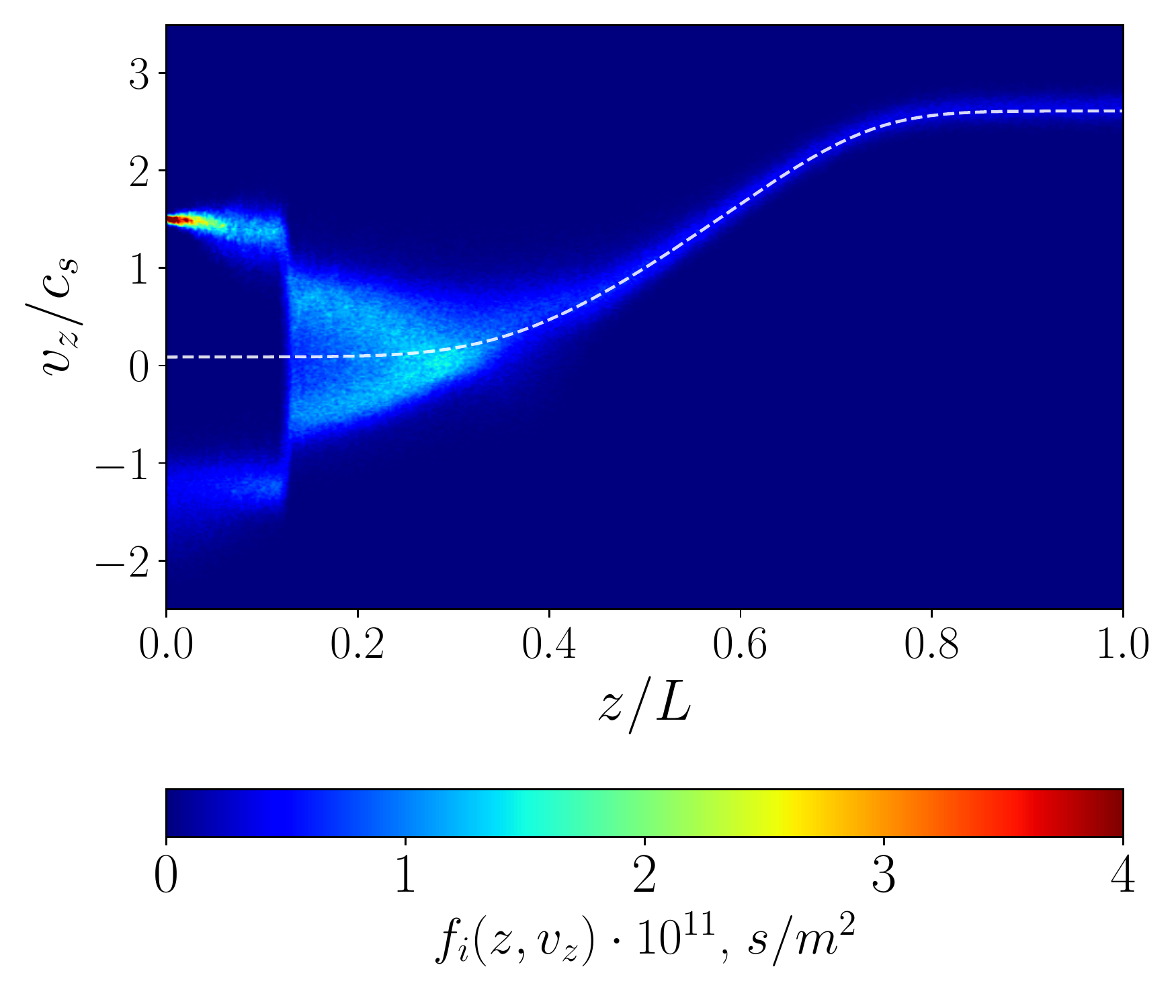}\label{ivdf_mono_cold15}} \\

    \subfloat[$v_0 = 2 c_s$]{\includegraphics[width=0.45\linewidth]{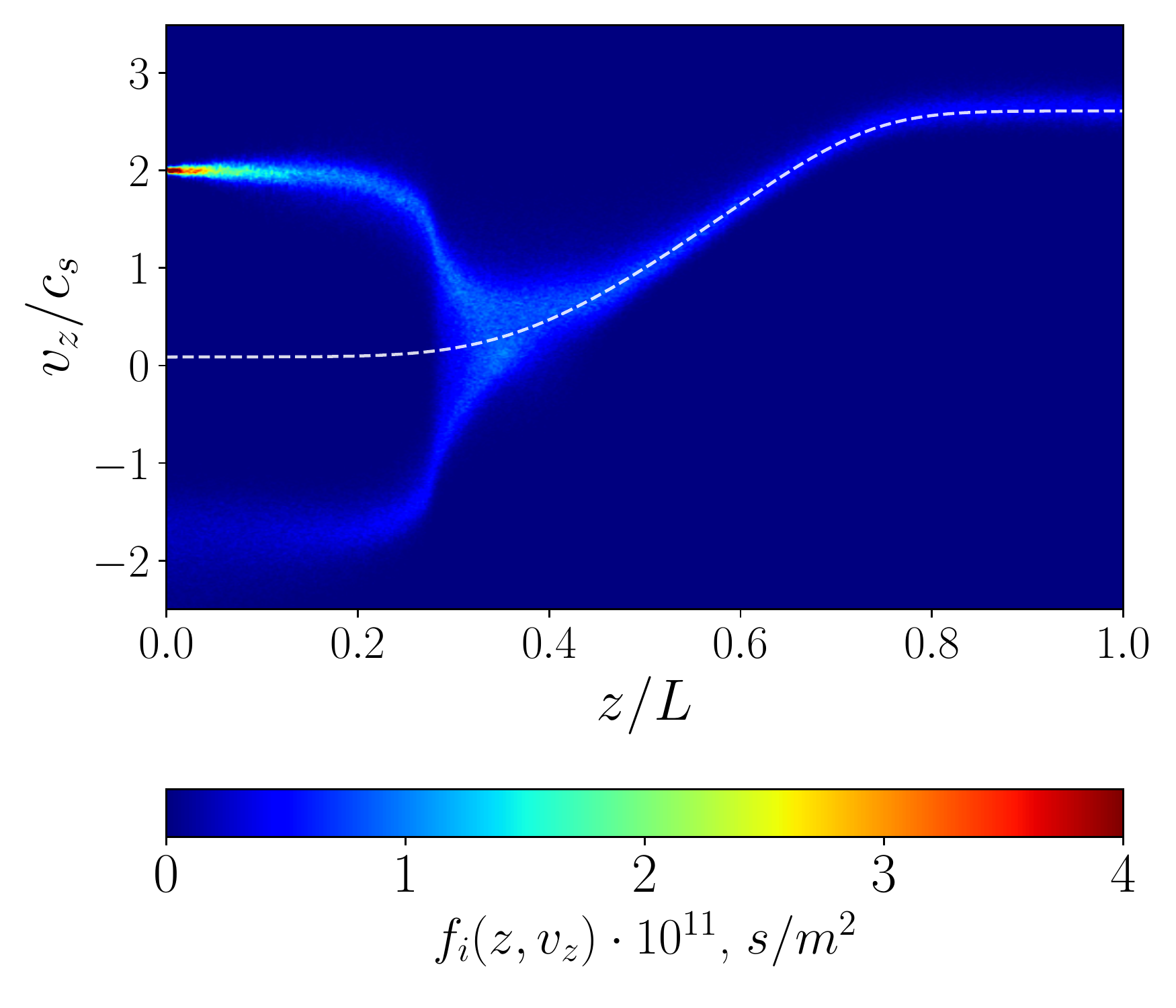}\label{ivdf_mono_cold2}}
    \subfloat[$v_0 = 2.5 c_s$]{\includegraphics[width=0.45\linewidth]{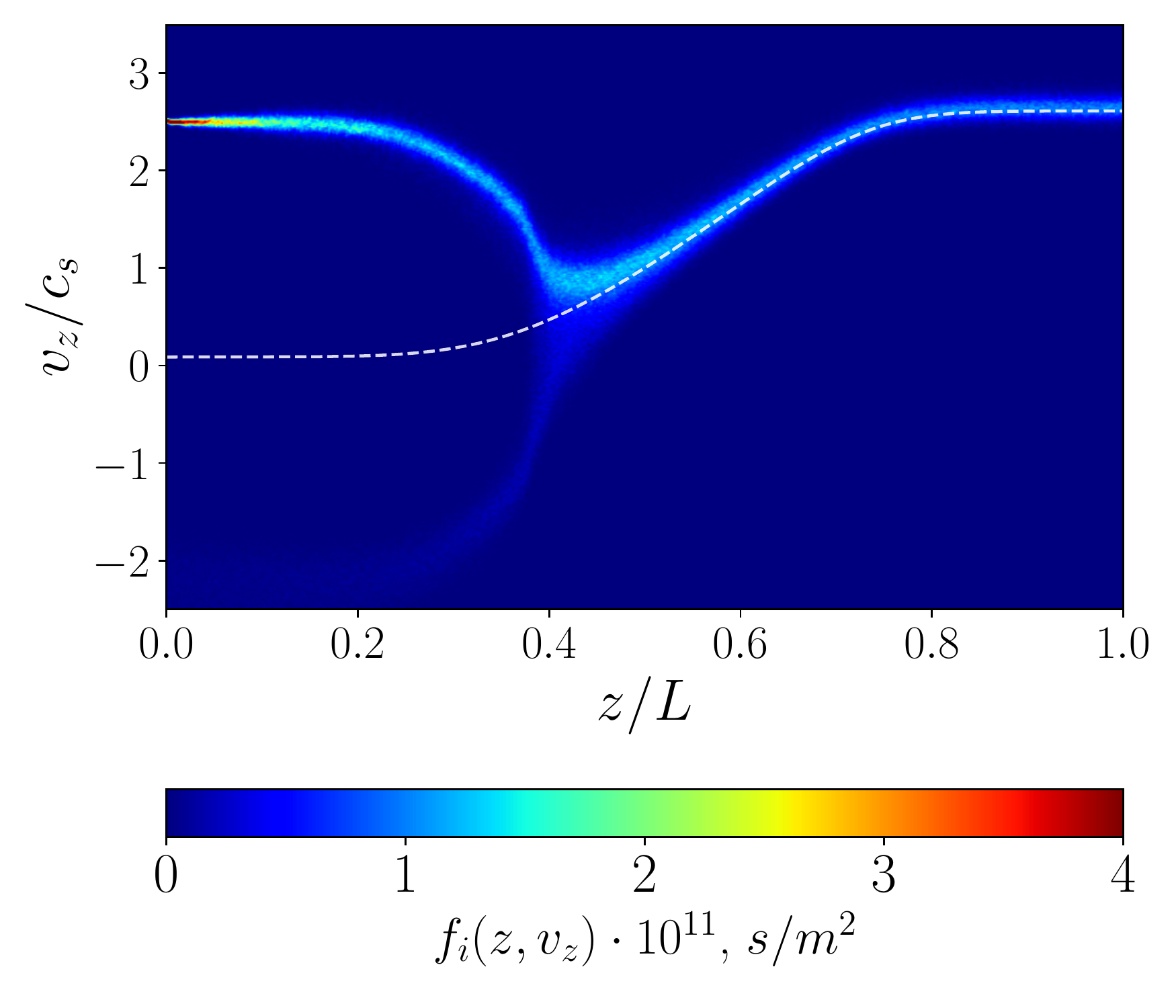}\label{ivdf_mono_cold25}}
   % \subfloat[]{\includegraphics[width=0.45\l%inewidth]{pic/}\label{cold4}}
    \caption{Ion distribution function in ($z-v_z$)-space for monoenergetic injection with $v_\perp=0$. The transonic accelerating profile from the fluid theory is shown by white dashed line. }     
    \end{figure}

It is found that monoenergetic injection with zero perpendicular velocity has two distinct regimes depending on the injection velocity $v_{0z}$. In the ``prohibited'' region $M_a < v_{0}/c_s < M_{b}$ we observe excitation of the electrostatic potential fluctuations that reflect a significant fraction of injected particles. The reflections reduce the mean flow velocity eventually forcing it to the global accelerating profile of the fluid solutions,  as shown in Fig.~\ref{cold2}. Time averaged potential profile is non-monotonous with the region of the negative  electric field that stalls the flow,  Fig.~\ref{cold2}. The mean flow velocity is reduced so that the plasma density is increased,  Fig.~\ref{cold2}. It is important to note that potential fluctuations  are non-stationary and the profiles in Figs.~\ref{cold1}-\ref{cold3} are time averages.  The potential fluctuation slows down fast particles (and partially reflects), creating multivalued distribution in the velocity space, as shown in phase space distribution in Figs.~\ref{ivdf_mono_cold1}-\ref{ivdf_mono_cold25}.  The reflections and  mixing are decreasing toward the $M_a$ and $M_b$ boundaries of the ``prohibited'' region: the ratio of the reflected flux $\Gamma_r/\Gamma_0= 0.71, 0.65, 0.48, 0.17$, respectively for $v_0$= $(1, 1.5,2.0, 2.5) c_s$, respectively. 
We note that while the turbulent pulsations exist in most cases we have studied, the non-monotonous potential (in averaged profiles) appears with  the mono-energetic and anisotropic injection, Fig. 20, while for the isotropic injection the potential remains monotonous, as in Fig. 12. 

\begin{figure}[H]
    \centering
    \subfloat[]{\includegraphics[width=0.45\linewidth]{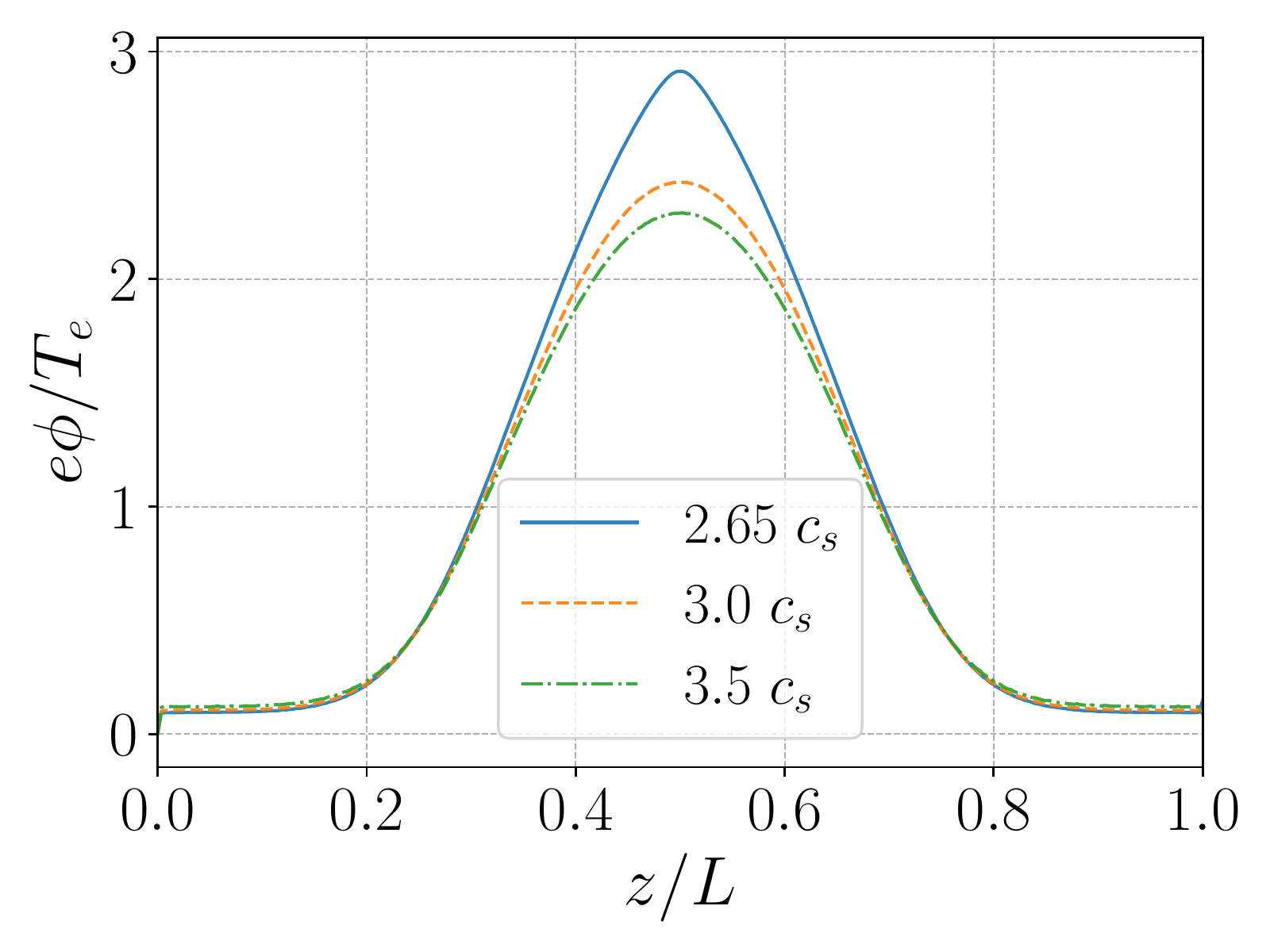}\label{cold4}}
    \subfloat[]{\includegraphics[width=0.45\linewidth]{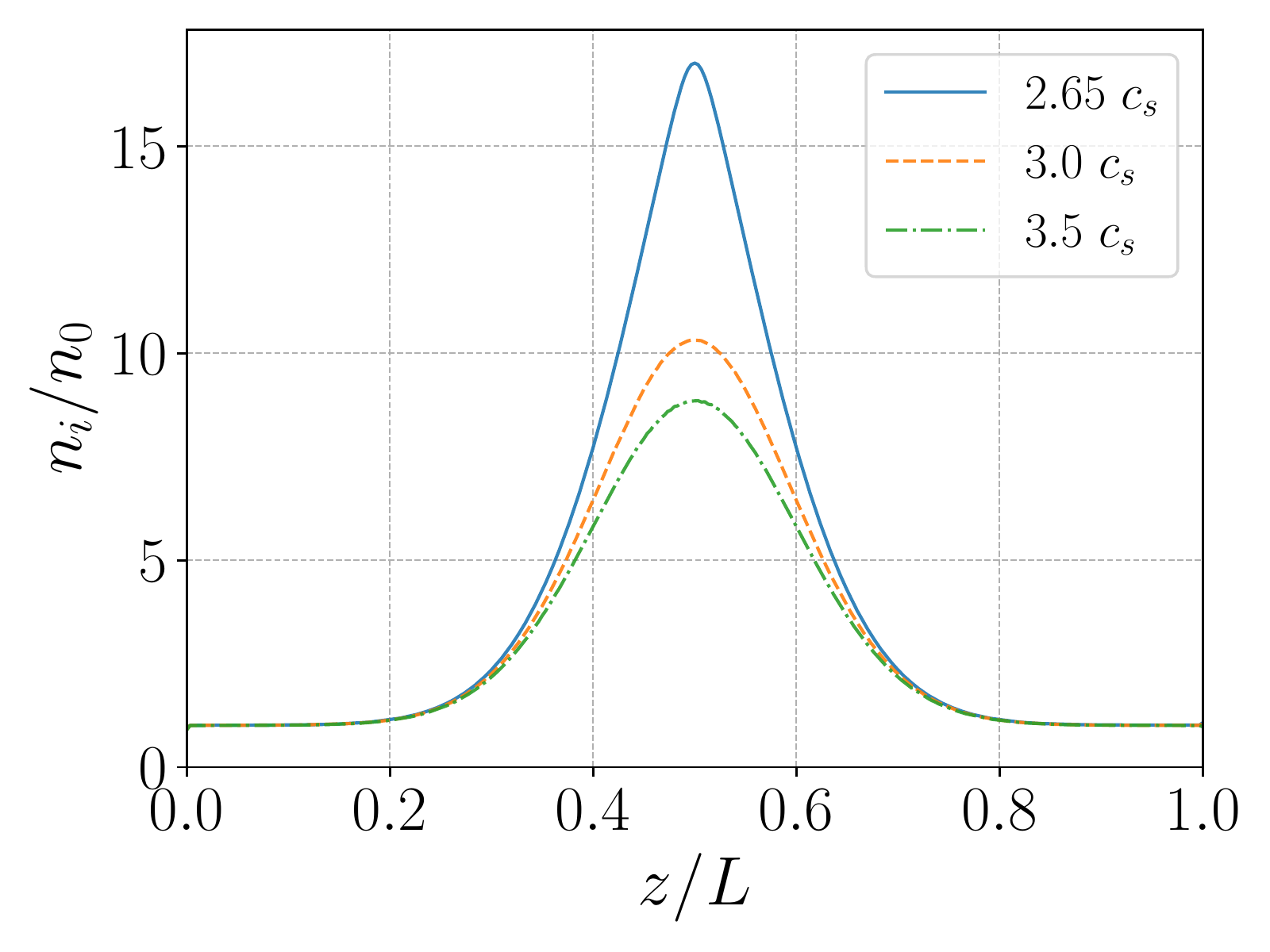}\label{cold6}} \\

  \subfloat[]{\includegraphics[width=0.45\linewidth]{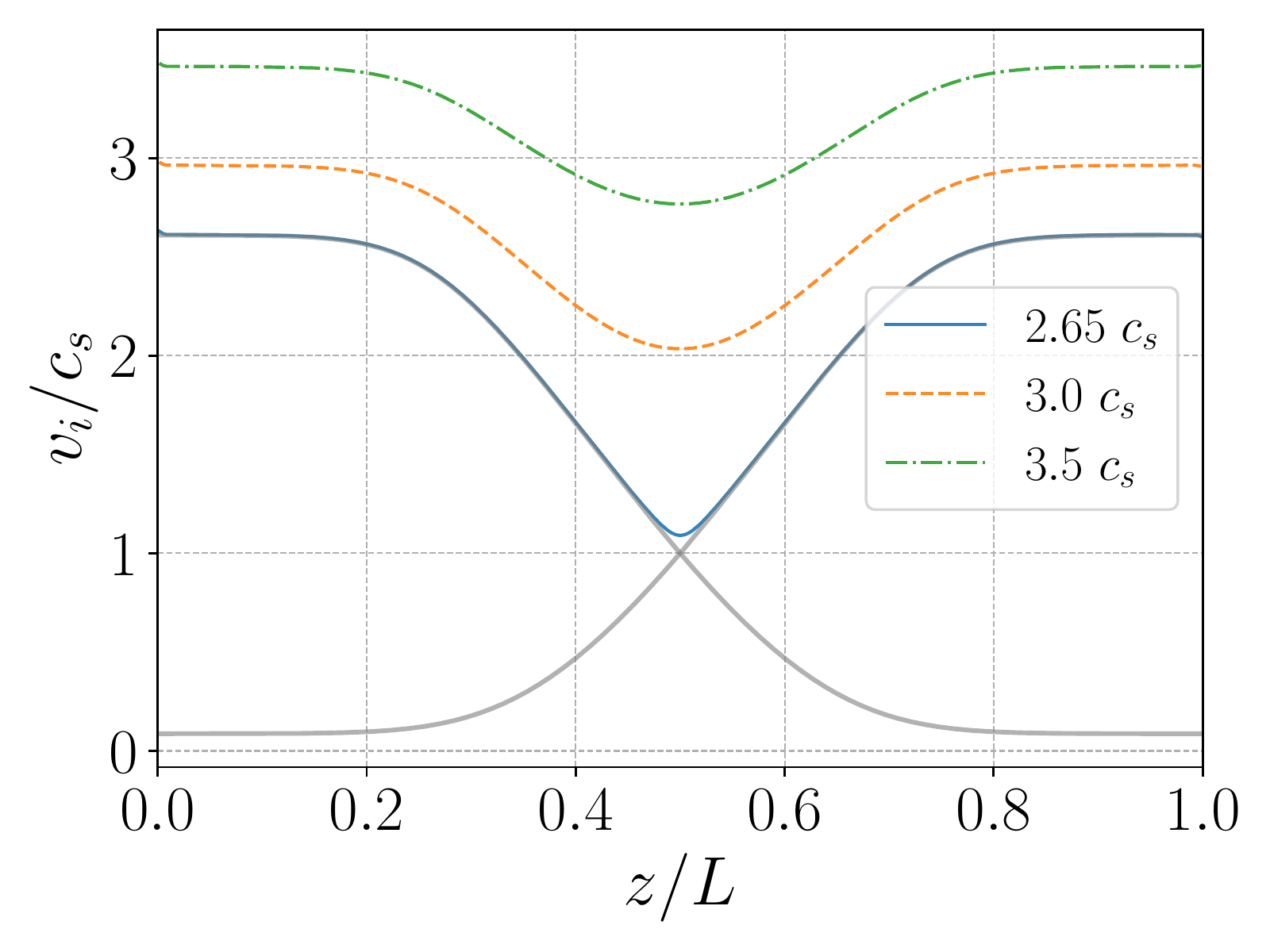}\label{cold5}}
\subfloat[]{\includegraphics[width=0.42\linewidth]{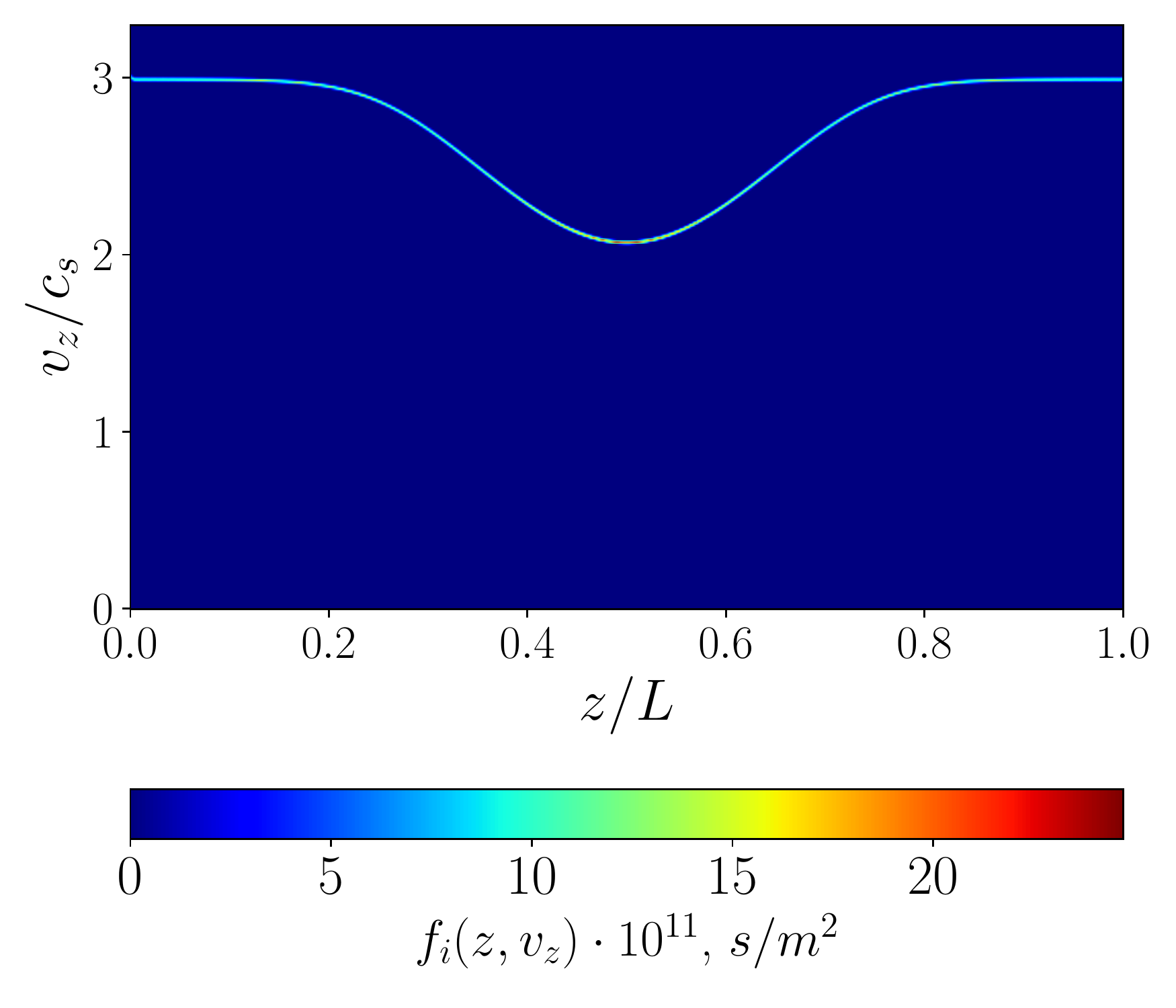}\label{super_phase}}
   % \subfloat[]{\includegraphics[width=0.45\l%inewidth]{pic/}\label{cold4}}
   \caption{Spatial profiles of electrostatic potential (a), ion density (b), flow velocity (c), for cases with the injection velocity above $\mathrm{M}_b=2.6096$, and ion distribution function in the ($z-v_z$)-space for $v_0=3 c_s$ (d).}
\end{figure}

The injections with $v_{0}/c_s > M_{b}$ result in smooth (laminar)  decelerating-accelerating flow which remains supersonic,  Fig.~\ref{cold5}.  There are no reflections in these cases and the velocity profile matches exactly to the fluid supersonic solution as shown in Fig.~\ref{fluid_solu} without reflections and phase mixing.

\begin{figure}[H]
    \centering
    \includegraphics[width=0.5\linewidth]{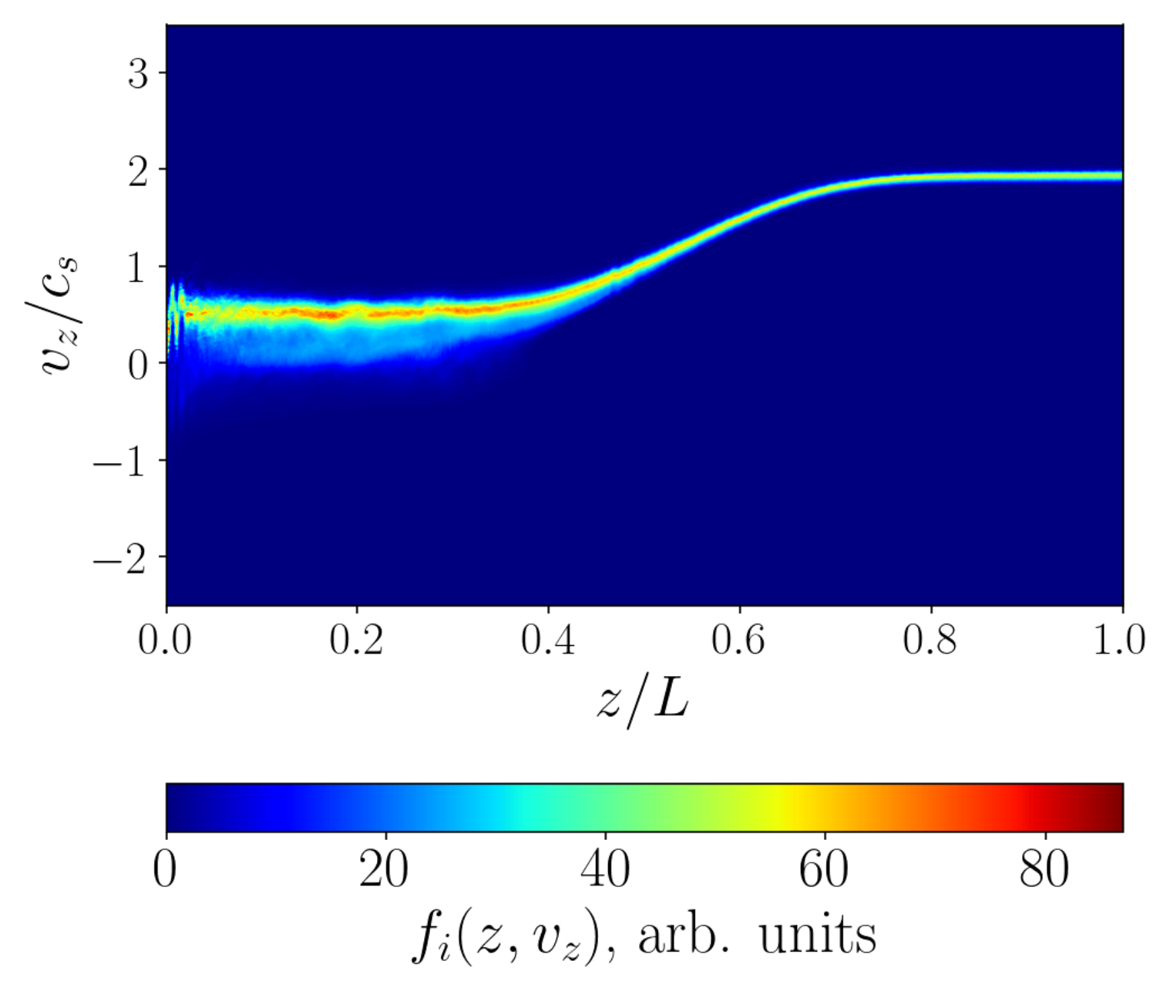}
    \caption{Ion distribution function in the ($z-v_z$)-space for  $v_0/c_s=M_a=0.319~c_s$, mirror ratio $R=2$.}
    \label{theory_phase}
\end{figure}

\begin{figure}[H]
    \centering
    \includegraphics[width=0.5\linewidth]{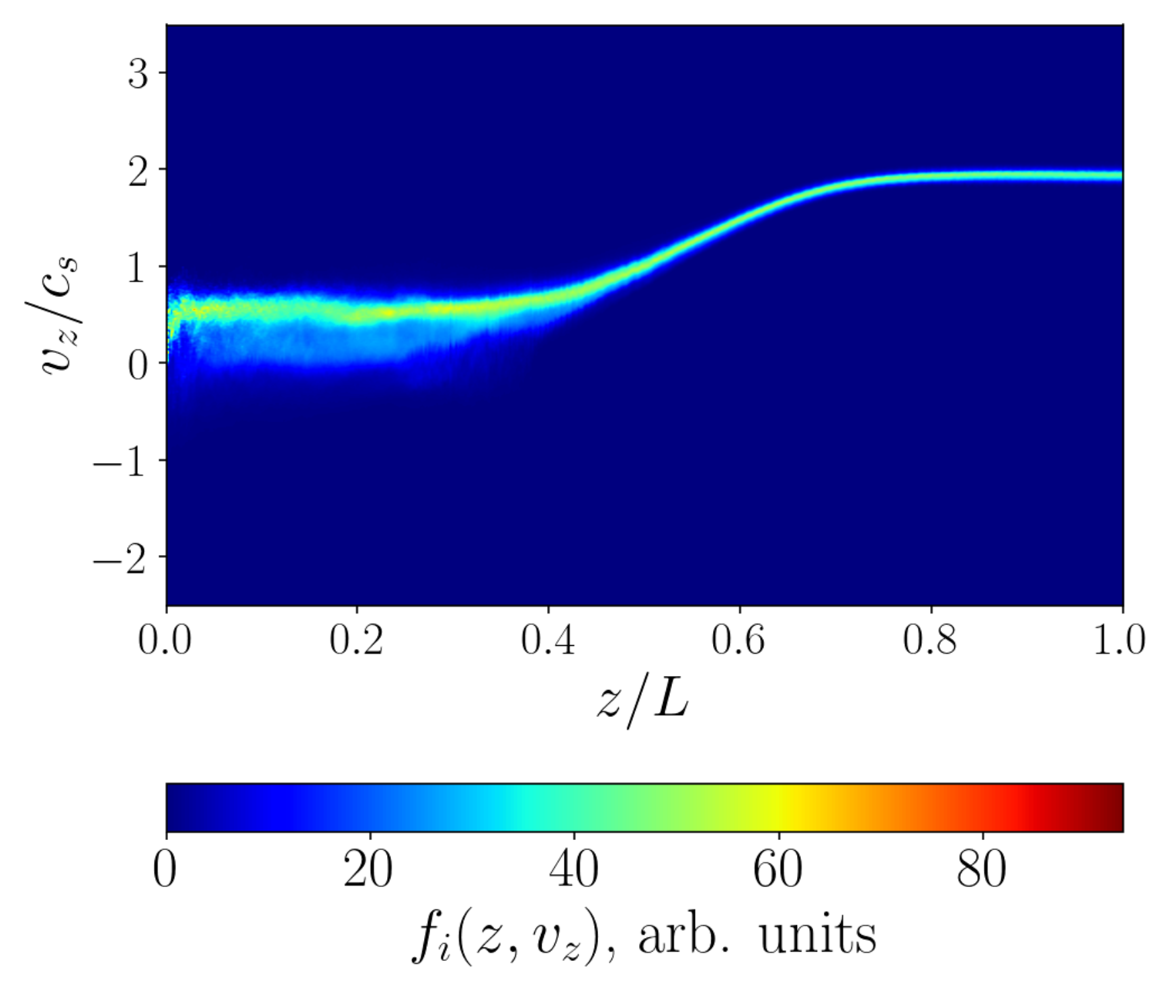}
    \caption{ Ion distribution function in the ($z-v_z$)-space for the subsonic injection   $v_0/c_s=0.2<M_a=0.319$,  mirror ratio $R=2$.}
    \label{low_phase}
\end{figure}

We were not able to recover the laminar subsonic solutions when injecting with $v_{0}/c_s < M_{a}$. For these cases,  we observe relatively large fluctuations of the potential in the converging part of the nozzle. As a result of the fluctuations,  the averaged velocity profile always switches into the accelerating mode  and no smooth subsonic solutions were obtained. Two examples of the injection with $v_{0}/c_s < M_{a}$ are shown in  Figs.~\ref{theory_phase}-\ref{low_phase}  for the mirror ratio  $R=2$ so that $M_a= 0.3198$ and $M_b= 1.9216$. We have considered the cases with  smaller values of $R$ to allow larger critical value $M_a$ and investigate how  it  affect the behaviour in the subsonic regimes. In all cases with different $R$ we observed  similar behavior as   in Figs.~\ref{theory_phase}-\ref{low_phase} with small amount of  reflections,  $\Gamma_r/\Gamma_0 \simeq 0.1$, and the solutions switching into the transonic accelerating mode.
\vspace{-20ex}

%\textbf{Supersonic injection}

%\begin{figure}[H]
%    \centering
%    \includegraphics[width=0.5\linewidth]{fig/ivdf_..pdf}
%    \caption{IVDF in $z-v_z$ phase space with injection energy, $E=3~c_s$.}
%    \label{super_phase}
%\end{figure}

%\subsection{Mono-energetic in $v_z$ and $v_{\perp}=0$.}

 \subsection{Effects of a finite ion temperature, isotropic injection.}
 
\begin{figure}[H]
    \centering
\subfloat[]{\includegraphics[width=0.5\textwidth]{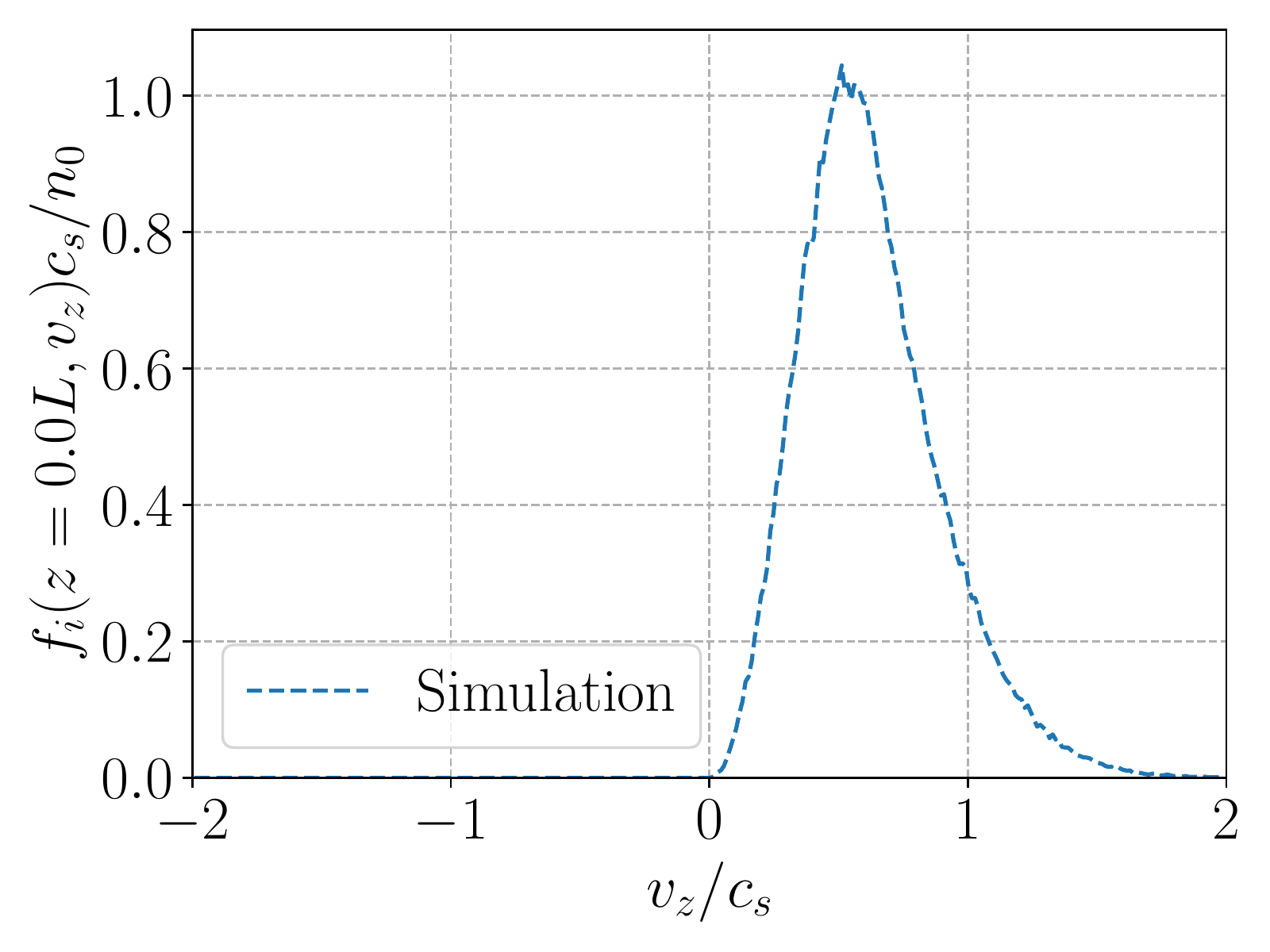}} 
\subfloat[]{\includegraphics[width=0.5\textwidth]{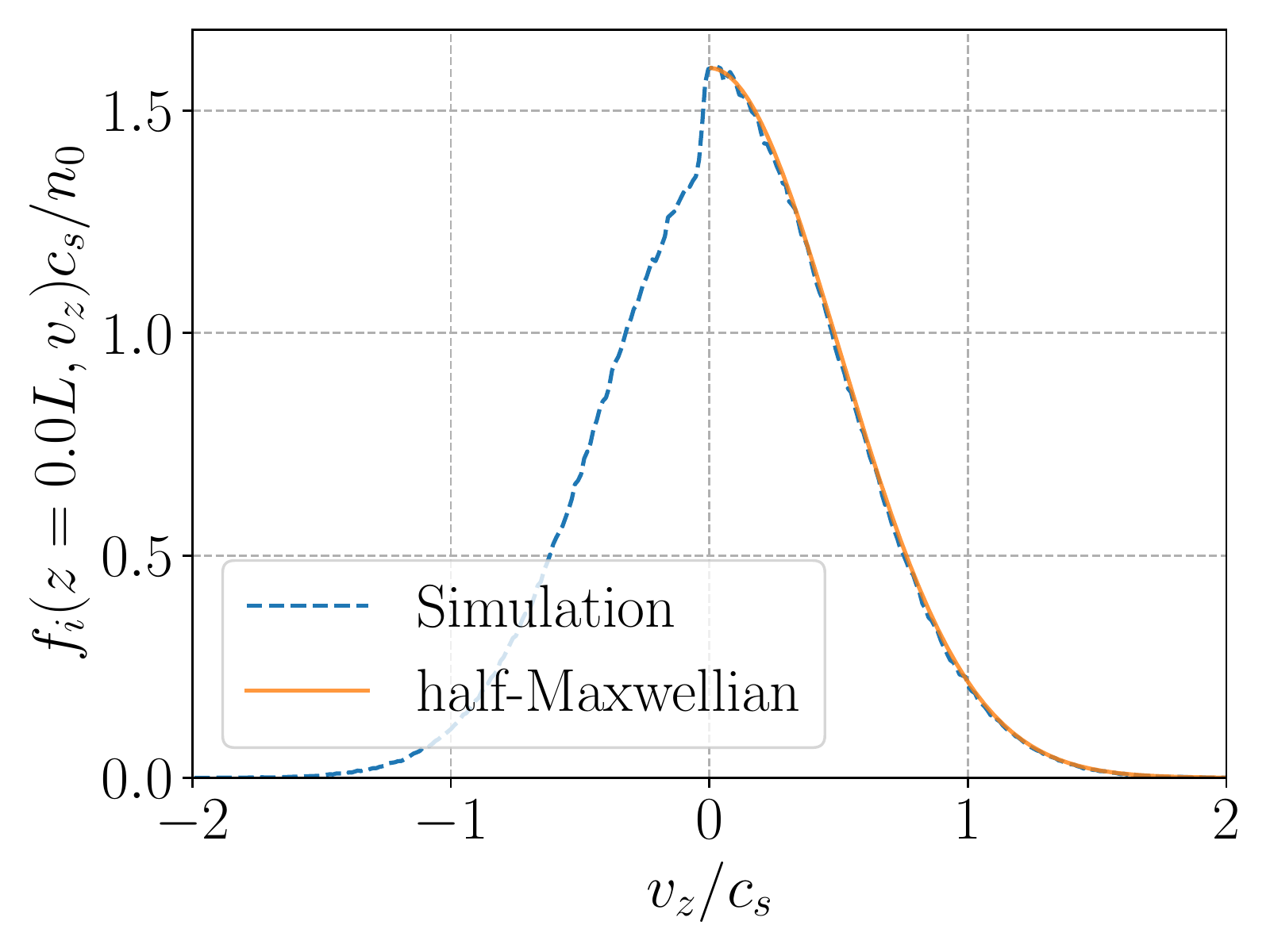}}
 \caption{Isotropic injection with Maxwellian flux at the source wall,  $z=0$ (a). The half-Maxwellian distribution for $v_z>0$ in the steady-state described by Eq.~(\ref{halfM_iso}) (b). Note the reflected particles in the  $v_z<0$ region.}
    \label{injection}
\end{figure}

\begin{figure}[H]
    \centering
    \includegraphics[width=0.5\linewidth]{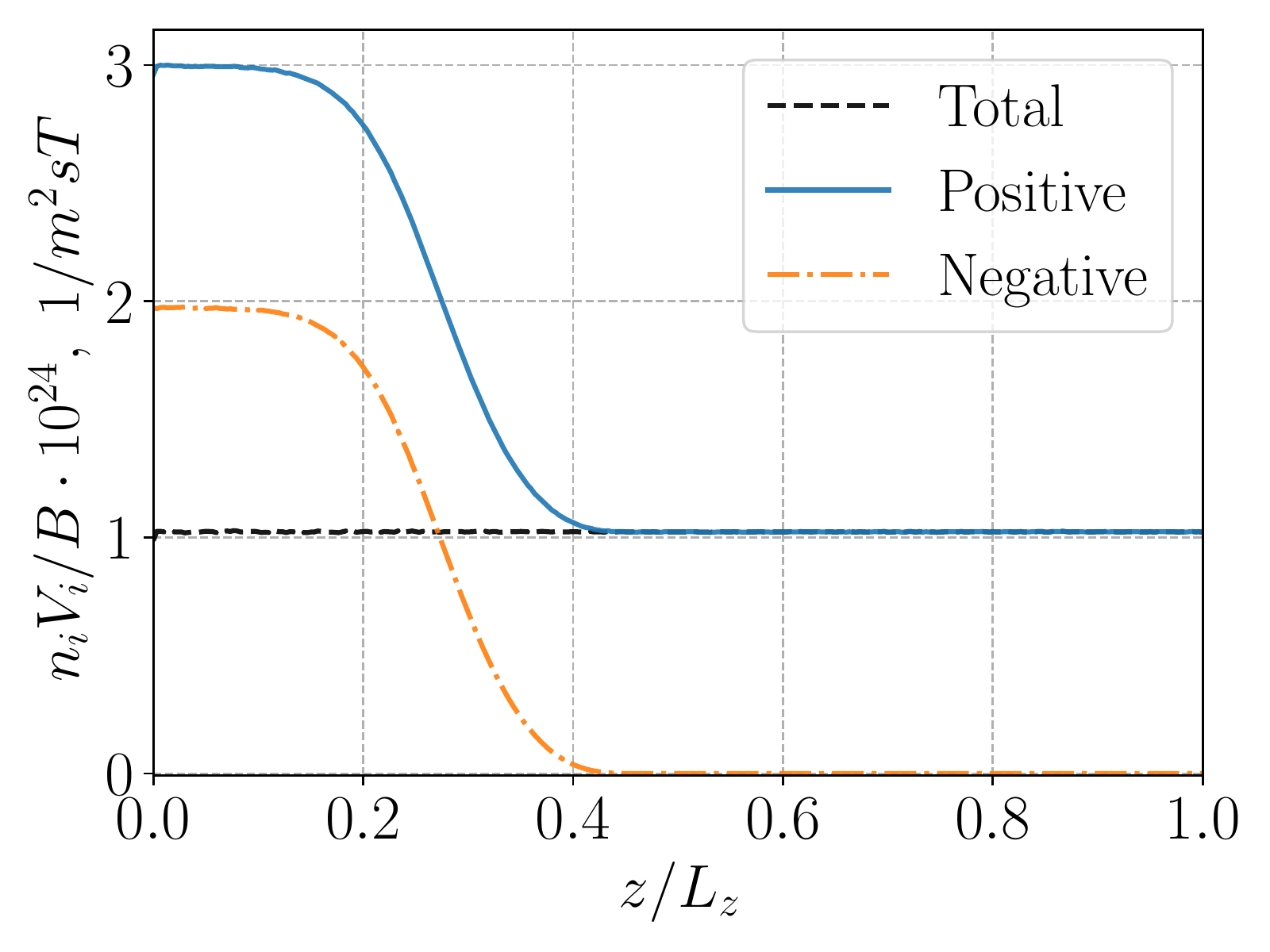}
    \caption{Flux conservation for isotropic injection, for the case with $T_i=100 \si{eV}$.}
    \label{injection2}
\end{figure}

To study the effects of a finite temperature spread,  we consider ion injection with a finite temperature from isotropic half-Maxwellian distribution in the parallel   and isotropic Maxwellian in the perpendicular directions: 
 \begin{equation}\label{halfM_iso}
f=\frac{2 n_{0} }{\pi^{3/2}V_{T}^{3}}  \exp \left( -\frac{v_{z}^{2}+v_{\bot}^{2}}
{V_{T}^{2}}\right),
\end{equation}
where the thermal velocity is $V_{T}= (2T_{}/m_i)^{1/2}$, $n_0$ is the density of  the injected beam, and only $v_z>0$ region is considered. In general anisotropic case, the corresponding temperatures are defined as:
\begin{eqnarray}
         T_z = \frac{1}{n_0} \int_{0}^{\infty} \int_{0}^{\infty} m v_{z}^{\prime 2} f \mathrm{d}v_z  2\pi v_{\perp} \mathrm{d} v_{\perp}, \\
         T_{\perp} = \frac{1}{2 n_0} \int_{0}^{\infty} \int_{0}^{\infty}  m v_{\perp}^{2} f \mathrm{d}v_z 2\pi v_{\perp} \mathrm{d}v_{\perp},
 \end{eqnarray}
 where $v_z^{\prime} = v_z - V_0$. For the isotropic half-Maxwellian distribution  $V_0 = V_{T}/\sqrt{\pi}$. Here we take $T_\perp=T_\Vert=T$. The distribution~(\ref{halfM_iso}) is sampled from the Maxwellian flux \cite{cartwright2000loading} in the parallel direction, Fig.~\ref{injection}a. 
After the steady state is reached,  the $v_z>0$ part gives the half-Maxwellian  distribution function, the $v_z<0$ corresponds to the reflected particles, Fig.~\ref{injection}b. Fig.~\ref{injection2} demonstrates the balance of the positive,  $v_z>0$, and negative fluxes, $v_z<0$, confirming the flux conservation. 

\begin{figure}
    \centering
    \subfloat[]{\includegraphics[width=0.45\linewidth]{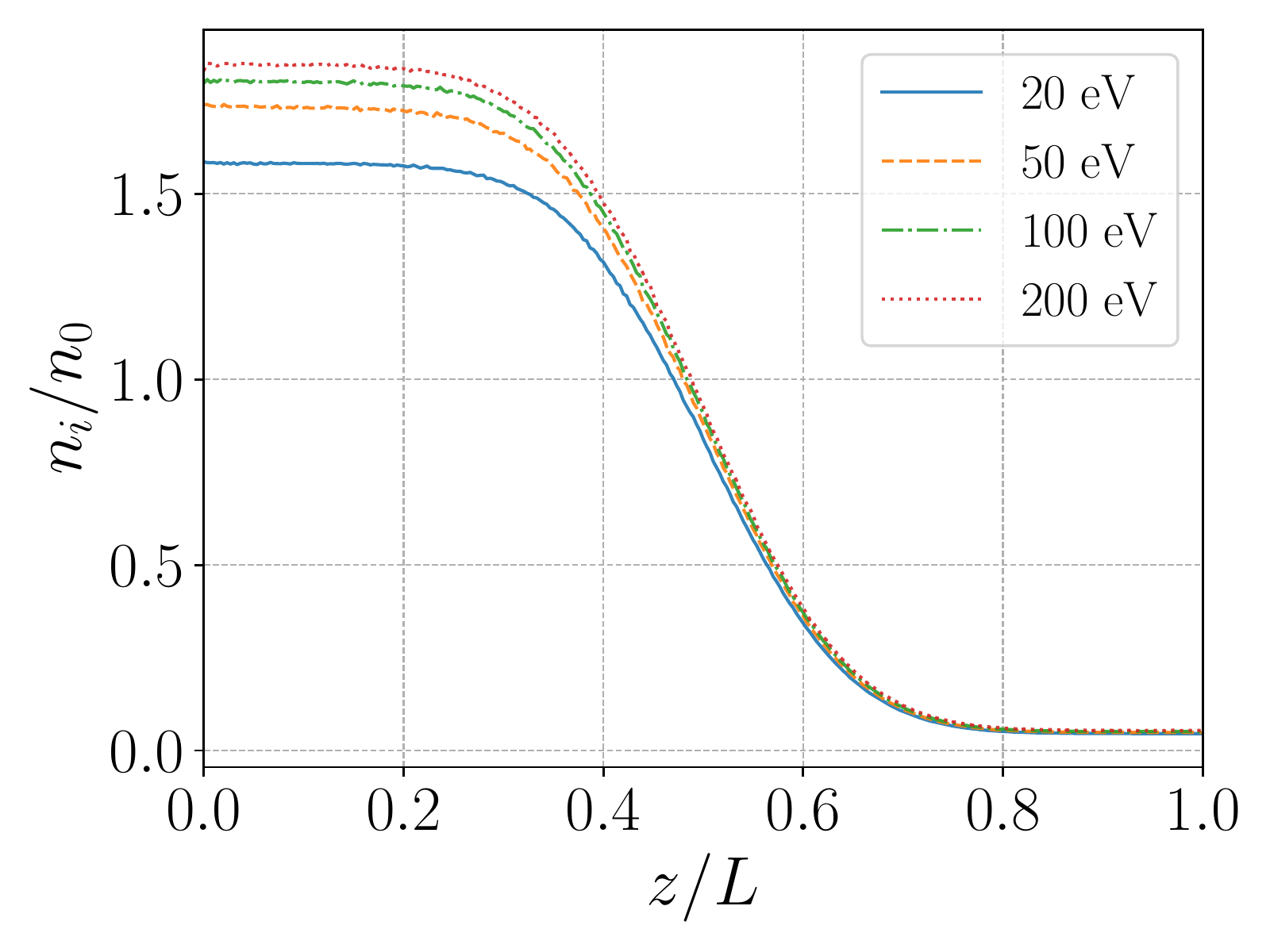}\label{iso-n}}
    \subfloat[]{\includegraphics[width=0.45\linewidth]{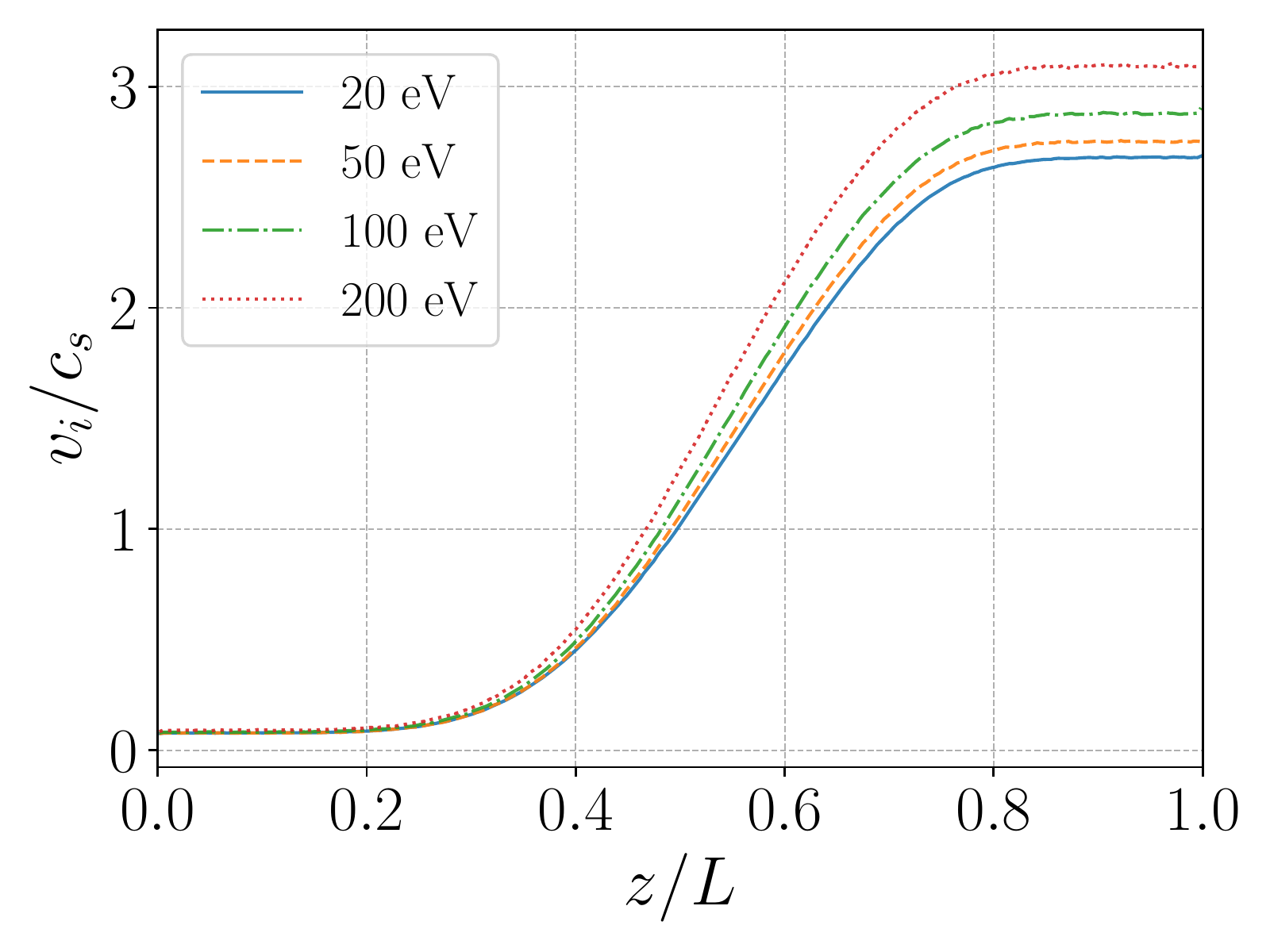}\label{iso-vi}} \\

    \subfloat[]{\includegraphics[width=0.45\linewidth]{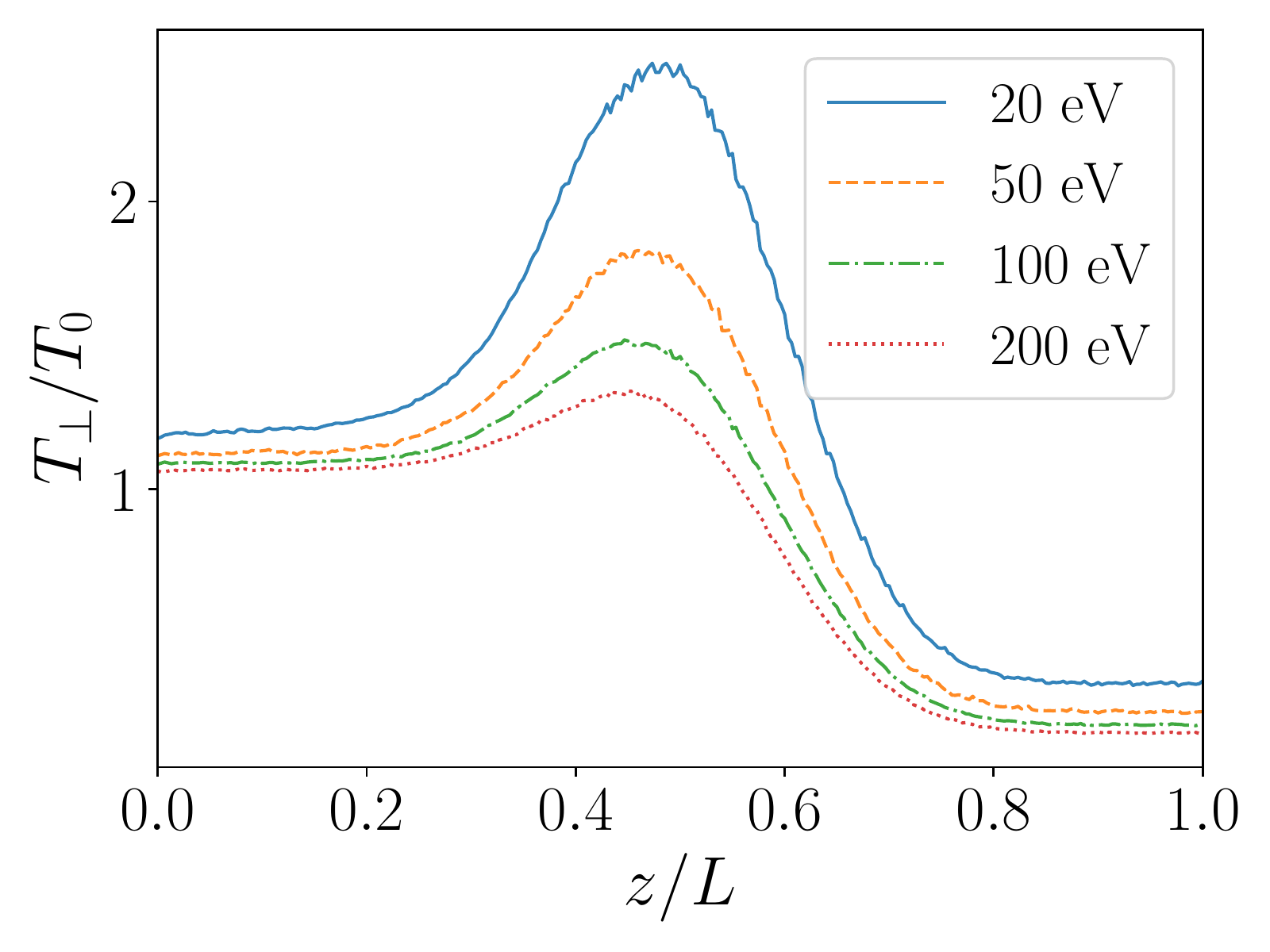}\label{iso-temp_perp}}
    \subfloat[]{\includegraphics[width=0.45\linewidth]{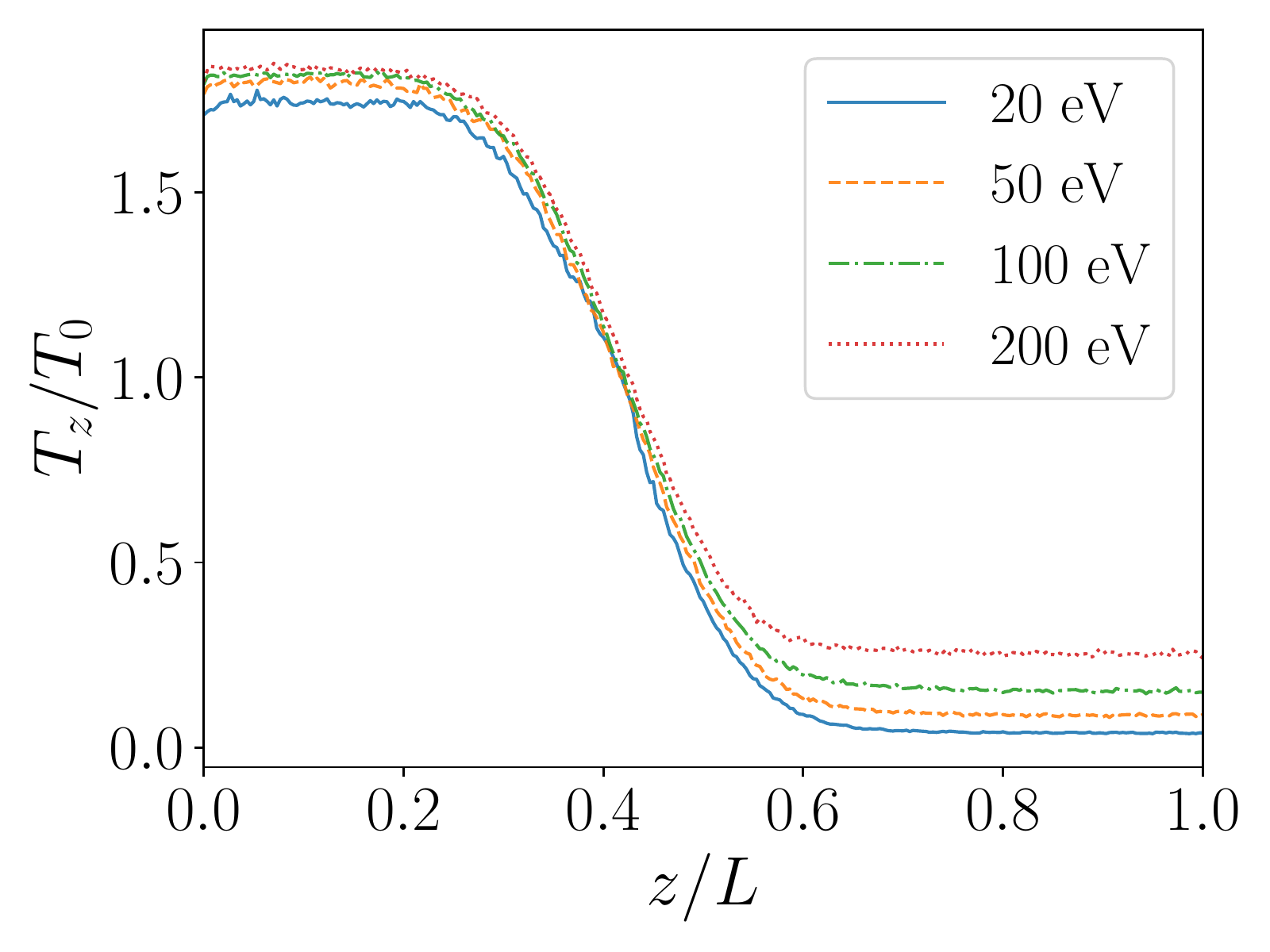}\label{iso-temp_par}}
    \caption{ Spatial profiles of plasma density (a), ion velocity (b), perpendicular temperature (c), parallel temperature (d) for isotropic injection.}
\end{figure}

\begin{figure}[H]
    \centering
   \subfloat[] {\includegraphics[width=0.5\linewidth]{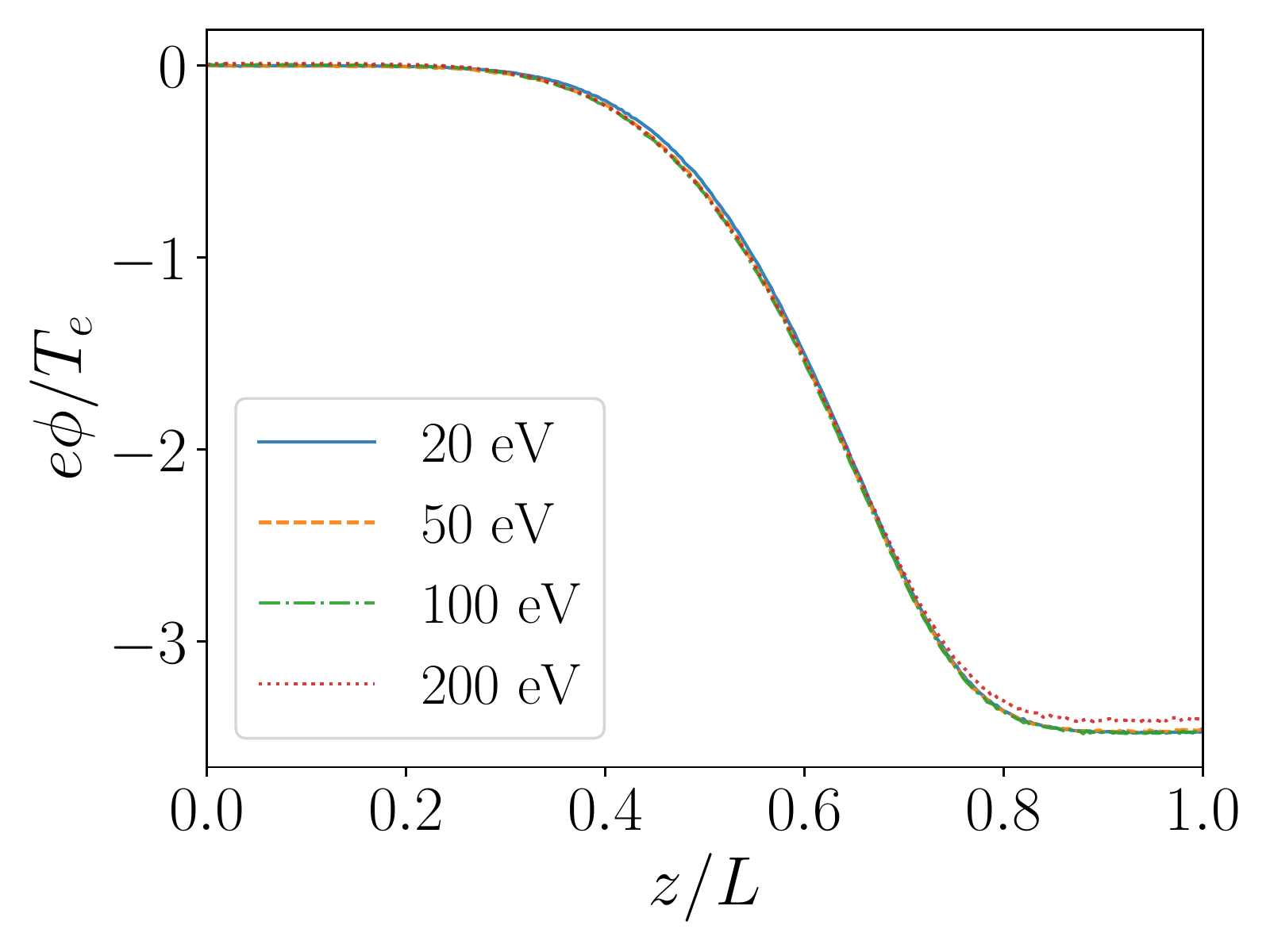}\label{iso-phi}}
    \subfloat[]{\includegraphics[width=0.5\linewidth]{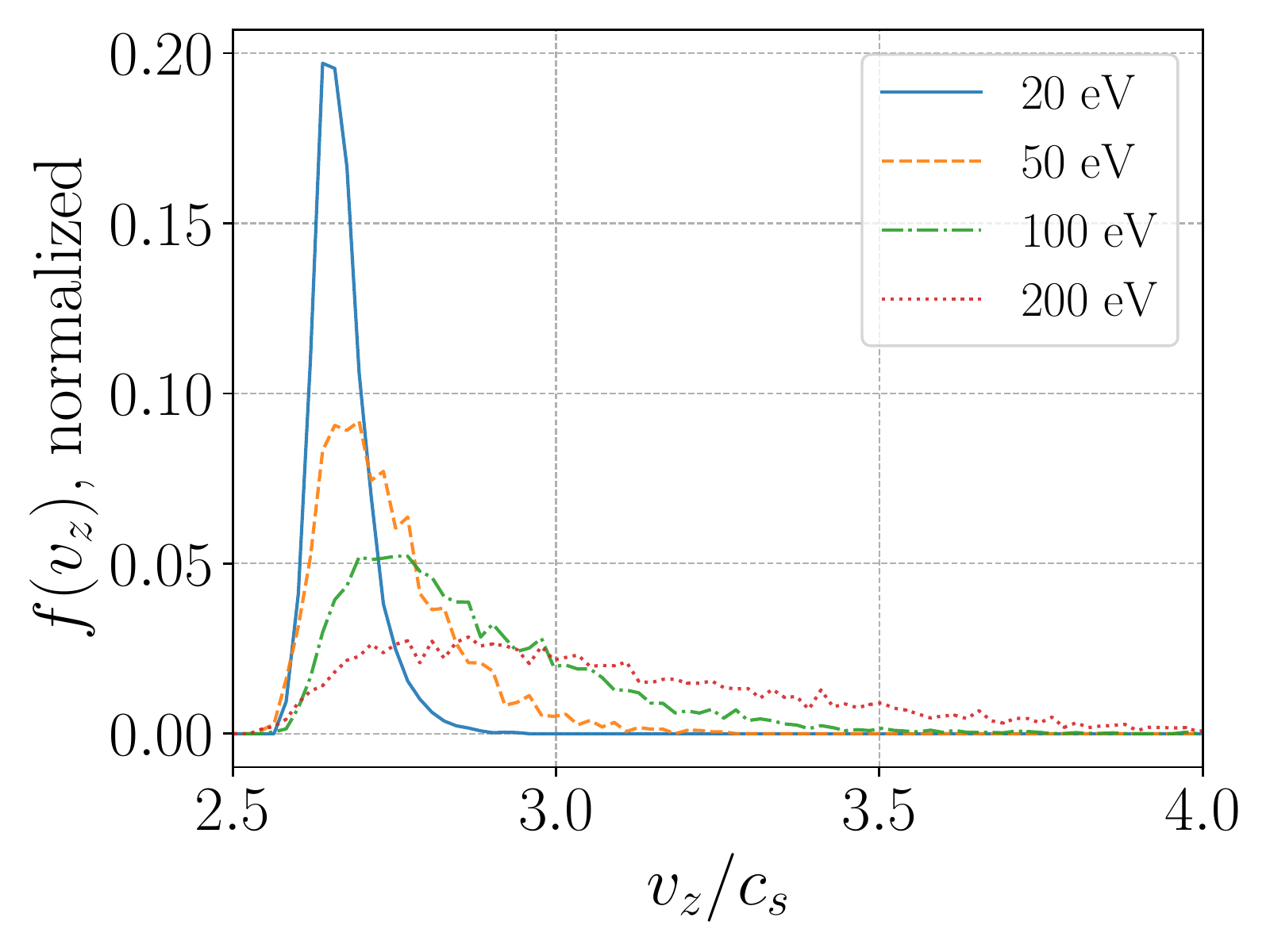}\label{iso-distri}}
    \caption{Spatial profile of electrostatic potential for isotropic injections (a), and ion distribution function near the exit $z=L$ (b).}
\end{figure}

The flux into the loss cone, $\Gamma_\Omega$, i.e.\ the flux of passing particles in absence of the electric field is   
\begin{equation}
    \Gamma_{\Omega}= \int_{0}^{+\infty} \int_{0}^{v_z \tan\theta }v_{z}f dv_z 2 \pi v_{\perp} \mathrm{d}v{_\perp}=n_0\frac{V_{T_z}} { \sqrt{\pi}} (1-\cos^2\theta),
\end{equation}
where $ \sin{\theta}=(B_{0}/B_{max})^{1/2}$ is the loss cone angle.
In absence of the accelerating potential the ratio of the  particle in the loss cone $\Gamma_{\Omega}/\Gamma_0 = 0.14$, for  $R=7$, which does not depend on the value of the  ion temperature in the isotropic case. Therefore, the ratio of reflected particles is $\Gamma_{r}/\Gamma_0 = 1 - (\Gamma_{\Omega}/\Gamma_0) = 0.86$.
The self-consistent accelerating electrostatic potential increases the fraction of passing particles.  
We consider five injection cases with isotropic  ions temperature  $T_i = (5, 20, 50, 100, 200)~\si{eV}$.
  A purely accelerating (monotonous) electrostatic potential is established for all  isotropic cases. Resulting reflection rates $\Gamma_r/\Gamma_0$ for cases with $T_i = (5, 20, 50, 100, 200)~\si{eV}$ are 0.47, 0.52, 0.68, 0.76, 0.81, respectively. Note that in all cases the reflections are reduced compared to the value  $\Gamma_{r}/\Gamma_0 = 0.86$ without the effect of the electric field.   However reflections are increased with temperature since the number of particles  in the ``prohibited'' region increases with $T_i$. 
 Larger  ion temperatures result in the  increase of the exhaust velocity at the nozzle exit, Fig.~\ref{iso-vi}, the result which is consistent  with fluid theory for warm ions \cite{SaboPoP2022}. There is little change in the velocity and potential  profiles which remain very similar for all temperatures and close to the fluid theory result, Figs.~\ref{iso-vi}-\ref{iso-phi}.  However the plasma density near the nozzle entrance is increased with temperature, Fig.~\ref{iso-n},  consistently with the increasing fraction of reflected particles. Plasma density shown in Fig.~\ref{iso-n} is normalized to the injected density $n_0=\Gamma_0/V_0$.  It is important to note that reflections occur from fluctuating potential so the parallel energy of reflected particles is larger. As a result   so the parallel temperature near the entrance is larger than the injection temperature $T_0$, Fig.~\ref{iso-temp_par}.

\begin{figure}[H]
\centering
\subfloat[]
{\includegraphics[height=0.37\textwidth]{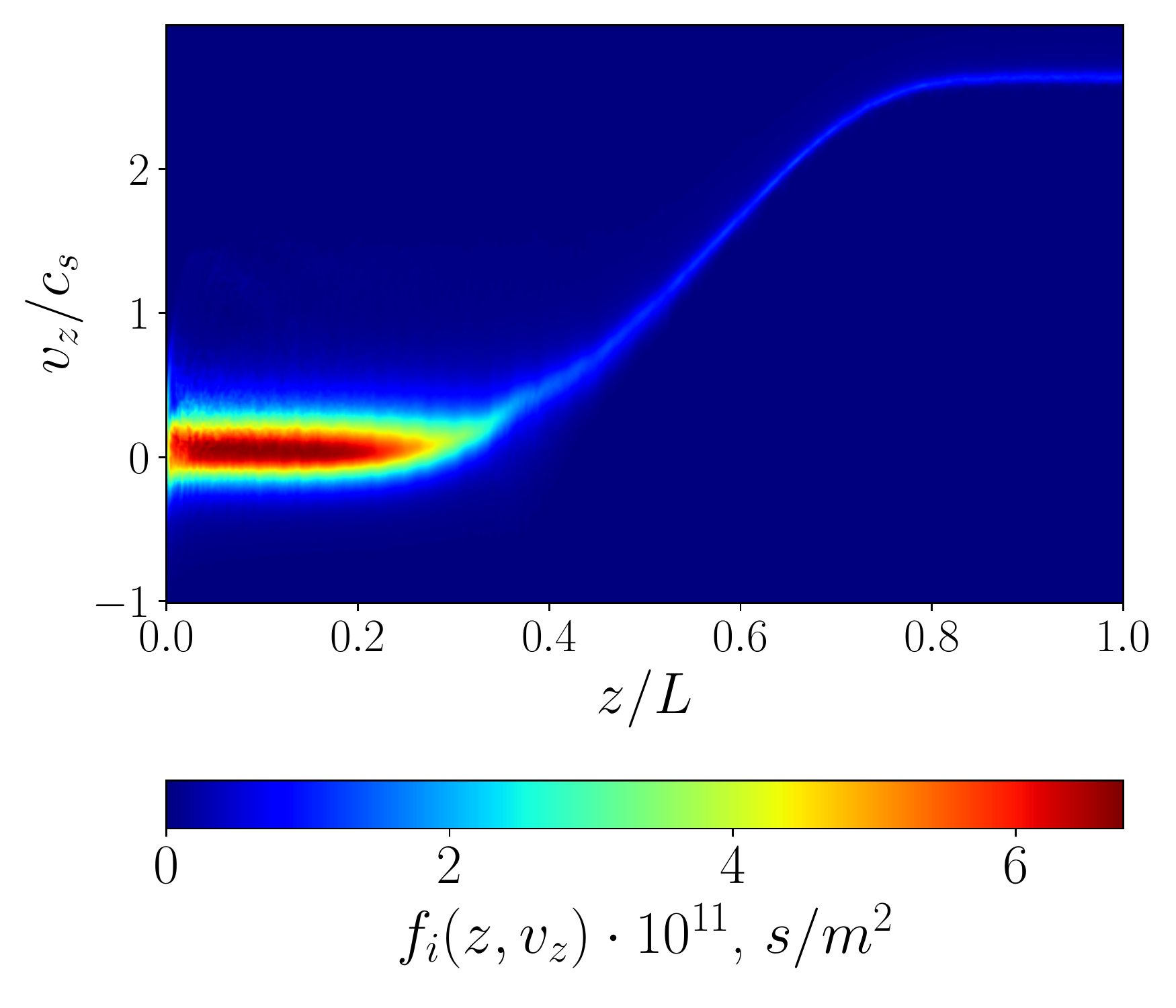}}    
\subfloat[]
{\includegraphics[height=0.37\textwidth]{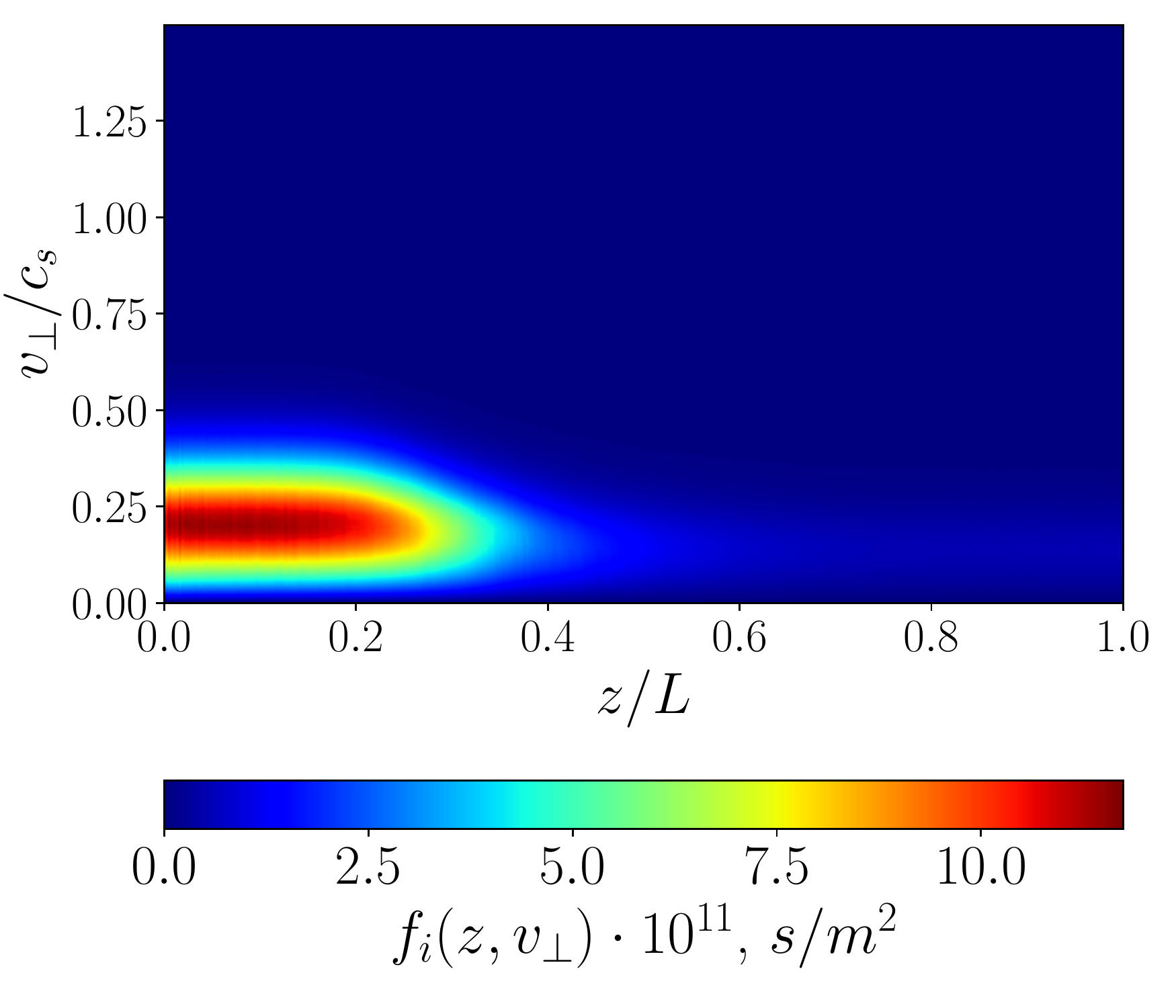}} 
\caption{Ion velocity distribution function in the phase space for isotropic injection, $T_i=\SI{5}{eV}$. The ($v_z-z$)-space, averaged over $v_\perp$ (a);
the ($v_\perp-z$)-space,  averaged over $v_z$ (b).}
 \label{ph5}
\end{figure}

\begin{figure}[H]
\centering
\subfloat[]
{\includegraphics[height=0.37\textwidth]{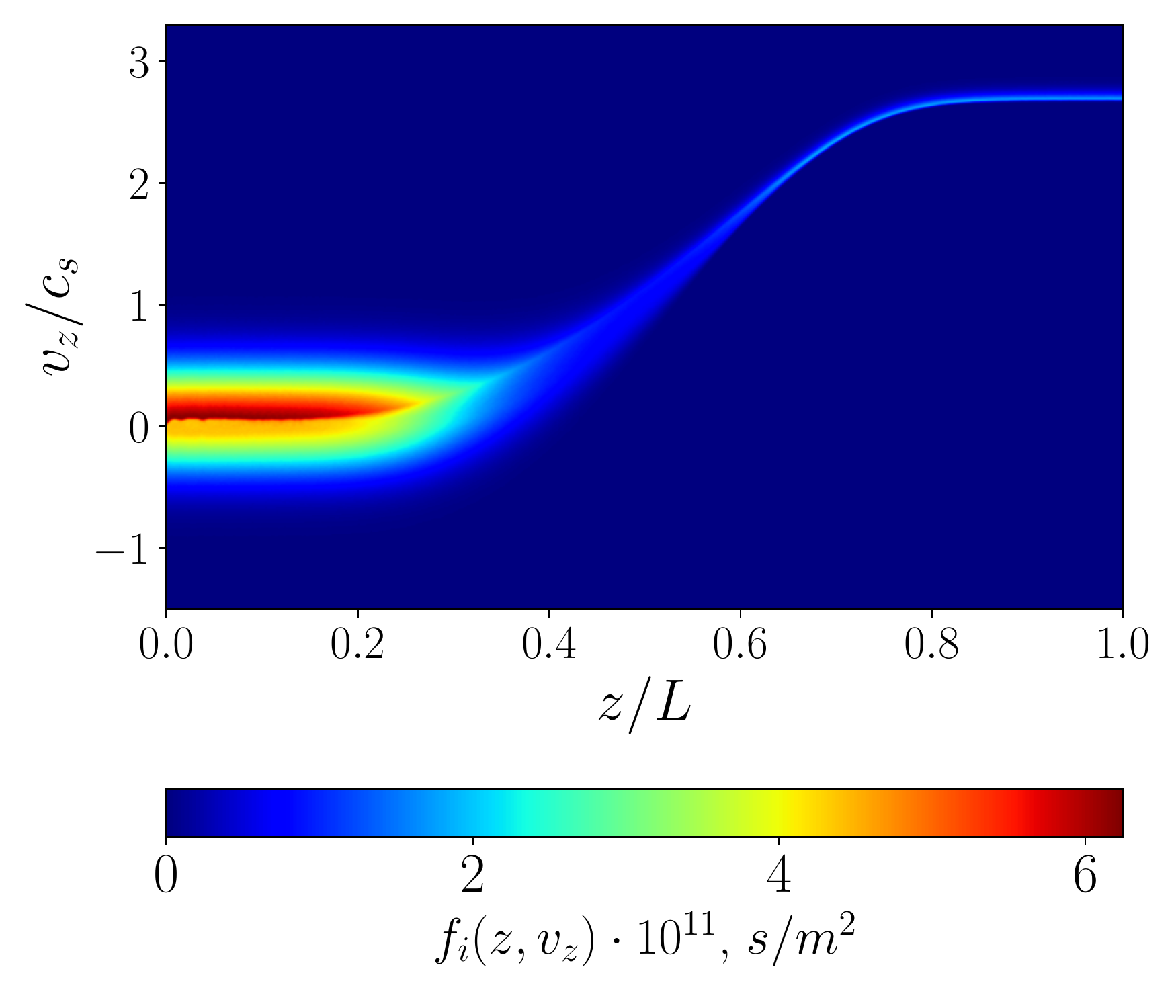}}    
\subfloat[]{\includegraphics[height=0.37
\textwidth]{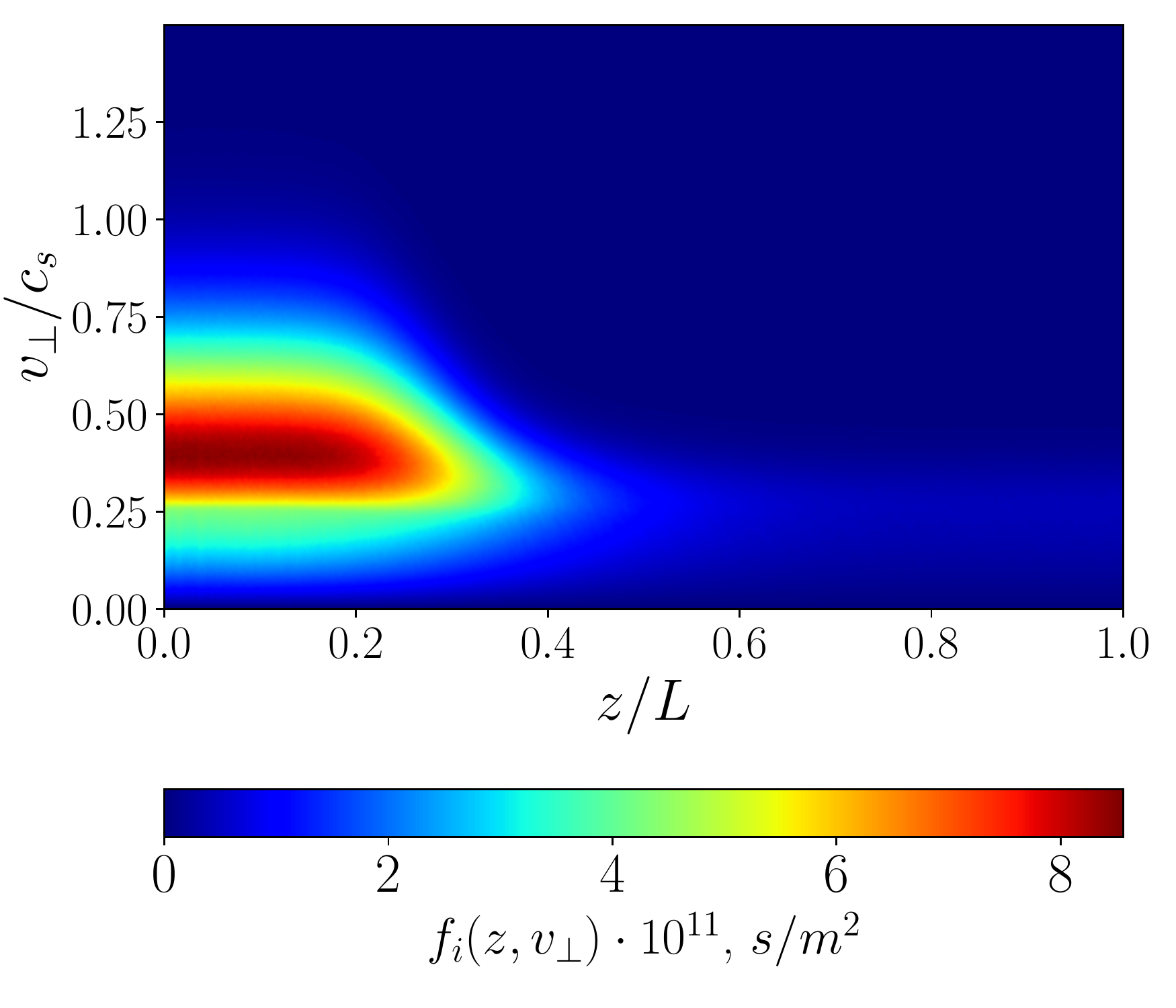}}    
\label{ph20}
\caption{Ion velocity distribution function for isotropic injection, $T_i=\SI{20}{eV}$. The ($v_z-z$)-space,  averaged over $v_\perp$ (a), the ($v_\perp-z$)-space, averaged over $v_z$ (b).}
\end{figure}

\begin{figure}[H]
\centering
\subfloat[]
{\includegraphics[height=0.35\textwidth]{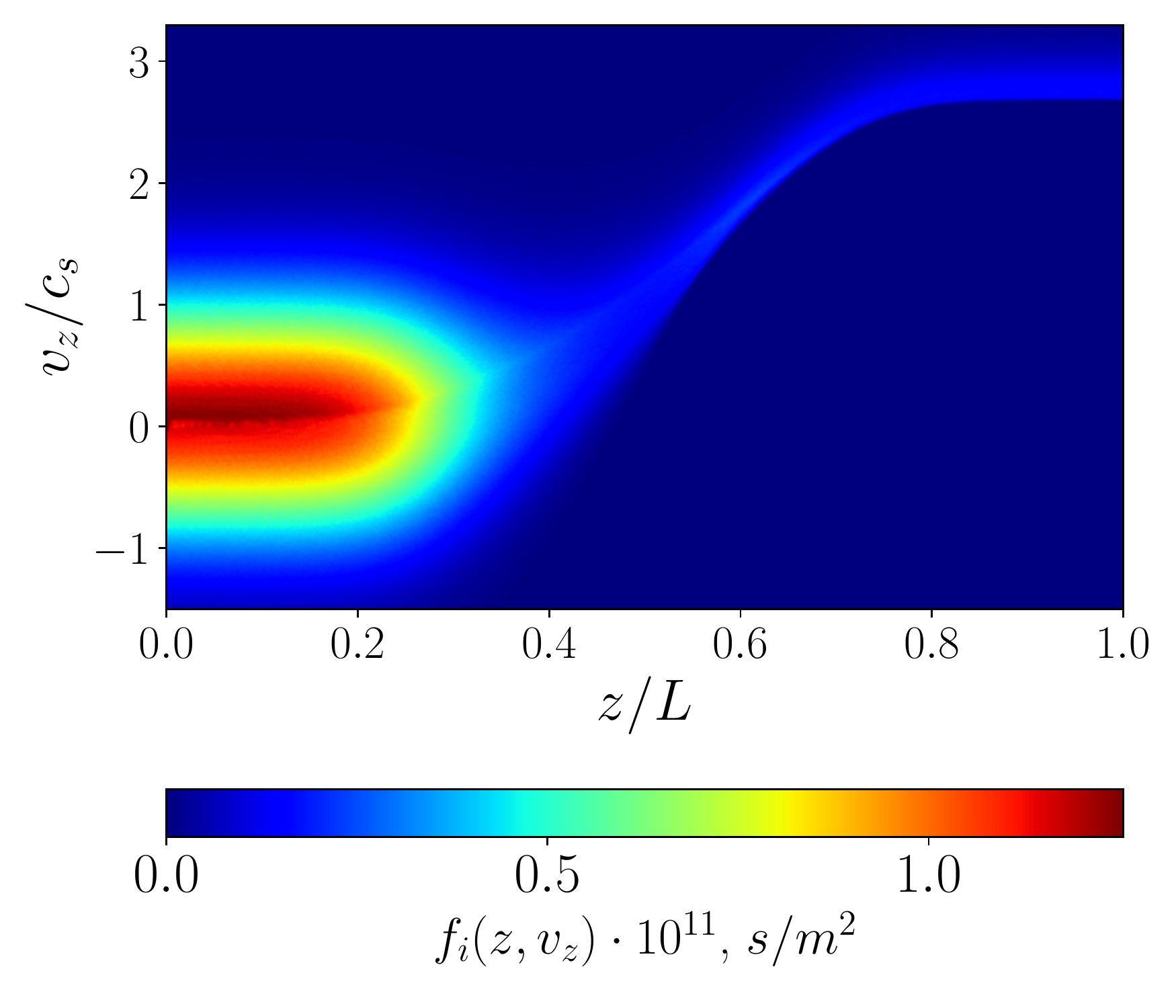}}    
\subfloat[]{\includegraphics[height=0.35
\textwidth]{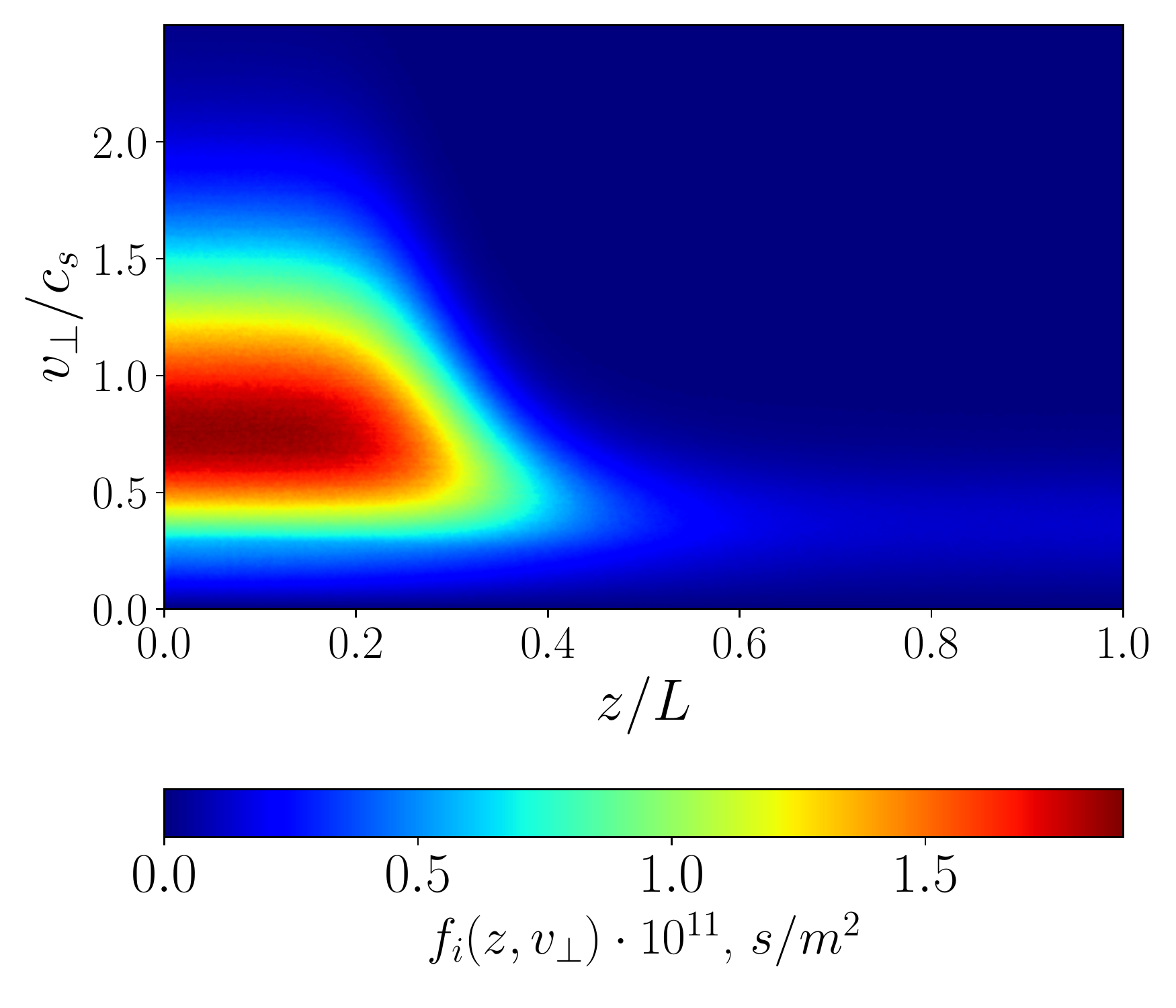}}    
\caption{Ion velocity distribution function for isotropic injection, $T_i=\SI{100}{eV}$. The ($v_z-z$)-space,   averaged over $v_\perp$ (a),
 the ($v_\perp-z$)-space, averaged over $v_z$(b).}
\label{ph100}
\end{figure}

This also explains the substantial drop  of the parallel energy in the diverging part of the nozzle, $z/L>0.5$, Fig.~\ref{iso-temp_par}. The perpendicular temperature, Fig.~\ref{iso-temp_perp}, is also reduced in that region compared to the adiabatic moment conservation law, $T_\perp \varpropto B$. The modification of the ion distribution function as a result of reflections can be seen in Fig.~\ref{iso-distri} and Figs.~\ref{ph5}-\ref{ph100}. 

Note that effective thermalization of $f_i$ in $v_z$-space of injected half-Maxwellian distribution function is achieved by reflections due to magnetic mirror force and the fluctuating potential in the first half of the system (converging region), Fig.~\ref{ph5}a. Finally, thermalization in the exhaust region is due to reflections of slower ion velocity component in the initial half-Maxwellian distribution, shown in Fig.~\ref{iso-distri}.

% The reflection ratio is:
% $ \Gamma_r/ \Gamma_0= 0.52 $ and $ \Gamma_{t}/ \Gamma_0=0.48.~
%  \Gamma_{\Omega}/\Gamma_0=0.14 $.\\

%  \begin{figure}[H]
%     \centering
%     \includegraphics[width=0.7\linewidth]{img/profs_20ev.png}
%     \caption{ Profiles, $T_{iz}=20~\si{eV}$, $T_{i\perp}=20~\si{eV}$.}
%     \label{fig:profB1}
% \end{figure}

%  \begin{figure}[H]
%     \centering
%     \includegraphics[width=0.70\linewidth]{img/profs_200.pdf}
%     \caption{Profiles for 200 \si{eV} ion temperature.}
%     \label{fig:profileB}
% \end{figure}

%\begin{figure}[H]
%\centering
%\subfloat[$f_i(z,v_z), \ \si{s/m^2}$.]
%{\includegraphics[height=0.35\textwidth]{img/ivdf_vz200.pdf}}    
%\subfloat[$f_i(z,v_{\perp}), \ \si{s/m^2}$.]{\includegraphics[height=0.35
%\textwidth]{img/ivdf_.perp_200.pdf}}    
%\caption{Ion distribution function for isotropic injection, $T_i=200$ eV. The $v_z-z$ space,   averaged over $v_\perp$ (a) ;
% the $v_\perp-z$ space,  averaged over $v_z$ (b).}
% \label{ph200}
%\end{figure}
\subsection{Effects of the anisotropic injection}
Here we consider the effects of the anisotropic injection  with two-temperature  Maxwellian distribution in the form 
 \begin{equation}\label{anis}
f=\frac{2 n_{0} }{\pi^{3/2}V_{Tz} V_{T\bot }^2}  \exp \left( -\frac{\ v_{z}^{2}}{V_{Tz}^{2}}\right) \exp \left( -\frac{v_{\bot }^{2}}{V_{T\bot }^{2}}\right),
\end{equation}
where  $V_{Tz}= (2T_{iz}/m_i)^{1/2}$ and $V_{T\perp}=(2T_{i\perp}/m_i)^{1/2}$, $n_0$ is the plasma density.
For anisotropic distribution function in Eq.~(\ref{anis}) the flux into the loss cone (in absence of the electric field) is  
 \begin{equation}\label{gama_omega_an}
    \Gamma_{\Omega}= n_0\frac{V_{T_z}} { \sqrt{\pi}} \left(1-\frac{\alpha^2}{\alpha^2+\tan^2\theta} \right),
\end{equation}     
where $\alpha=V_{T\perp}/V_{T_z}$, and $\theta= \sin^{-1}(\sqrt{1/R}).$ The analytical reflections rates $\Gamma_r/\Gamma_0$ given by Eq.~(\ref{gama_omega_an}) for the cases with fixed $T_{iz}=20~ \si{eV}$ and varying $T_{i\perp} = (0.5, 20, 50, 100, 200) ~\si{eV}$ are 0.13, 0.85, 0.93, 0.96, 0.98, respectively.

\begin{figure}[H]
    \centering
\subfloat[]{
    \includegraphics[width=0.45\linewidth]{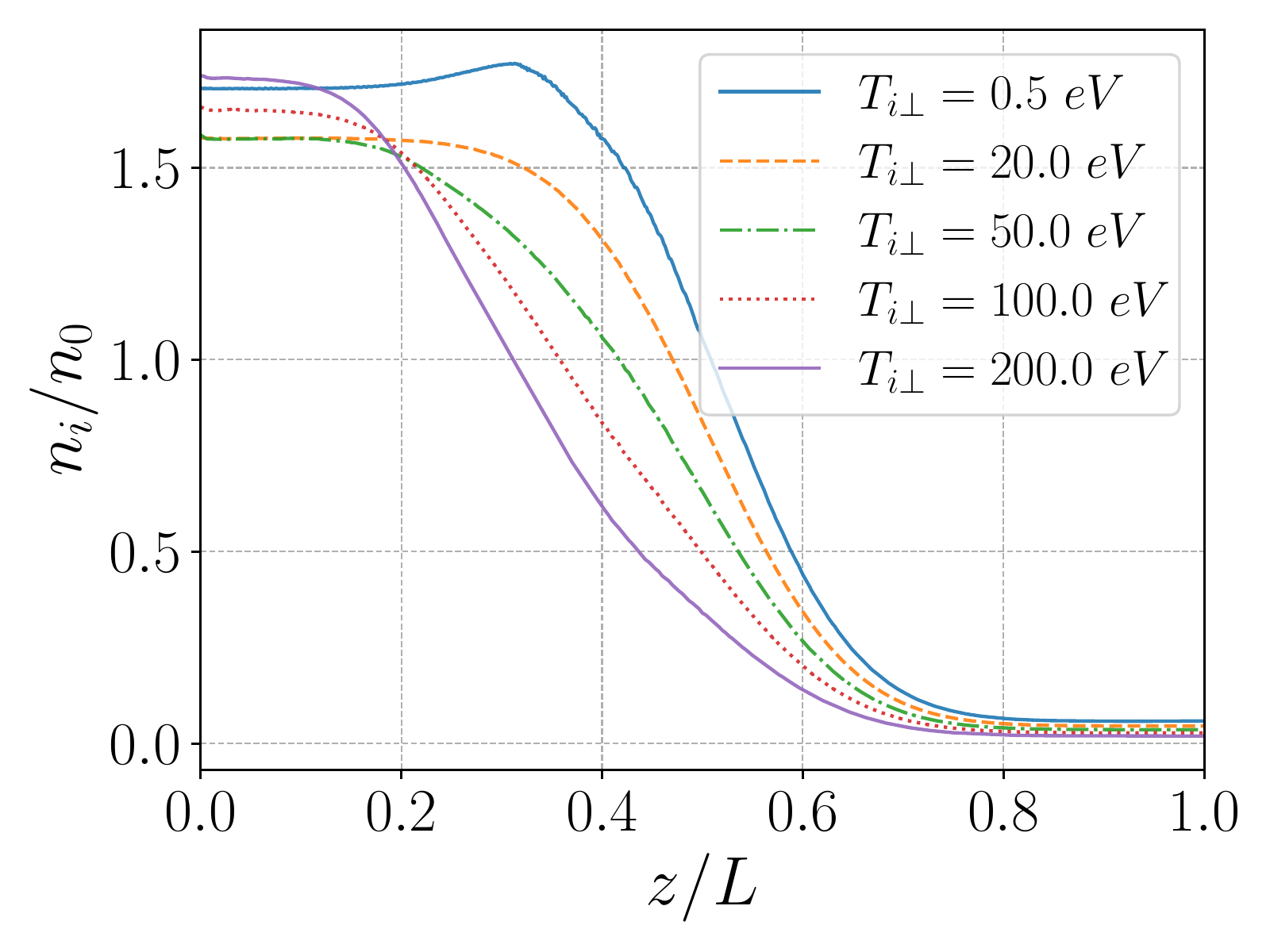}\label{ni-tperp}}
\subfloat[]{\includegraphics[width=0.45\linewidth]{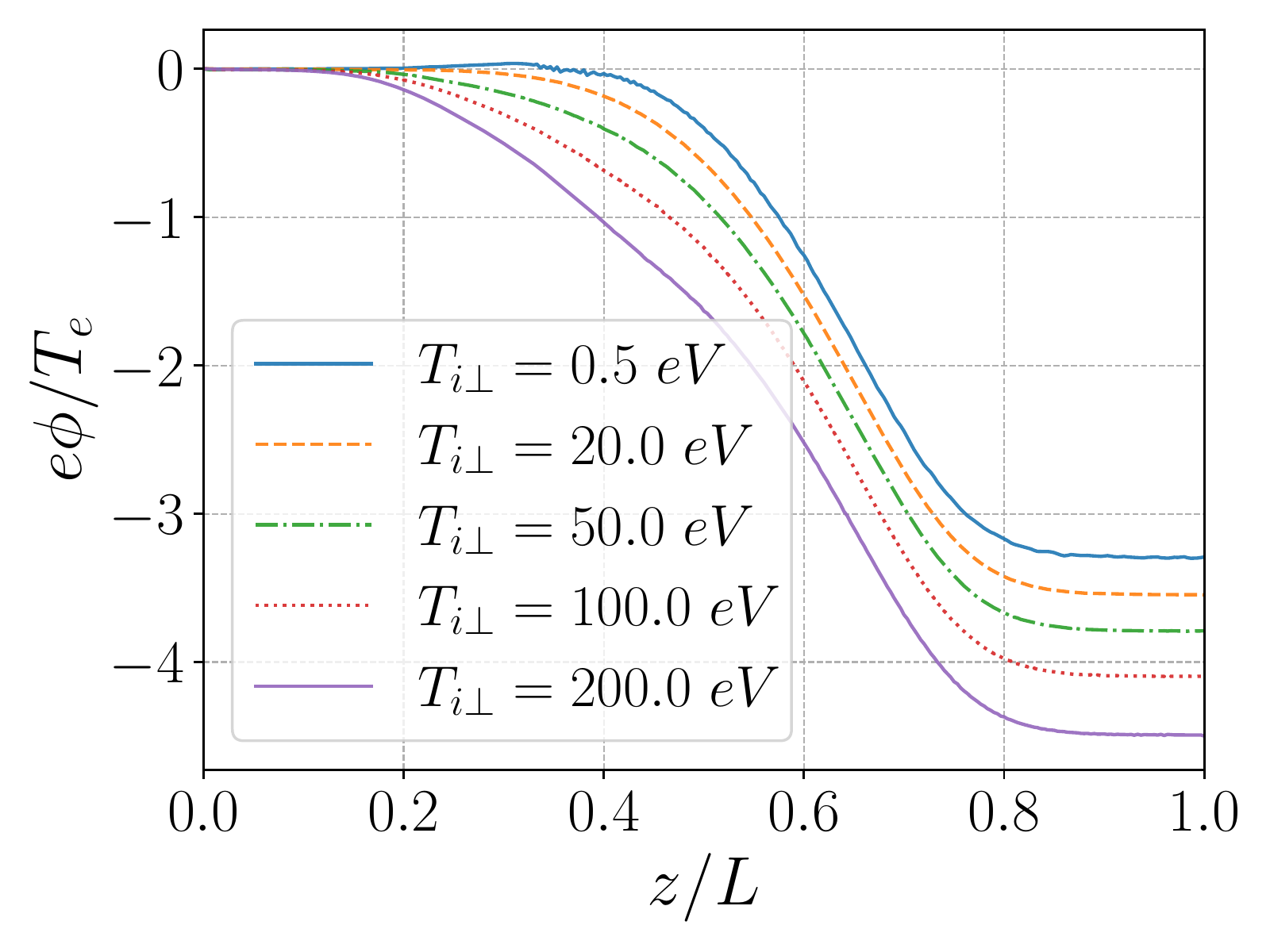}\label{phi-tperp}}
   \caption{Spatial profiles of plasma density(a), and electrostatic potential (b), for anisotropic injection, $T_{iz}=20$ eV.}
    \label{comp1}
\end{figure}

\begin{figure}[H]
    \centering
    \includegraphics[width=0.55\linewidth]{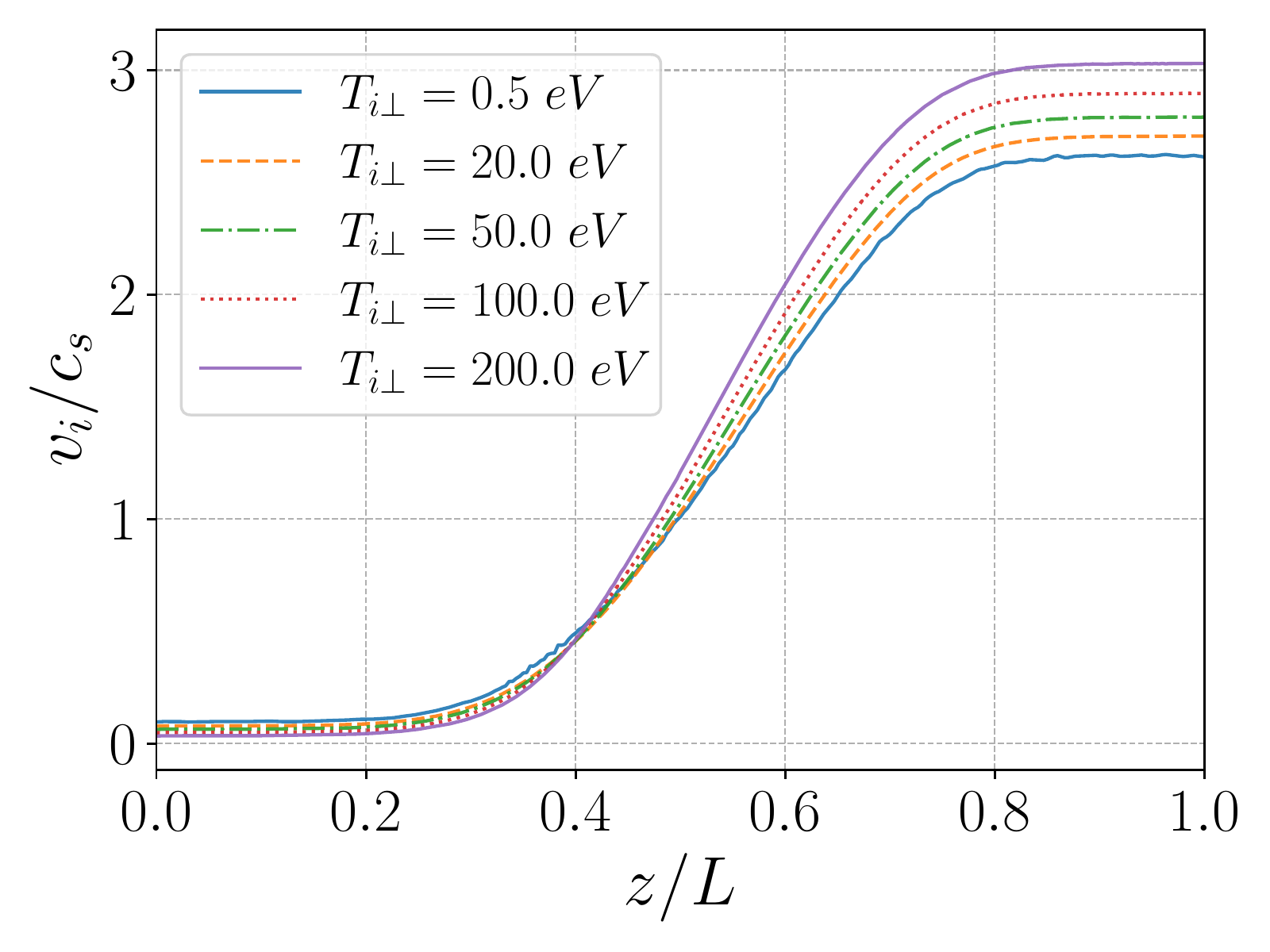}
   \caption{Ion velocity for anisotropic injection, $T_{iz}=20$ eV. }
    \label{fig18}
\end{figure}

\begin{figure}[H]
    \centering
  \subfloat[]{
    \includegraphics[width=0.45\linewidth]{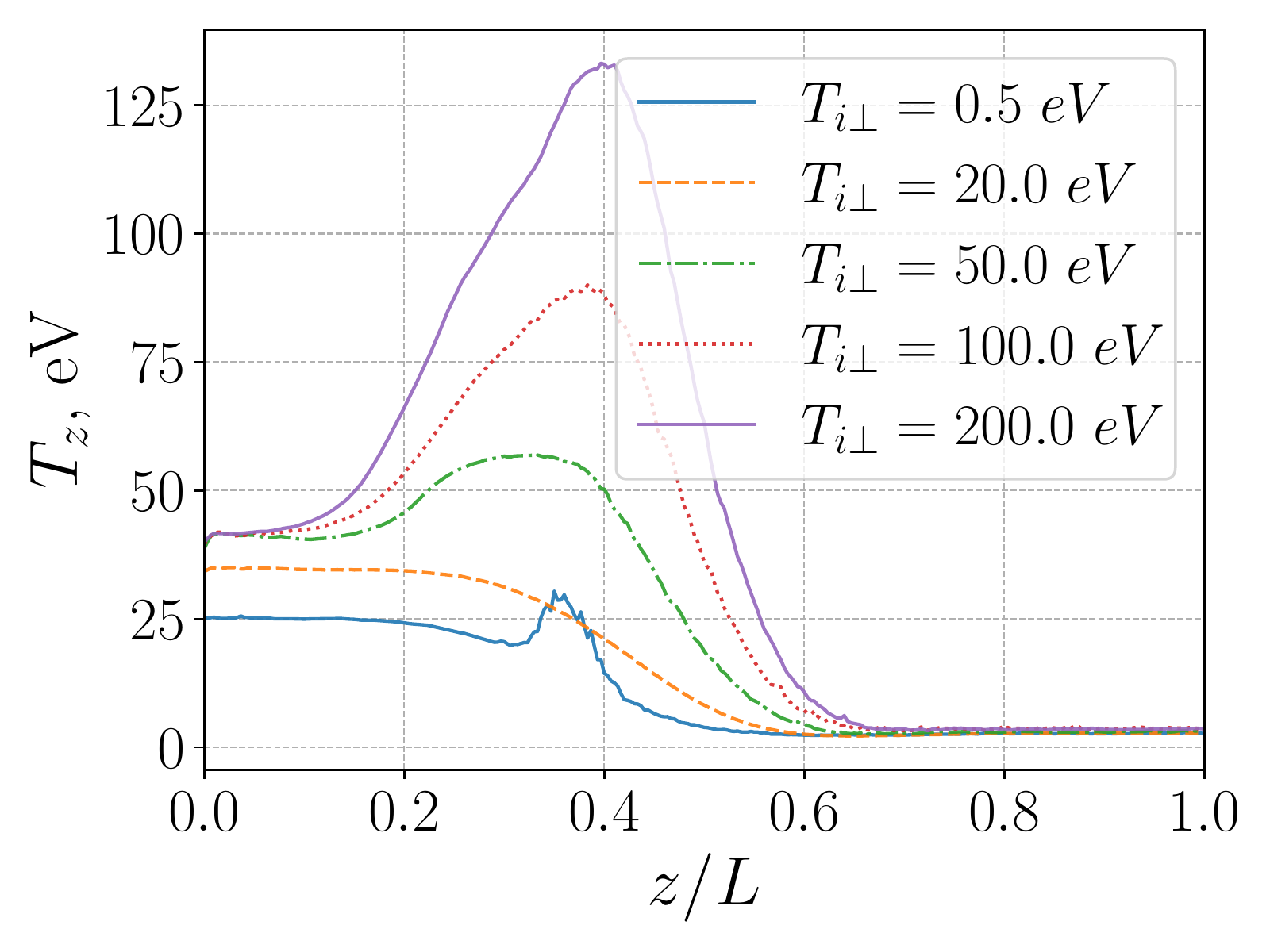}\label{tix-perp}}
\subfloat[]{
    \includegraphics[width=0.45\linewidth]{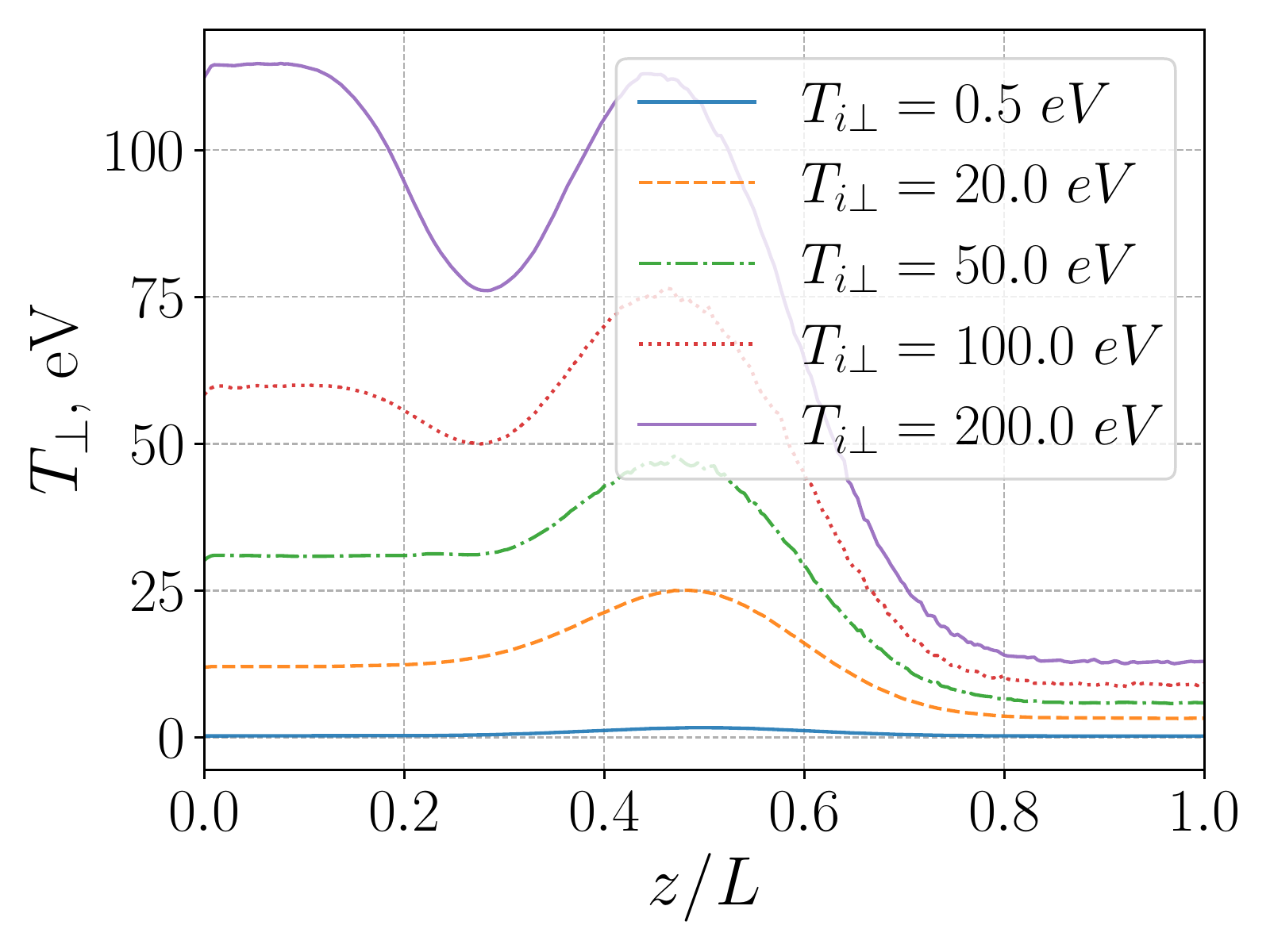}\label{tiperp-perp}}
   \caption{Spatial profiles of parallel (a), and perpendicular (b) temperature for anisotropic injection, $T_{iz}=20$ eV.}
    \label{fig19}
\end{figure}

Consistent with the above results, the measured reflections rates $\Gamma_r/\Gamma_0$ for cases with fixed $T_{iz}=20~ \si{eV}$ and varying $T_{i\perp} = (0.5, 20, 50, 100, 200)~\si{eV}$ are 0.37, 0.51, 0.61, 0.68, 0.77, respectively. Therefore, when increasing the perpendicular temperature the reflection rate increases as expected, however, due to the self-consistent electric field, these reflections are lower than what predict the analytical values. Note that for $T_{i\perp} = \SI{0.5}{eV}$ the value is not lower than the theoretical value, due to non-monotonic behaviour of the electrostatic potential Fig.~\ref{ni-tperp}. The ion velocity and temperature profiles for these injections are shown in Figs.~\ref{fig18}-\ref{fig19}.

The analytical reflections rates $\Gamma_r/\Gamma_0$ given by Eq.~(\ref{gama_omega_an}) for the cases with fixed $T_{i\perp}=20~ \si{eV}$ and varying $T_{iz} = (0.5, 20, 50, 100, 200) ~\si{eV}$ are $0.99,0.85,0.70.0.54,0.37$, respectively. As expected, when increasing the parallel temperature, the reflection rate decreases. 

\begin{figure}[H]
    \centering
 \subfloat[]{
     \includegraphics[width=0.45\linewidth]{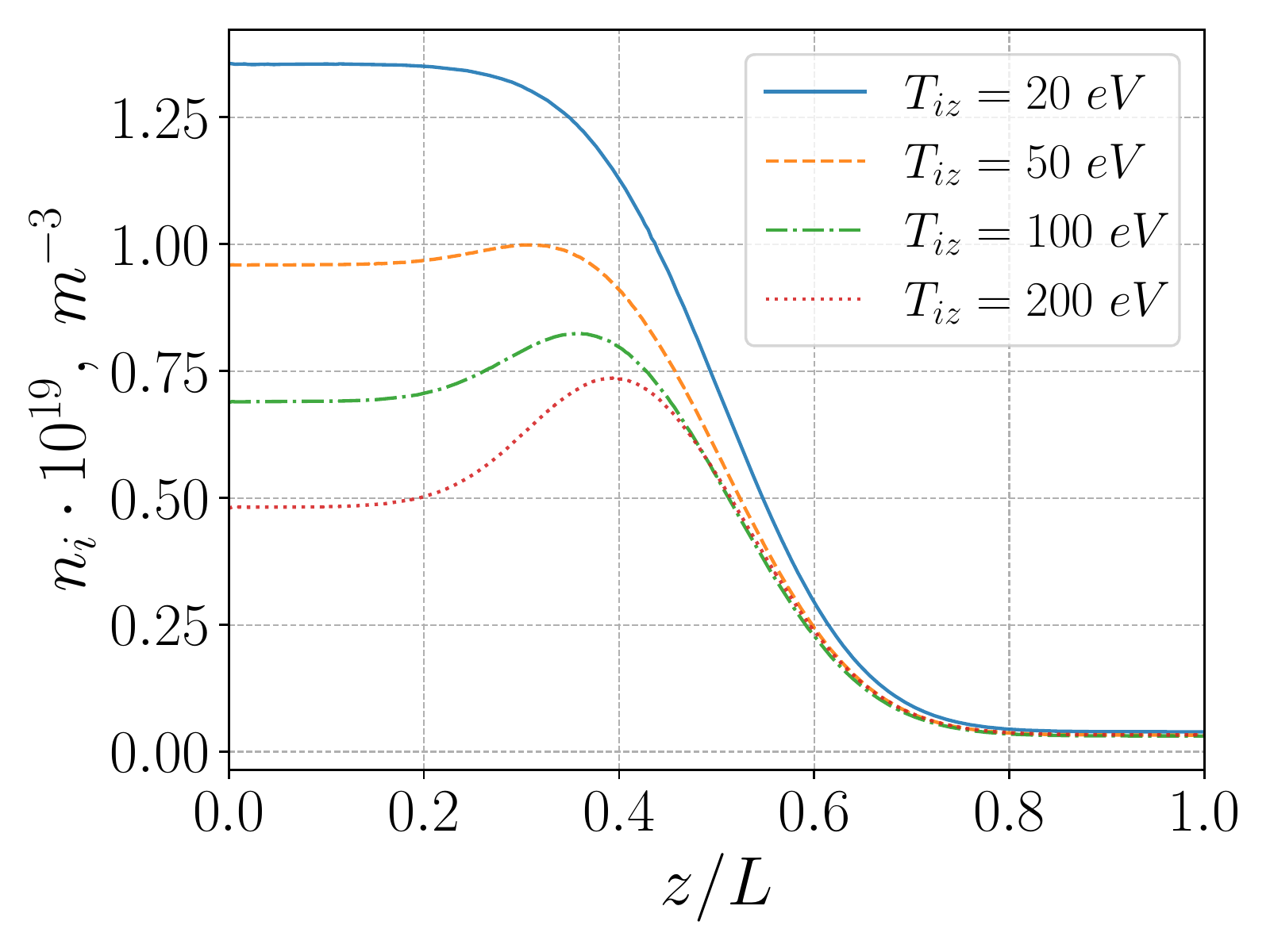}\label{comp6a}}
    \subfloat[]{
    \includegraphics[width=0.45\linewidth]{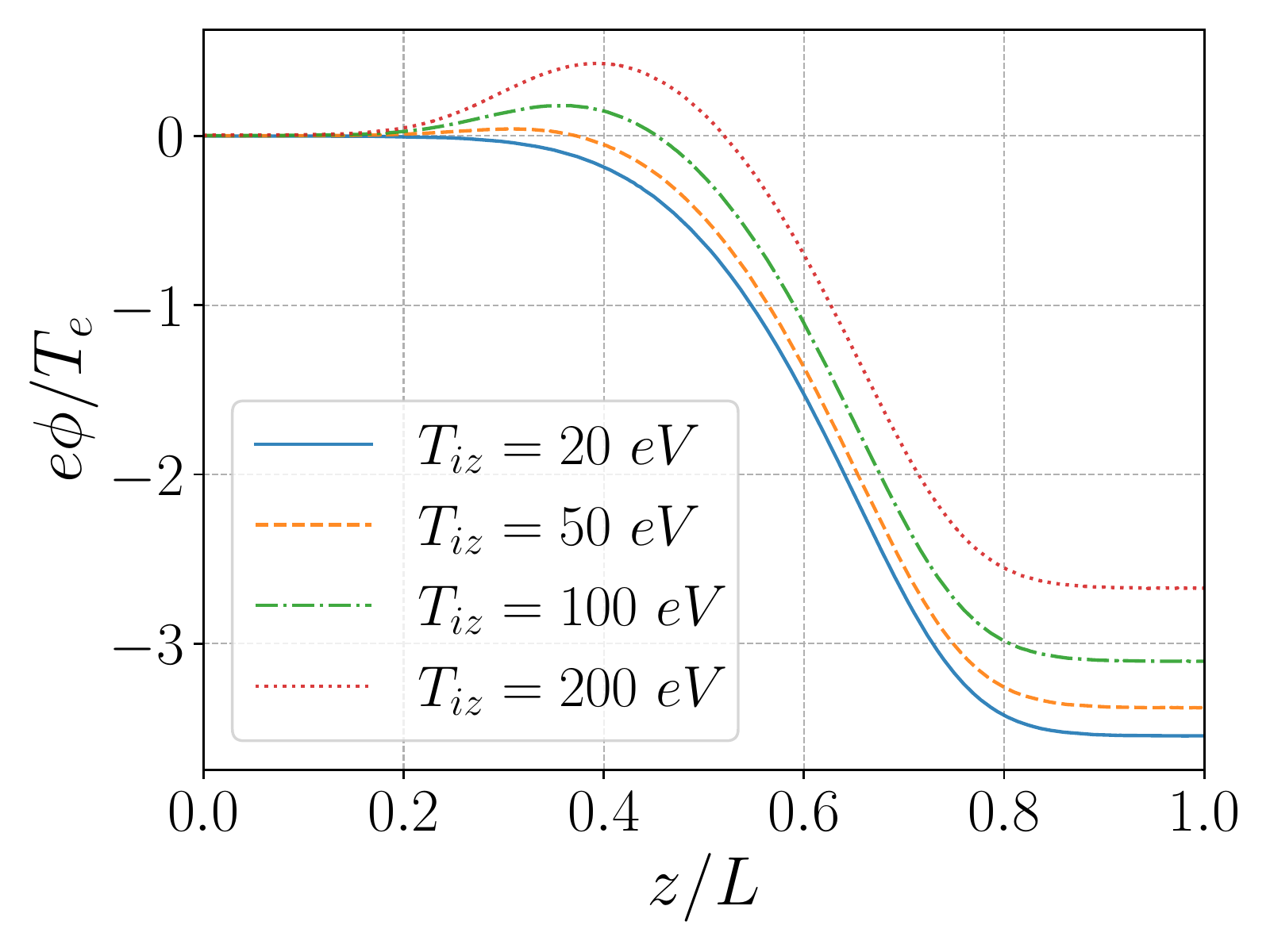}\label{comp6b}}
   \caption{Spatial profiles of plasma density (a) and electrostatic potential (b) for anisotropic injection, $T_{\perp}=\SI{20}{eV}$.}
    \label{comp6}
\end{figure}

\begin{figure}[H]
    \centering
    \includegraphics[width=0.55\linewidth]{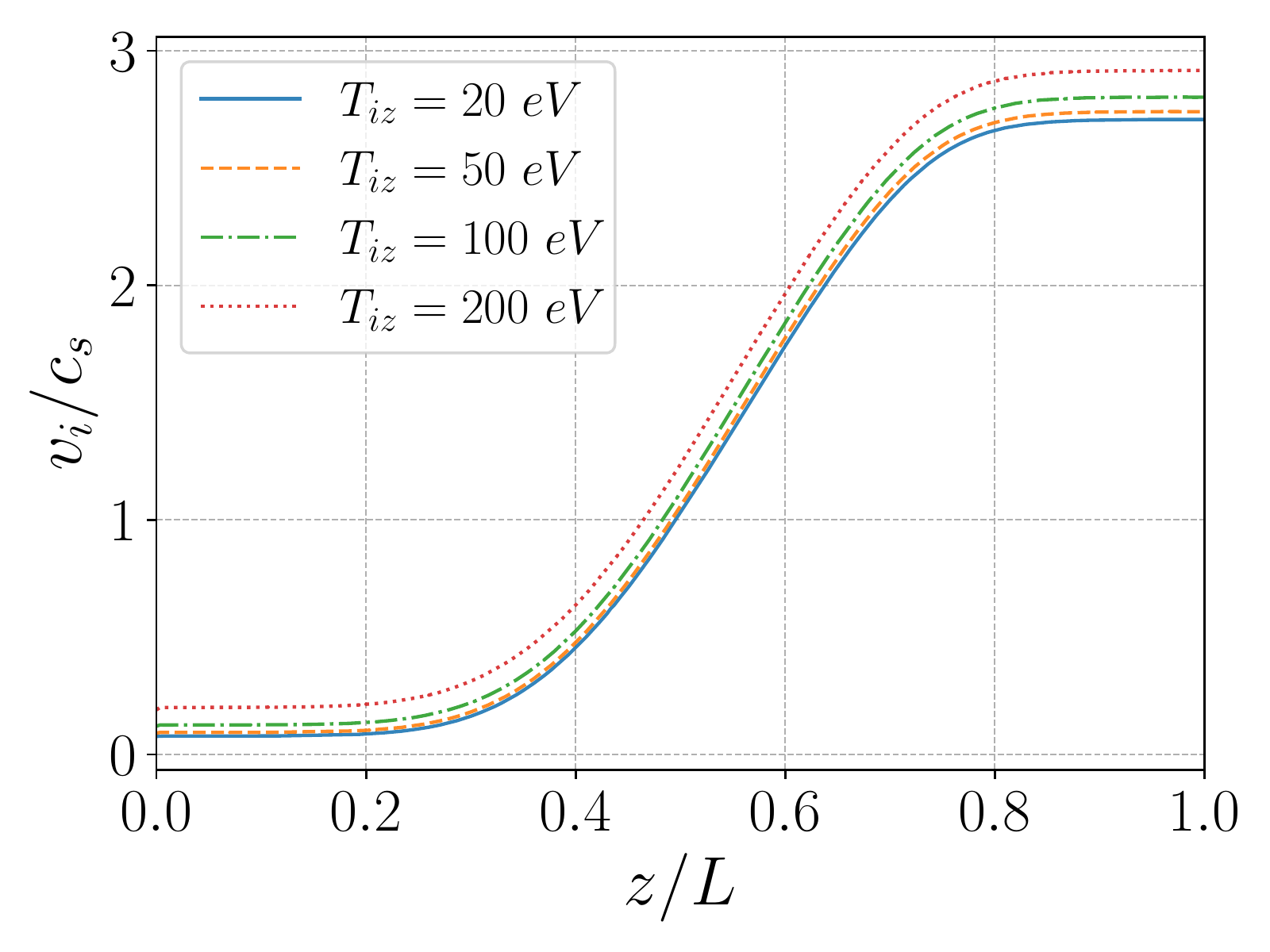}
   \caption{Spatial profile of flow velocity for anisotropic injection, $T_{\perp}=\SI{20}{eV}$.}
    \label{comp7}
\end{figure}

The measured reflection rates $\Gamma_r/\Gamma_0$ for cases with fixed $T_{i\perp}=20~ \si{eV}$ and varying $T_{iz} = (20, 50, 100, 200)~\si{eV}$ are $0.51, 0.59, 0.61, 0.56$ respectively. In these cases, the resulting reflections rate do not decrease as the analytical values, instead, the values are approximately stay the same due to the appearance of the non-monotonic electrostatic potential structures. The plasma density, potential, and ion velocity for these injections are shown in Figs.~\ref{comp6}-\ref{comp7}.  

Anisotropic injection reveals non-monotonic behavior in density profile  due to the appearance of a negative electric field structure in the convergent region, see Figs.~\ref{comp6}. These features are generally  observed when the perpendicular ion temperature is lower than the parallel ion temperature ($T_{i\perp}< T_{iz}$). This effect occurs as  a result of large number of particles within the prohibited region, $v_0>c_s M_a$, and are similar to the ion flow stalling effects observed with cold injection, cf. Figs.~\ref{cold1}-\ref{cold4}.

%Flux $\Gamma_{0} = n V_0 = \SI{5.655e23}{m^{-2} s^{-1}}$.

%  \begin{figure}[H]
%     \centering
%  \subfloat[]{
%      \includegraphics[width=0.45\linewidth]{comp/tix_tperp20_tz_eff.pdf}\label{}}
% \subfloat[]{
%     \includegraphics[width=0.45\linewidth]{comp/tiperp_tperp20_tz_eff.pdf}\label{}}
%   \caption{Spatial profiles of parallel (a), and perpendicular (b) temperatures for anisotropic injection, $T_{\perp}=20$ eV.}
%     \label{comp10}
% \end{figure}
% \begin{figure}[H]
%     \centering
% \subfloat[]{
%     \includegraphics[width=0.45\linewidth]{comp/ivdf_x_vx_f200_tz20_tperp100.pdf}}
% \subfloat[]{
%  \includegraphics[width=0.45\linewidth]{comp/ivdf_x_vperp_f200_tz20_tperp100.pdf}}
%     \caption{Ion velocity distribution function for anisotropic injection, $T_{iz}=20$ eV and  $T_{\perp}=100$ eV. The $v_z-z$ space,   averaged over $v_\perp$ (a),
%  the $v_\perp-z$ space,   averaged over $v_z$ (b).}
%     \label{fig:my_label}
% \end{figure}
% \begin{figure}[H]
%     \centering
% \subfloat[]{
%     \includegraphics[width=0.45\linewidth]{comp/ivdf_x_vx_f100_tz100_tperp20.pdf}}
% \subfloat[]{
%  \includegraphics[width=0.45\linewidth]{comp/ivdf_x_vperp_f100_tz100_tperp20.pdf}}
%     \caption{Ion velocity distribution function for anisotropic injection, $T_{iz}=100$ eV and  $T_{\perp}=20$ eV. The $v_z-z$ space,   averaged over $v_\perp$ (a),
%  the $v_\perp-z$ space,   averaged over $v_z$ (b).}
%     \label{fig:my_label}
% \end{figure}

\begin{figure}[H]
\centering

\subfloat[$z/L=0$]{
    \includegraphics[width=0.36\linewidth]{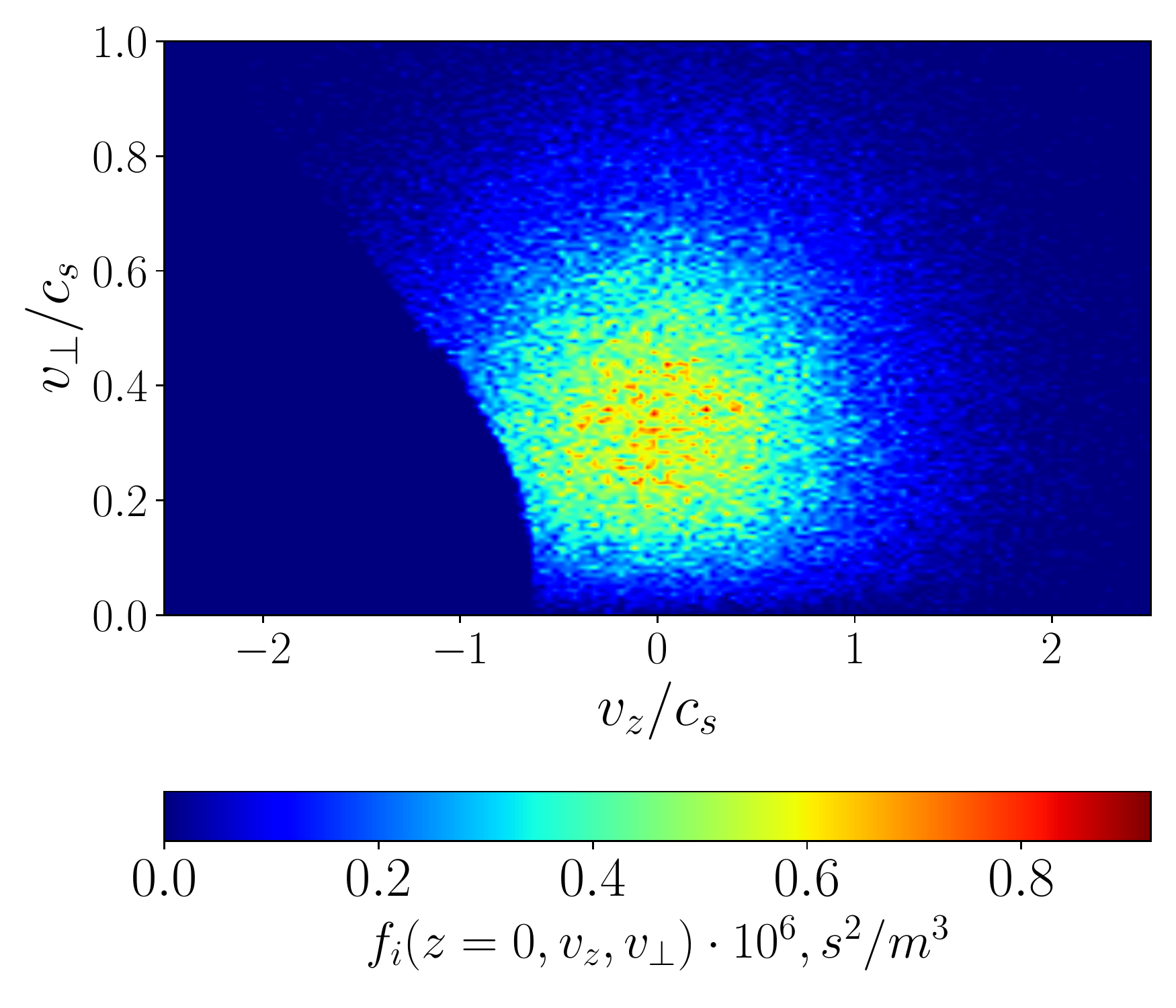}}
\subfloat[$z/L=0.25$]{
 \includegraphics[width=0.36\linewidth]{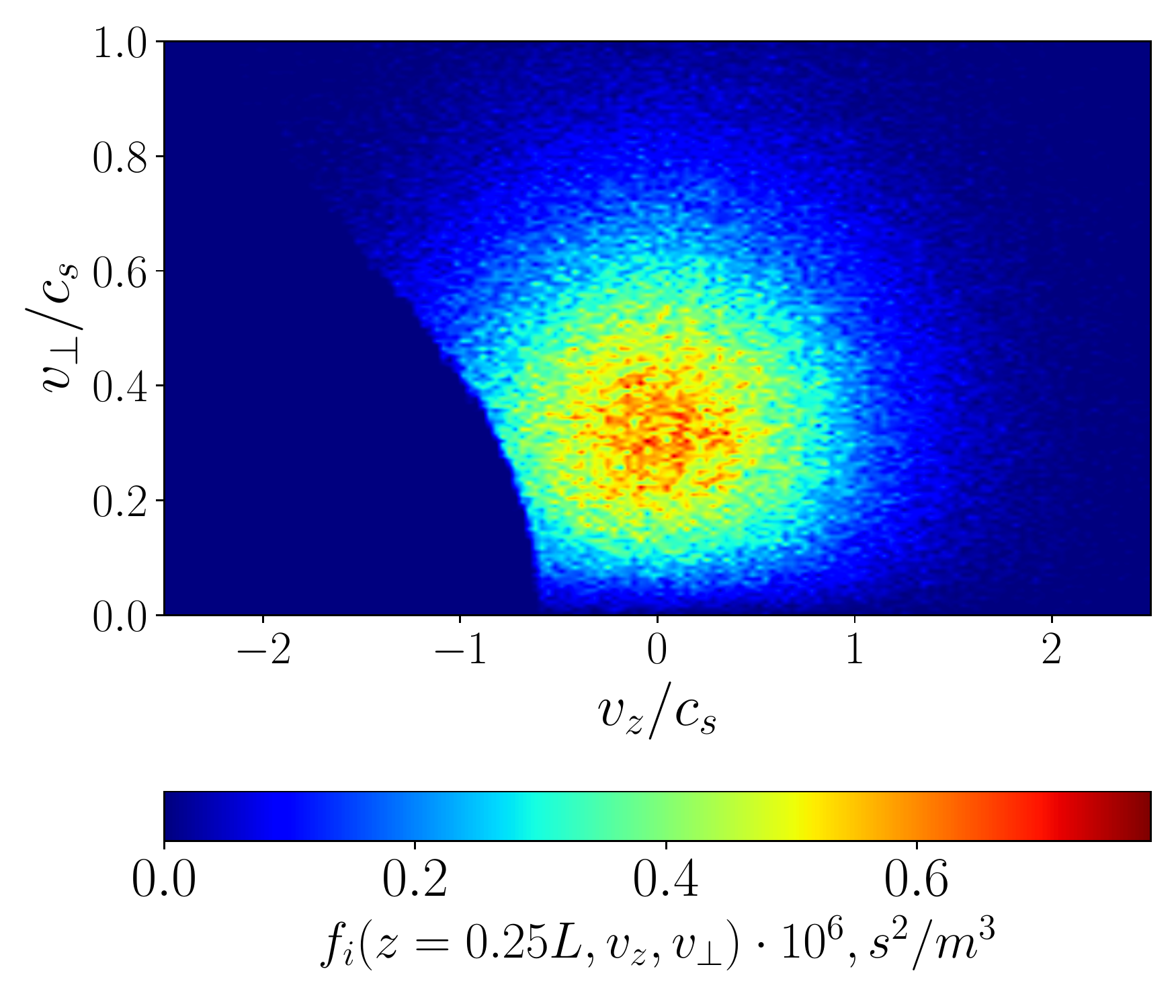}}
 \subfloat[$x/L=0.4$]{
 \includegraphics[width=0.36\linewidth]{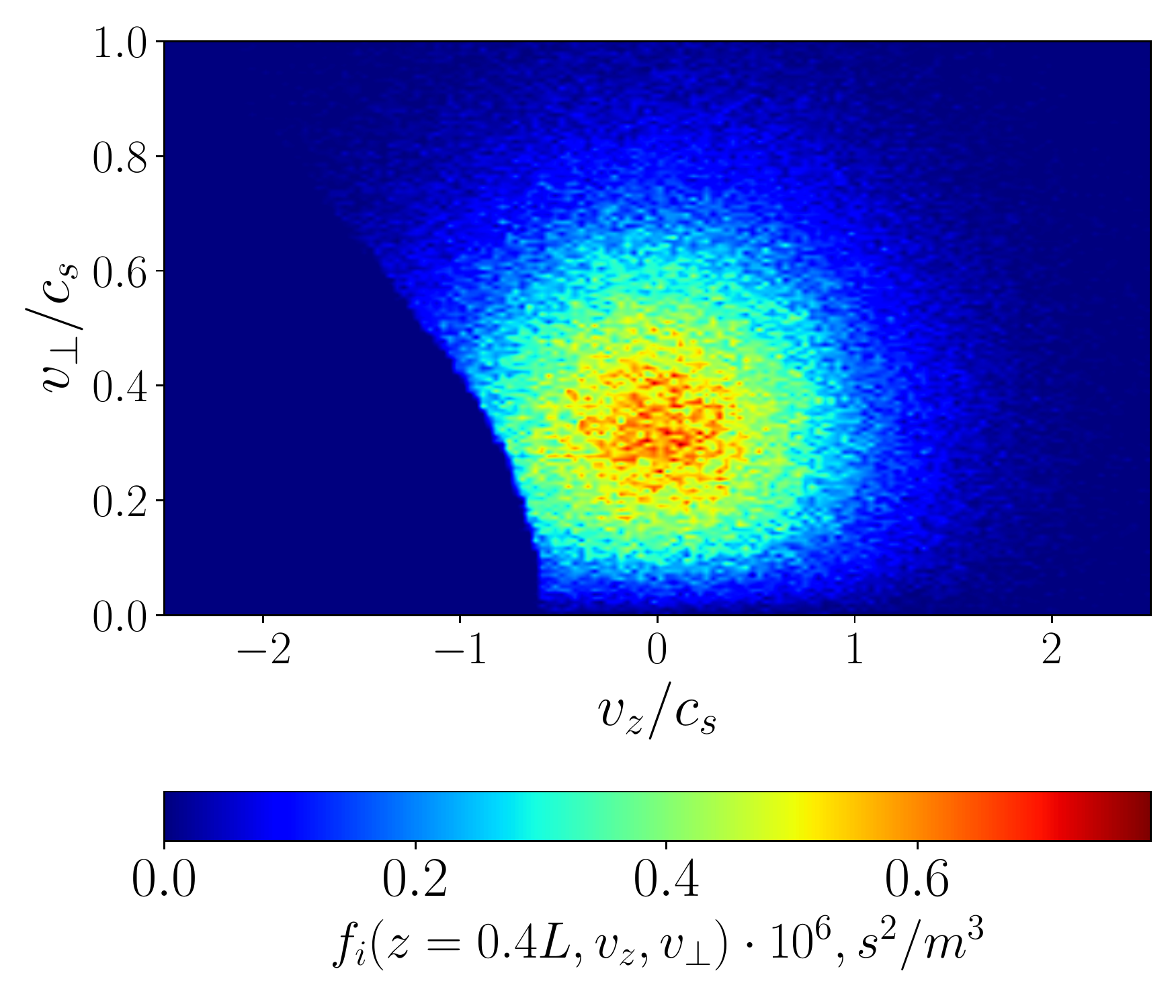}}
    \caption{Ion velocity distribution function for anisotropic injection, $T_{iz}=\SI{100}{eV}$ and $T_{\perp}=\SI{20}{eV}$ at different locations.}
    \label{anisotr_100_20}
\end{figure}

\begin{figure}[H]
    \centering
\subfloat[$z/L=0$]{
    \includegraphics[width=0.36\linewidth]{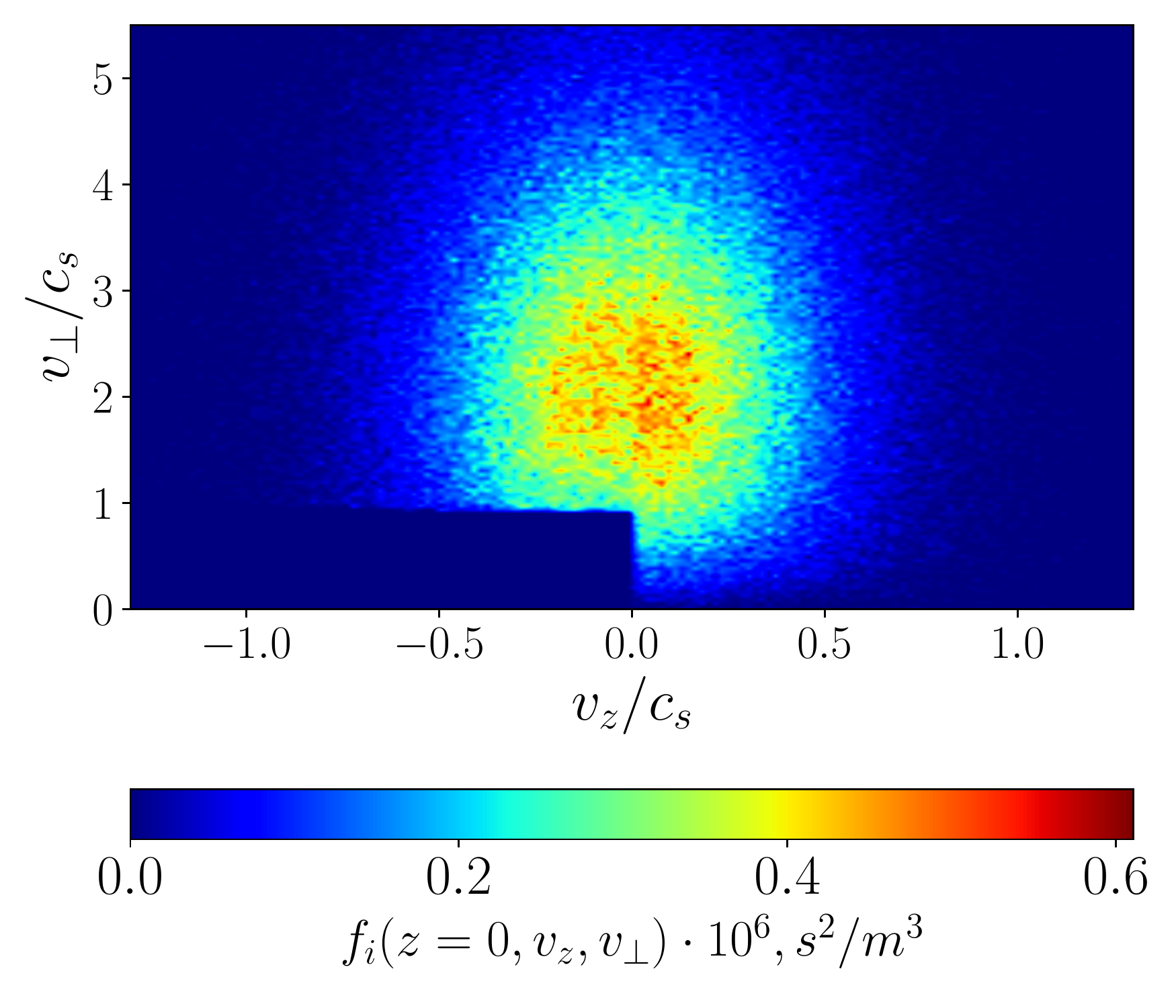}}
\subfloat[$z/L=0.25$]{
 \includegraphics[width=0.36\linewidth]{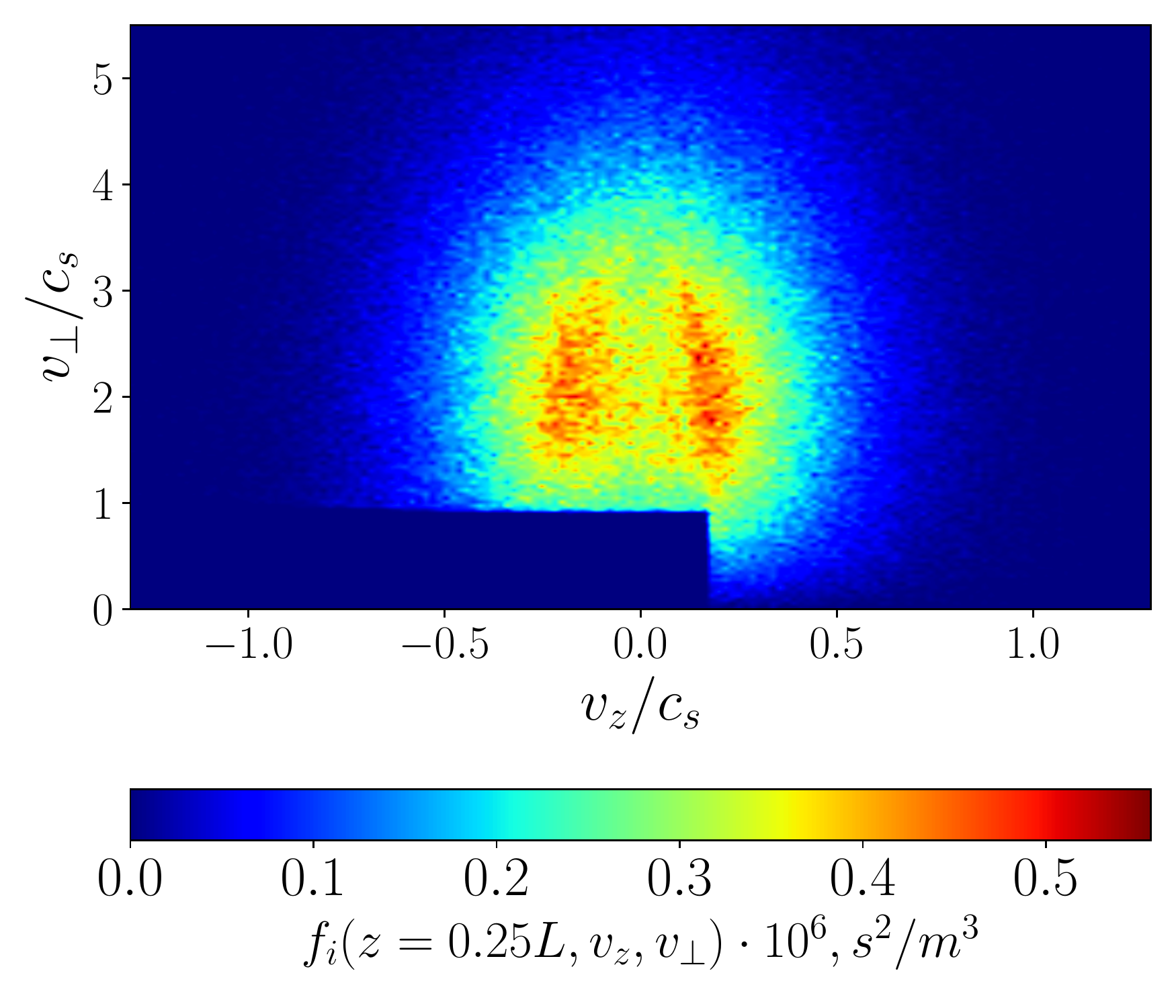}}
 \subfloat[$z/L=0.4$]{
 \includegraphics[width=0.36\linewidth]{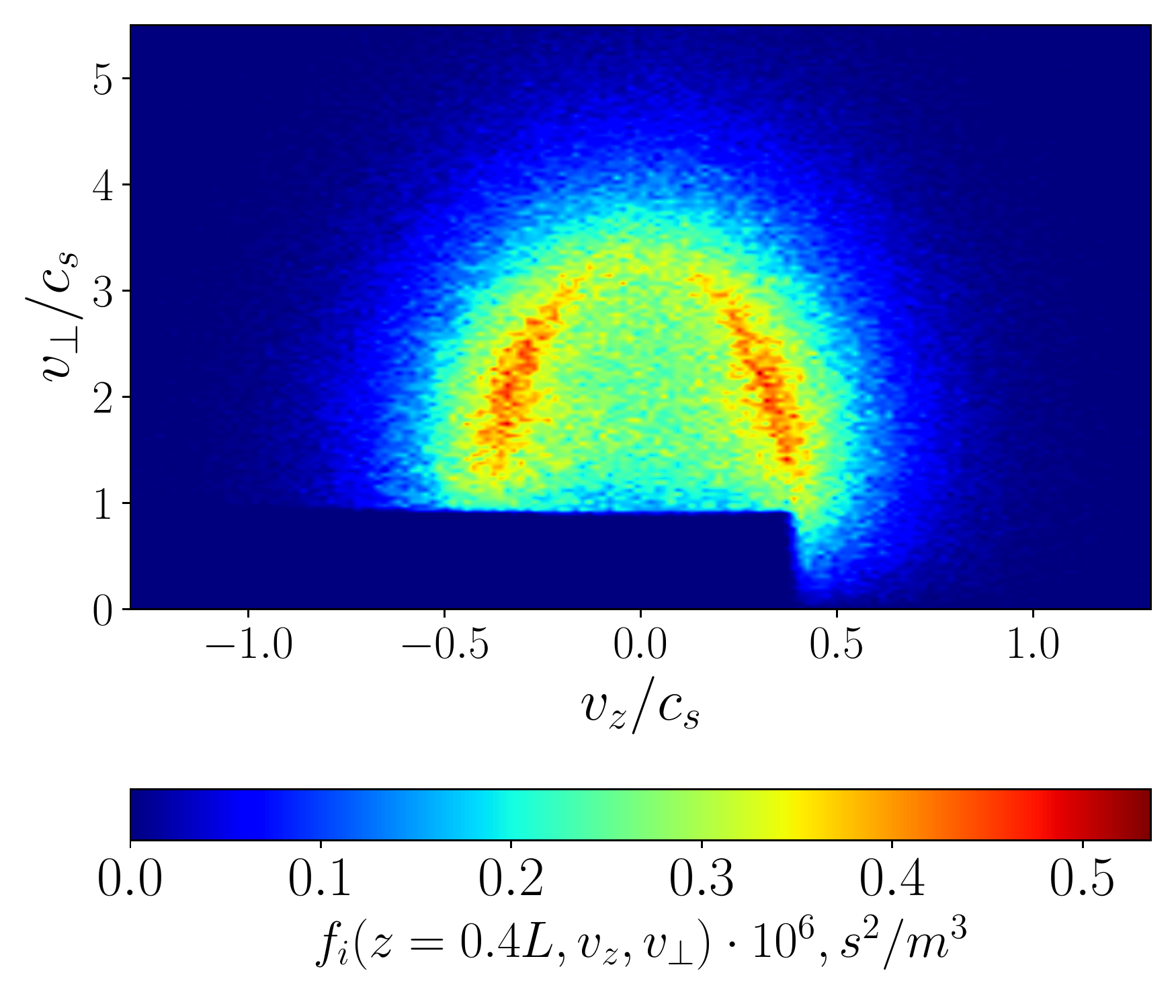}}
    \caption{Ion velocity distribution function for anisotropic injection, $T_{iz}=\SI{20}{eV}$ and $T_{\perp}=\SI{100}{eV}$ at different locations.}
    \label{anisotr_20_100}
\end{figure}

For the case with $T_{i\perp} < T_{iz}$, the appearance of non-monotonic potential structure (negative electric field) before the nozzle throat leads to reflections of ions with the low perpendicular energies, see Figs.~\ref{anisotr_100_20}.The Ion Velocity Distribution Function (IVDF) is compressed in parallel direction due to the negative electric field.
In case of low parallel temperature, $T_{i\perp} > T_{iz}$, due to acceleration in purely positive electric field, the initial IVDF spreads in $v_z$-direction, Fig.~\ref{anisotr_20_100}.

\section{Plasma source effects and transitions between  trapped and passing particles}\label{sec4}

\begin{figure}[H]
    \centering
\subfloat[]{
    \includegraphics[width=0.45\linewidth]{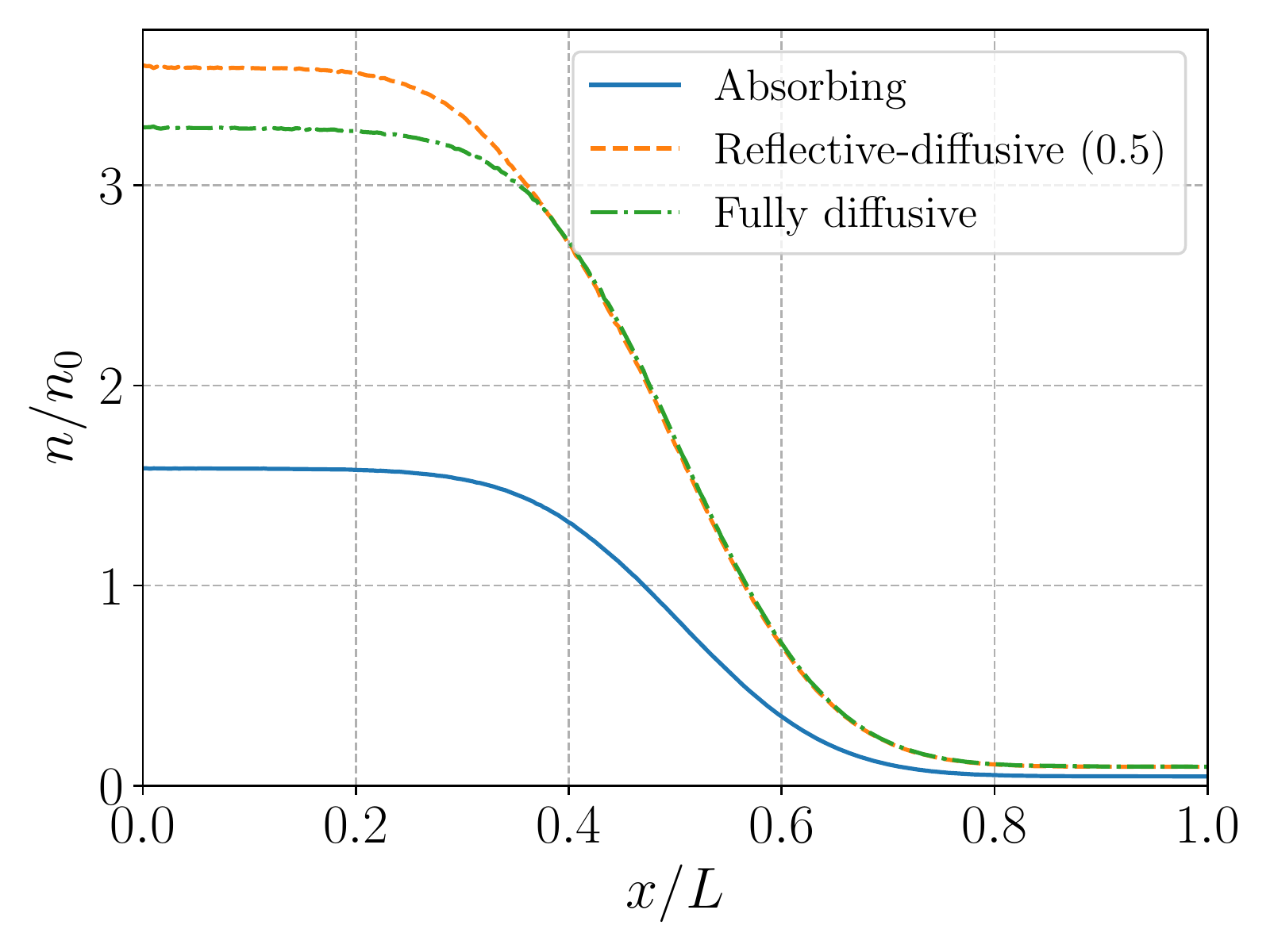}}
\subfloat[]{
    \includegraphics[width=0.45\linewidth]{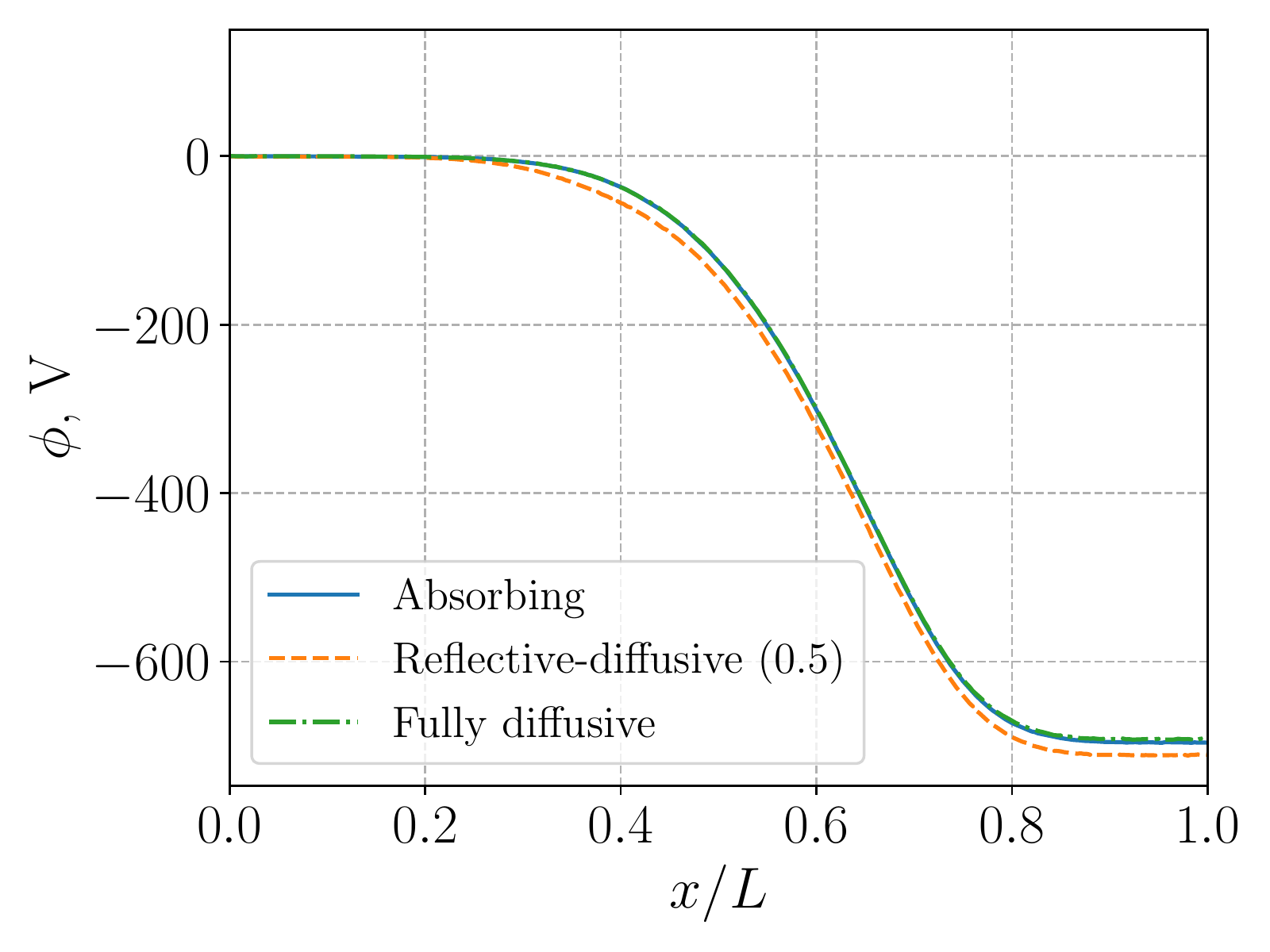}}
  \caption{ Spatial profiles of plasma density (a), and electrostatic potential (b) for different models of the source wall.}
    \label{we_nphi}
\end{figure}

\begin{figure}[H]
    \centering
\subfloat[]{
    \includegraphics[width=0.45\linewidth]{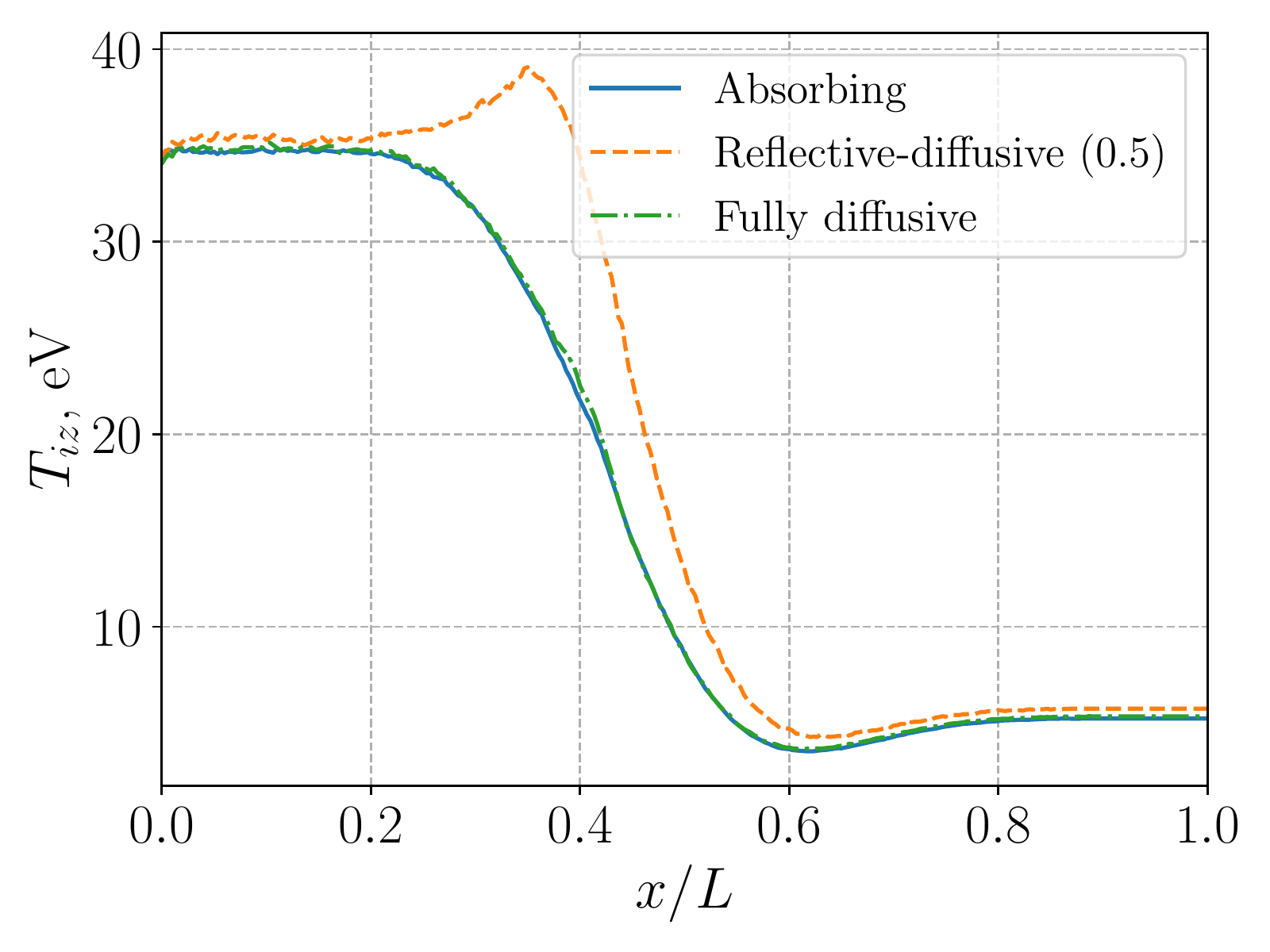}}
\subfloat[]{
    \includegraphics[width=0.45\linewidth]{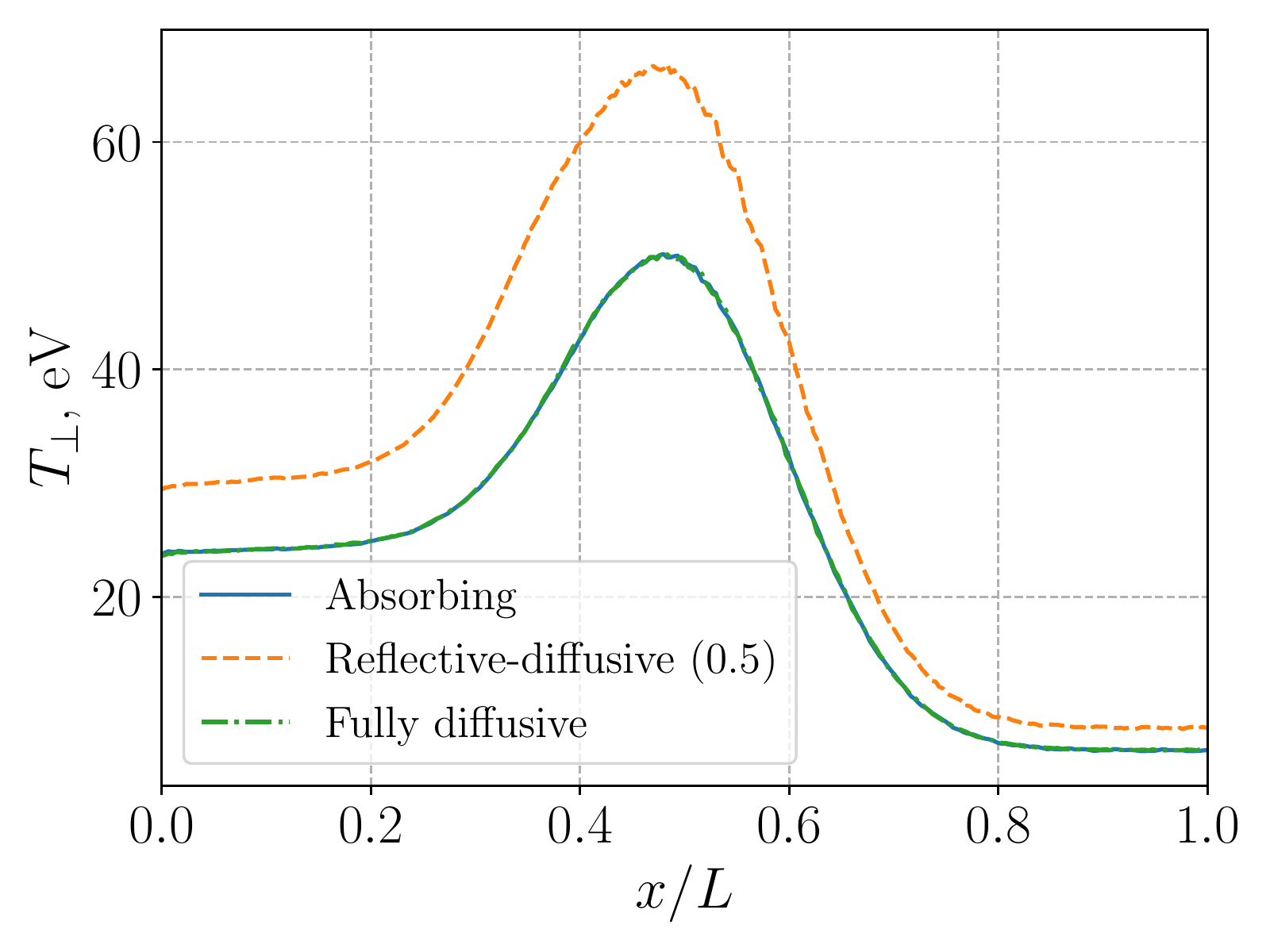}}
  \caption{Spatial profiles of parallel (a), and perpendicular (b) temperatures for different models of the source wall.}
    \label{we_temps}
\end{figure}

\begin{figure}[H]
    \centering
    \includegraphics[width=0.55\linewidth]{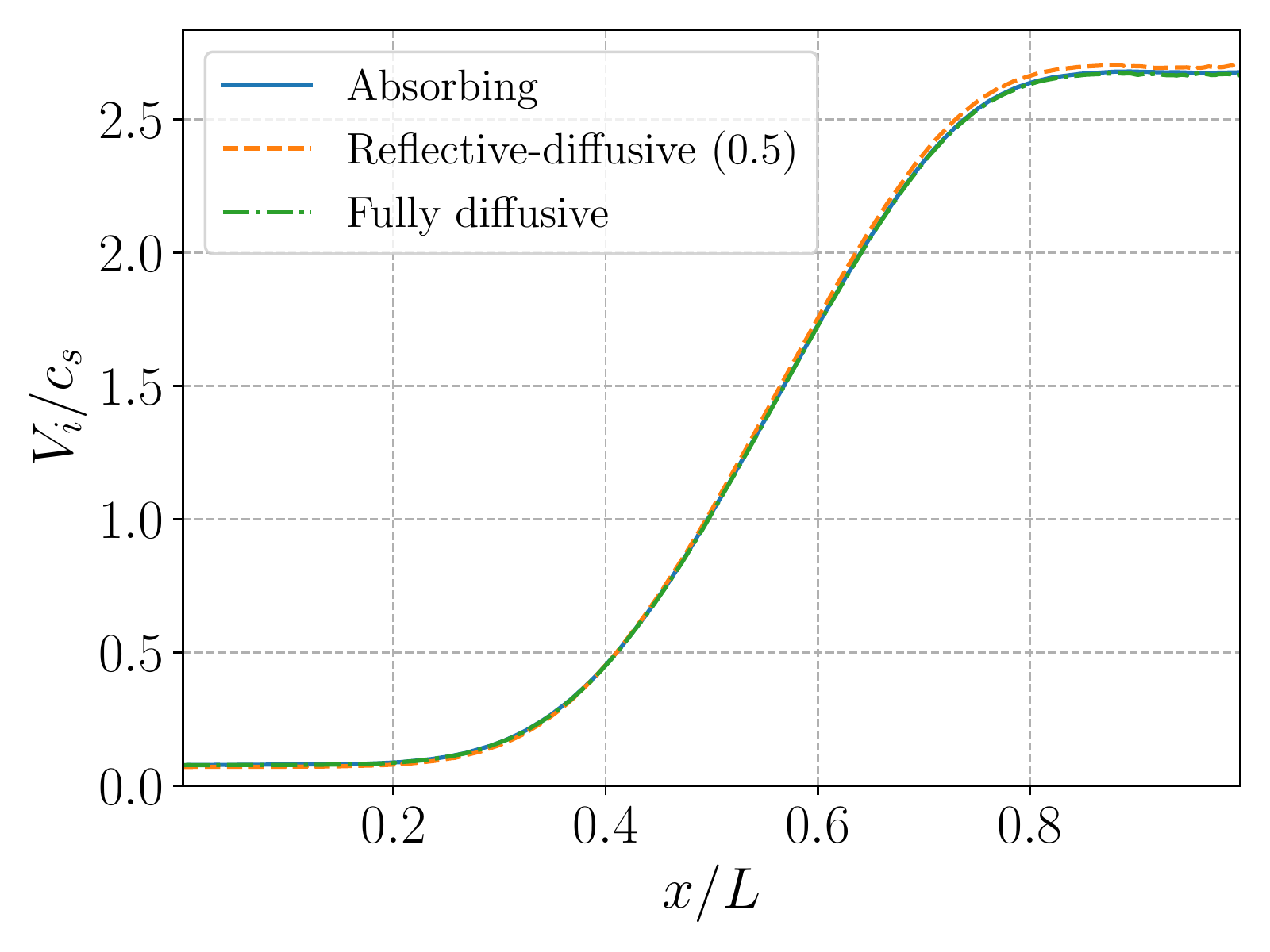}
  \caption{Spatial profile of flow velocity for different models of the source wall.}
    \label{we_vi}
\end{figure}

Our simulations are collisionless and do not include the transitions between trapped and passing particles due to collisions that are important to fill the loss cone \cite{BaldwinNF1972}.  Neither, we have included a self-consistent plasma  source in our simulations. Normally, the distribution function in the source would be established self-consistently as a result of the interactions of the specific heating mechanism, losses of passing particles, reflection of trapped particles and transitions between trapped  and passing particles. 
To mimic these processes we modify  the reflections at the left wall. When particles are reflected inside the system (e.g.\ by magnetic mirror force and/or electric potential fluctuations) and return to the left wall we can reflect a given fraction  of such particles. The reflections can be  diffusive or specular (mirror reflection). In case of diffusive reflections, particles are re-injected into the system with a randomly sampled reflection angle.  
With purely specular reflections we observe continuous increase  particles density near the source wall due to the increase in  number of trapped particles with continuous injections.  
Adding some diffusive reflections allows the transitions between trapped and passing particles therefore establishing the equilibrium  balance between injection, transitions to the passing region, and particle losses. This effect is illustrated in Fig.~\ref{we_nphi} which shows plasma density increase in the source region for the case when the combination of reflective and diffusive reflection is used  at the left wall in the ratio 0.5 to 0.5; half particles experience diffusive reflections and the other half are reflected specularly. When all particles are reflected diffusively, the density  is lower, but obviously larger compared to the case when full absorbing conditions are used.    
Figs.~\ref{we_nphi}-\ref{we_vi} show plasma parameters profiles for three cases:  fully diffusive,  half-and-half  specular and diffusive, and fully absorbing condition. In all cases the isotropic injection temperature  $\SI{20}{eV}$ is used. One should note that the potential, plasma density and ion velocity remain very similar in all cases as expected. Temperature profiles however are modified due to additional heating in diffusive reflections. 
% \begin{figure}[H]
%     \centering
%     \includegraphics[width=0.40\linewidth]{comp/navg_no_satur.png}
%   \caption{Plasma density, fully specular reflective wall.}
%     \label{specwall_dens}
% \end{figure}

% \begin{figure}[H]
%     \centering
%     \includegraphics[width=0.44\linewidth]{comp/dens_ti20_spec_refl_cidt40.pdf}
%     \includegraphics[width=0.45\linewidth]{comp/vi_ti20_spec_refl_cidt40.pdf} \\
%     \includegraphics[width=0.46\linewidth]{comp/phi_ti20_spec_refl_cidt40.pdf} 
%      \includegraphics[width=0.45\linewidth]{comp/temp_ti20_spec_refl_cidt40.pdf}
%   \caption{Plasma density and ion velocity a), electrostatic potential and temperatures b) profiles}
%     \label{spe1}
% \end{figure}

\section{Summary and discussion}\label{sec5}

In this work we have presented a quasineutral  hybrid model with kinetic ions, modeled with the particle-in-cell (PIC) approach, and fluid  electrons to describe plasma flow and acceleration in the paraxial magnetic mirror. Ions are described by the drift-kinetic equation while electrons follow Boltzmann distribution which is used to find the electric field.

Analytical fluid  theory \cite{smolyakov2021quasineutral} predicts that the global accelerating  velocity profile formed by the magnetic barrier is unique. The transonic solution crossing the sonic point is fully determined by the magnetic field profile and therefore has fixed velocity at the nozzle entrance boundary (as well as at any other point along the nozzle). Fully subsonic and supersonic solutions  are obtained  for low and large entrance velocities, while for some intermediate values, $M_a<v_0/c_s<M_b$, Fig.~\ref{fluid_solu}, solutions are multivalued and therefore do not exist in the fluid theory. Our  kinetic simulations reveal that the unique global transonic solution robustly persists in the kinetic regimes. For the boundary conditions with the velocities in the region of multi-valued solutions,  between $M_a$ and $M_b$, we observe development of turbulent fluctuations that reflect some incoming ions, while  passing ions are accelerated. There exist several mechanisms of the fluctuations. One is related to the break-up of the flow in the region when fluid solutions are  multi-valued and therefore develop the infinite derivative singularity in the velocity profile. Additional mechanism is provided by streaming type instabilities of counter propagating ion beams due to the reflections of some particles by the magnetic mirror force. Such instabilities are generally of the Buneman and ion-sound types as it is confirmed by the observed frequencies in the ion-sound range. 

It is interesting, that the stationary turbulent state is reached where the net flow velocity profile is in agreement with the prediction of the fluid theory.  With the boundary condition at the left boundary corresponding to  supersonic regimes,  $v_0/c>M_b$,  we obtain a smooth laminar  solution which remain supersonic across the nozzle in accordance with the fluid result. 

We were not able to obtain  the laminar solutions in the subsonic regimes with $v_0/c<M_a$: some level of  turbulent fluctuations always exist  in these regimes resulting in some reflections of ions and switching the solution into the transonic accelerating  mode. Our preliminary studies have shown little  sensitivity of these  fluctuations to the numerical parameters such as time step and grid size, therefore suggesting   that the nature of these fluctuations is not purely numerical. Stability  of fluctuations in the accelerated flow in the nozzle were discussed in various conditions \cite{ParkerAJ1966,VelliAJ1994,GrappinAJ1997,WebbJPP1999}. While globally accelerated flows, such as solar wind,  are considered to be stable \cite{ParkerAJ1966}, it was shown that background fluctuations can convectively grow exponentially in the subsonic (converging) part of the nozzle  \cite{ParkerAJ1966,GrappinAJ1997,GuioPhysRevE2020}. We conjecture that the fluctuations we observe is a result of such convective amplification of the inherent noise present in PIC simulations. The detailed analysis of such effects is left for future studies.

We have investigated the acceleration of warm plasmas with isotropic and anisotropic ion distribution function. We have shown that finite ion temperature, in particular, finite perpendicular temperature, increase  acceleration and final ion exhaust velocity from the nozzle. This result is consistent with the predictions of fluid theory for warm ions with anisotropic ion pressure \cite{SaboPoP2022}.  In all situations, the global transonic accelerating velocity profile appears as a global constrain on plasma flow in the nozzle, also revealing  the  generic mechanism  of the instabilities exciting turbulent fluctuations that force the net ion flow into a robust accelerating mode profile. The ``forcing'' occurs as a  time dependent process, where a fraction of ions are reflected by fluctuations in  the stationary turbulent state. In some regimes with a prescribed ion distrubution function, such as monoenergetic injection in the multi-valued region,  Fig. 5a, and anisotropic  injection, as in Figs 17b and 20b, the time average profile of the potential become non-monotonous.  Peaked non-monotonous profiles of the potential in the mirror-trapped plasma were shown to occur\cite{TurlapovPRE1998,SemenovFT1999} due to the anisotropic electron temperature, $T_{\perp e} \neq T_{\Vert e}$. Here, the non-monotonous potential is induced by anisotropy in the ion distribution function.  In general, the ion distribution function has to be determined taking into account the filling of the loss cone due to collisions \cite{BaldwinNF1972} and self-consistent electric field.

The existence of the ``universal'' constrain on the velocity in the magnetic nozzle has been demonstrated in our simulations with the absorbing-reflective source boundary (imitating the plasma source region) when a  fraction of returned  particles is diffusively reflected by the wall (at the source) therefore allowing transitions between trapped and passing particles. We observe formation of the global  ``universal'' transonic accelerating velocity and density profiles with the absolute value of the density determined self-consistently by the balance between particle injection rate  and the losses through the nozzle. The shape of the ``universal'' velocity profile is fixed by the magnetic field profile  and the value of the maximum velocity by the electron and ion temperature \cite{smolyakov2021quasineutral,SaboPoP2022}. In certain sense, the constrain on the velocity can be used similar to the Bohm condition on the ion velocity at plasma boundary that is often use to define equilibrium plasma density from the energy balance. More generally, full particle and energy balance should be considered together with the velocity constrain to define plasma parameters inside the source region.  

Acceleration of plasma flow in the nozzle/mirror magnetic field configurations has long  been studied with fluid \cite{Morozov_V8,WeitznerPF1977,RognlienPF1981,WangPoP2001,AndreussiAIP2010} and kinetic theory \cite{RamosPoP2018,MartinezPoP2011,AhedoPoP2001,AhedoPSST2020,WethertonPoP2021,SainiPoP2020}, see also references in Ref.~\onlinecite{KaganovichPoP2020}. Although the role of the sonic point in the formation of the global accelerating solution and some exact solutions were known \cite{FruchtmanPRL2006}, the constraints that follow from the existence of the unique solution and its consequences were not discussed. We have shown here that these constraints play important role determining the flow in the nozzle and therefore affecting overall particle and energy balance. These results are  important to determine the plasma and energy losses in fusion mirror systems and  efficiency of the propulsion systems using the magnetic nozzle.   
Our study does not consider the detachment problem \cite{HooperAIAA1993,BreizmanPoP2008,MerinoPSST2014} and we do not include electron anisotropy  and electron trapping effects \cite{WethertonPoP2021,MartinezPoP2011,AbramovNF2019} which potentially may be included with fluid type closures proposed for space plasmas and reconnection problems \cite{BoldyrevPNAS2020,WethertonGRL2019,EgedalPoP2013,LePoP2010}. The investigations of the role of these effects on the formation and stability of transonic flows are left for future studies.

\appendix

\section{Numerical details}
Unless stated otherwise, particles are injected form the left boundary (source wall) at $z=0$ with the flux-Maxwellian distribution parallel to the magnetic field lines and Maxwellian for its perpendicular components. The initial flux $\Gamma_0= n_0 V_0$, and the number of injected macroparticles per time step $C_{\Delta t}$, are specified as the input parameters. The particle weight is evaluated as
\begin{equation}
    w=\frac{\Gamma_0 A_0 \Delta t} {C_{\Delta t}},
\end{equation}
where $A_0=\SI{4e-3}{m^2}$ is the cross-sectional area at entrance, and $\Delta t$ is the time step. Assuming that the particles are perfectly magnetized and following the magnetic field lines, the varied cross-section area of  is used $A(z)B(z)=\text{const}$ to describe  two-dimensional effects\cite{ebersohn2017kinetic} . 

The leapfrog method is used for particle integration. The electric potential is evaluated from the Boltzmann relation~(\ref{boltzmann}) for electrons and full quasineutrality assumption, thus $\phi = T_e \ln n_i$ (where electron temperature $T_e$ is in \si{eV} units). The electric field is defined at cell centers and found via central difference scheme, with a free boundary condition at both ends:  $\partial_z^2 E = 0$.
The uniform mesh with $n_z+1$ nodal points is assumed, thus the cell size is $\Delta z = L / n_z$; the time step is set to $\Delta t = {\Delta z} / {v_{\text{max}}}$, where $v_{\text{max}} = N_{\text{max}} c_s$, with $N_{\text{max}} = 10$.
Lengths and time are normalized to $\Delta z$ and $\Delta t$, respectively, and other scaling parameters are:
\begin{eqnarray}
v_0 = \frac{\Delta z}{\Delta t}, E_0 = \frac{m_i v_0^2}{e \Delta z}, \ \phi_0 = E_0 \Delta z. \nonumber
\end{eqnarray}

Message Passing Interface (MPI) is used for parallel implementation, where the whole number of injected particle is distributed evenly among all the processors. Thus, each processor handles just a fraction of the total number of particles in the whole spatial domain. The communication at each time step occurs when the total plasma density is gathered in order to calculate the electric field and to integrate the particles motion for the next time step.

\begin{figure}[H]
    \centering
    \includegraphics[width=0.65\linewidth]{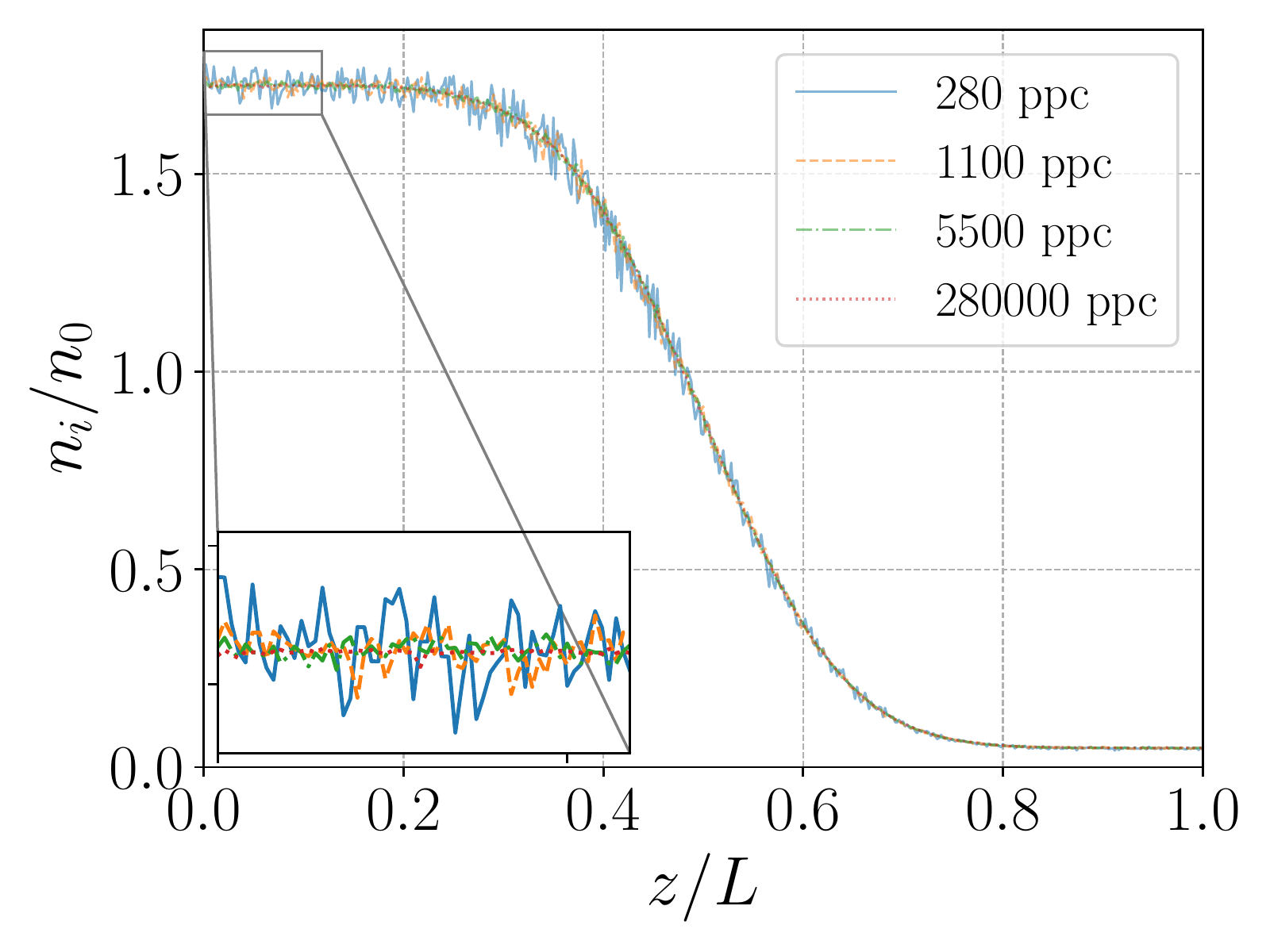}
    \caption{Spatial profile of plasma density for different values of PPC.}
    \label{conv_ni}
\end{figure}

The convergence with number of macroparticles per cell (PPC) for the isotropic case with $T_i=\SI{50}{eV}$ is shown in Fig.~\ref{conv_ni}. Four cases, using different number of PPC are performed in a whole domain of 500 cells, with a fixed $\Delta t$. For the simulations in this paper we used the parameter $C_{\Delta t} = \num{4000}$, which in this test converged to \num{280000} PPC (denoted by red line in the profile shown in Fig.~\ref{conv_ni}).
%to a total number of 140 927 630 macroparticles in the system or  .

\begin{acknowledgements}

This work was supported in part by NSERC Canada and the U.S. Air Force Office of Scientific Research FA9550-15-1-0226 and FA9550-21-1-0031. Computational resources were provided by Compute Canada/Digital Research Alliance of Canada. The authors thank the TAE Team and the investors of TAE Technologies for the discussions and their support. A.S. would like to acknowledge illuminating discussions with S.I. Krasheninnikov.

\end{acknowledgements}

 \section*{Data availability} Data generated in this study is available from the authors upon reasonable request.

\bibliography{REF}
 
\end{document}